\documentclass[a4paper,11pt]{article}
\usepackage{jheppub}
\usepackage{lineno}
\usepackage{subfigure}
\usepackage{graphicx,color,slashed}
\usepackage{xcolor}
\usepackage{tikz}
\usepackage{verbatim}% Include figure files
\usepackage{hyperref}
\usepackage{booktabs}
\usepackage{multirow}
\hypersetup{colorlinks=True, linkcolor=blue,citecolor=green}

\def\be{\begin{eqnarray}}
\def\ed{\end{eqnarray}}
\def\beq{\begin{equation}}
\def\eeq{\end{equation}}
\def\bea{\begin{eqnarray}}
\def\eea{\end{eqnarray}}
\def\non{\nonumber}

\def\dg{\dagger}

\arxivnumber{2407.05831}
\date{\today}

\begin{document}

\title{\bf \Large Revisiting for maximal flavor violating $Z'_{ e \mu}$ and its phenomenology constraints}

\author[a,b]{Jia Liu}
\author[b,a]{Muyuan Song}
\author[a]{Haohao Zhang}
\affiliation[a]{School of Physics and State Key Laboratory of Nuclear Physics and Technology, Peking University, Beijing 100871, China}
\affiliation[b]{Center for High Energy Physics, Peking University, Beijing 100871, China}
\emailAdd{jialiu@pku.edu.cn}
\emailAdd{muyuansong@pku.edu.cn(corresponding author)} 
\emailAdd{2101110115@stu.pku.edu.cn}

\date{\today}% It is always \today, today,

%{\begin{flushright}{xxx}
%\end{flushright}}
\abstract{
Lepton flavor violation (LFV), observed conclusively in neutrino oscillations, remains a pivotal area of investigation due to its absence in the Standard Model (SM). Beyond the Standard Model (BSM) physics explores charged lepton flavor violation (CLFV), particularly through new particle candidates such as the $Z'$. This article focuses on maximal LFV interactions facilitated by the $Z'$ boson, specifically targeting its off-diagonal interactions with the first and second generations of charged and neutral leptons. In our ultraviolet (UV) model for the origin of the $Z'$, inspired by the work of [R.Foot \textit{et al.,}~\href{https://arxiv.org/abs/hep-ph/9401250}{Phys.Rev. D50 (1994) 4571-4580}], we utilize the discrete $Z_2$ symmetry to investigate the maximal LFV mediated by the $Z'$ between the muon ($\mu$) and electron ($e$) arising from the additional scalars. This symmetry prohibits flavor-conserving interactions between $Z'$ and $\mu^+\mu^-,\, e^+e^-$. In conjunction with collider, $(g-2)_{\mu}, (g-2)_e$, inverse $\mu$ decay, Muonium-to-Antimuonium conversion and LFV decay constraints, we provide forecasts for anticipated limits derived from processes such as $\nu_\mu N \to \nu_e \mu^+ e^- N$ in neutrino trident experiments like the DUNE search at the first time. These limits highlight the prospective scope and significance of LFV investigations within these experimental frameworks. Within the mass range of 0.01 GeV to 10 GeV, the most stringent limit arises from $\mathcal{B} (\mu \to e + X + \gamma)$ when $M_{Z'} < m_\mu$, while $\Delta a_e$ provides effective constraints as $M_{Z'}$ approaches 10 GeV. Looking ahead, the proposed Muonium-to-Antimuonium Conversion Experiment (MACE) is expected to impose the most stringent constraints on Muonium-to-Antimuonium oscillation, improving sensitivity by about one order of magnitude against $\Delta a_e$.
}

\maketitle
%{\hypersetup{hidelinks}
%\tableofcontents
%}
\newpage
\section{Introduction}

Lepton Flavor Violation (LFV) holds considerable significance in particle physics, offering insights into fundamental interactions beyond the Standard Model (SM). The theoretical motivation for searching for LFV processes stems from the possibility that new symmetries might suppress non-conservative leptonic flavor interactions, rendering these phenomena less apparent~\cite{Feinberg:1959ui,Feinberg:1961zz,Cirigliano:2005ck,Calibbi:2017uvl}. Investigating LFV in novel particle interactions continues to be a dynamic and compelling area of research, despite the yet unrealized expectations, such as searches for axion-like particles (ALPs)~\cite{Bjorkeroth:2018dzu, Cornella:2019uxs, Calibbi:2020jvd,Cheung:2021mol,Calibbi:2024rcm,Davoudiasl:2024vje}. Moreover, LFV studies involving right-handed neutrinos with non-zero mass in SUSY~\cite{Hisano:1995cp, Abada:2014kba}, little Higgs models~\cite{Blanke:2007db,Han:2011aq}, seesaw (with EFT)~\cite{Ardu:2023yyw}, SMEFT~\cite{Altmannshofer:2023tsa,Altmannshofer:2022fvz}, B-LSSM~\cite{Huo:2024puy} and dark matter~\cite{Toma:2013zsa,Tapender:2024ktc} have spurred investigations into various lepton flavor violating interactions. In contrast, exploring smaller masses around the electroweak scale has prompted investigations both with and without new gauge symmetries, as highlighted in recent studies such as~\cite{xiaogang:1991a, xiaogang:1991b,Foot:1994vd,Iguro:2020rby,AtzoriCorona:2022moj,Crivellin:2023sig,Espinosa-Gomez:2023xrq,Eguren:2024oov,marin:2024LFV}, even offering potential explanations for discrepancies observed in quantities like $(g-2)_\mu$~\cite{Altmannshofer:2016brv,Kang:2019vng,Cheng:2021okr,Eijima:2023yiw}. There has been significant emphasis on new gauge interactions involving $U(1)'$, with researchers actively searching for flavor-conserving interactions between the $Z'$ boson and leptons, from theoretical studies to experimental investigations~\cite{Buras:2021btx,ATLAS:2018sky,Araki:2022xqp,ding2024study}. Special attention has been given to maximal flavor violation, where off-diagonal couplings are considered dominant, as highlighted in references~\cite{Foot:1994vd,CDF:2008zud,Foldenauer:2016rpi,Kriewald:2022erk}\footnote{Indeed, maximal flavor violation has been explored extensively in the quark sector, even in the absence of maximal flavor violation in the lepton sector~\cite{Bar-Shalom:2007xeu}.}. For experiment searches, neutrino experiments have undertaken several LFV searches, as labeled by references such as \cite{CHARM-II:1990dvf,CCFR:1991lpl, NuTeV:1999wlw}. In neutrino experiments, the new gauge interaction undergoes testing through a process known as neutrino trident production~\cite{Czyz:1964zz,Fujikawa:1971nx}, where an initial neutrino collides with stable nuclei, resulting in the production of a pair of charged leptons. Previous searches have primarily focused on pairs of $\mu$ with opposing charges~\cite{CHARM-II:1990dvf,CCFR:1991lpl, NuTeV:1999wlw}, extending to potential combinations with LHC energy regimes~\cite{Francener:2024wul,Altmannshofer:2024hqd}  and exploring new physics effects from atmospheric neutrinos~\cite{Ge:2017poy}. Meanwhile, investigations of neutrino energies ranging from TeV to PeV, such as those conducted by the IceCube detectors, have explored the dominant on-shell W boson and trident processes~\cite{Zhou:2019vxt,Zhou:2019frk}. Nevertheless, maximal mixing between $\mu$ and $e$ flavors could produce a clear signal without background interference. Such distinct flavor signatures considerably simplify the analysis compared to previous searches. The proposed explorations on DUNE intend to investigate final states featuring muon-electron pairs within SM~\cite{Ballett:2018uuc}, thereby augmenting sensitivity to lepton flavor-violating interactions. From a theoretical standpoint, Ref.~\cite{Foot:1994vd}, emphasized maximal LFV involving the $Z'\bar{\mu}\tau$ interaction, suggesting equivalence in the effects on the Yukawa couplings $H\bar{\mu}\mu$ and $H\bar{\tau}\tau$. However, recent Higgs measurements contradict this assumption. On the other hand, the recent Higgs decay measurement still survives the scenario where $H\bar{\mu}\mu, H\bar{e}e$ couplings to be identical in the presence of LFV $Z' \mu^{\pm}e^{\mp}$ within similar models. Therefore, exploring $\mu$ and $e$ flavor interactions proves crucial for advancing LFV investigations from both experimental and theoretical perspectives.

In this study, we explore the implications of neutrino trident production limits on maximal lepton flavor violation (LFV) involving $Z'$ in the DUNE experiment. This is the first time that the maximal LFV of the gauge interaction between $Z'_{e\mu}$ within the neutrino trident context has been considered. We demonstrate that the coupling constraints imposed by the neutrino trident process on LFV interactions between $\mu$ and $e$ are ranged from $\mathcal{O} (10^{-5}) $ to $ \mathcal{O}(10^{-2})$ within $m_\mu < M_{Z'} < 10$ GeV. In the lower mass regime, the most restrictive limit originates from the inclusive search for the $\mathcal{B} (\mu \to e + \gamma + X)$ process when $M_{Z'} < m_\mu$ while the most stringent limit is imposed by the $\Delta a_e$ when $m_\mu < M_{Z'} < 10$ GeV.

The paper is structured as follows: Section \ref{sec:2} clarifies the simplified $Z'$ model, emphasizing its dominant new off-diagonal gauge interaction and discussing its implications for charged and neutral leptons. Additionally, we introduce a UV model in Section \ref{sec:2} that comprises three scalar doublets and an additional singlet, where the singlet is pivotal in providing mass for the maximal LFV $Z'$. Section \ref{sec:3} addresses the LFV constraints on $Z'$, incorporating findings from collider experiments, LFV $\mu$ decays, neutrino-charged lepton scattering (inverse $\mu$ decay), and the neutrino trident process. Additionally, we analyze the implications of LFV couplings for $(g-2)_{\mu,e}$. In Section \ref{sec:4}, we depict exclusion limit plots derived from the couplings between $Z'$ and leptons. Finally, we conclude in Section \ref{sec:5}.

\section{Maximal LFV $Z'$ model }
\label{sec:2}

In models incorporating an additional $Z'$ boson, maintaining a global anomaly-free symmetry is essential within the SM framework. Previous studies of such models have concentrated on interactions between the $Z'$ boson and leptons, highlighting scenarios where a discrete symmetry prevents diagonal couplings~\cite{Foot:1994vd}. This precaution is crucial to prevent mixing between the $A$ and $Z'$ bosons, as described by transformations within the kinetic Lagrangian,
\bea
\mathcal{L} = -\frac{1}{4} F_{\mu\nu} F^{\mu\nu} - \kappa F_{\mu\nu}F^{'\mu\nu} - \frac{1}{4} F'_{\mu\nu} F^{'\mu\nu},
\eea
where $B_\mu \to B_\mu, Z'_\mu \to - Z'_\mu$. Under the $Z_2$ discrete symmetry, the above mixing between $A$ and $Z'$ is forbidden.

To satisfy the anomaly-free requirements, we assign distinct $U(1)'$ charges to different generations of leptons. Specifically, we consider a scenario where these $U(1)'$ charge assignments apply to the first and second generations while the third generation remains uncharged under $U(1)'$. Alternatively, opposite charges could be assigned to any other pair of generations, such as $\mu/\tau$ or $e/\tau$. In such cases, the flavor states under $SU(3) \times SU(2) \times U(1)_Y \times U(1)'$ are structured as described in Ref.~\cite{Foot:1994vd},
\bea
\ell_{1L} &=& \frac{(e+\mu)_L}{\sqrt{2}}, \ell_{2L} = \frac{(e-\mu)_L}{\sqrt{2}}, e_{1R} = \frac{(e+\mu)_R}{\sqrt{2}},e_{2R} = \frac{(e-\mu)_R}{\sqrt{2}},\non \\
\ell_{1L} &\thicksim & (\mathbf{1},\mathbf{2})(-1,2a), \ell_{2L} \thicksim   (\mathbf{1},\mathbf{2})(-1,-2a), \non \\ e_{1R} &\thicksim & (\mathbf{1},\mathbf{1})(-2,2a), e_{2R} \thicksim (\mathbf{1},\mathbf{1})(-2,-2a),
\eea
where $a$ is a parameter for the $U(1)'$ charge, $\ell_{1L,2L}$ and $e_{1R,2R}$ are the flavor states, while $e$ and $\mu$ are the eigenstates in the mass basis.
With the discrete symmetry ($\ell_{1L} \to \ell_{2L}, e_{1R} \to e_{2R}, Z'_\mu \to - Z'_\mu$) in place, the interaction between $Z'$ and flavor states would demonstrate the maximal mixing feature for lepton mass states:
\bea\label{eq:ZemuL}
\mathcal{L}^{Z'}_{\rm{NC}}  &=& g_{e\mu}\,  Z'_\mu ( \bar{\mu} \gamma^\mu e + \bar{e} \gamma^\mu \mu  ) ,
\eea
and neutrino-$Z'$ interaction, 
\bea \label{eq:ZnuL}
\mathcal{L}^{Z'}_{\rm{NC}}  &=& g_{e\mu}\, Z'_\mu ( \bar{\nu}_{\mu L} \gamma^\mu \nu_{eL}  + \bar{\nu}_{eL} \gamma^\mu \nu_{\mu L}  ) ,
\eea
where the $Z'$ does not couple to the diagonal term in the mass basis.

\begin{figure}[!h]
    \centering
    \includegraphics[width=0.7\textwidth]{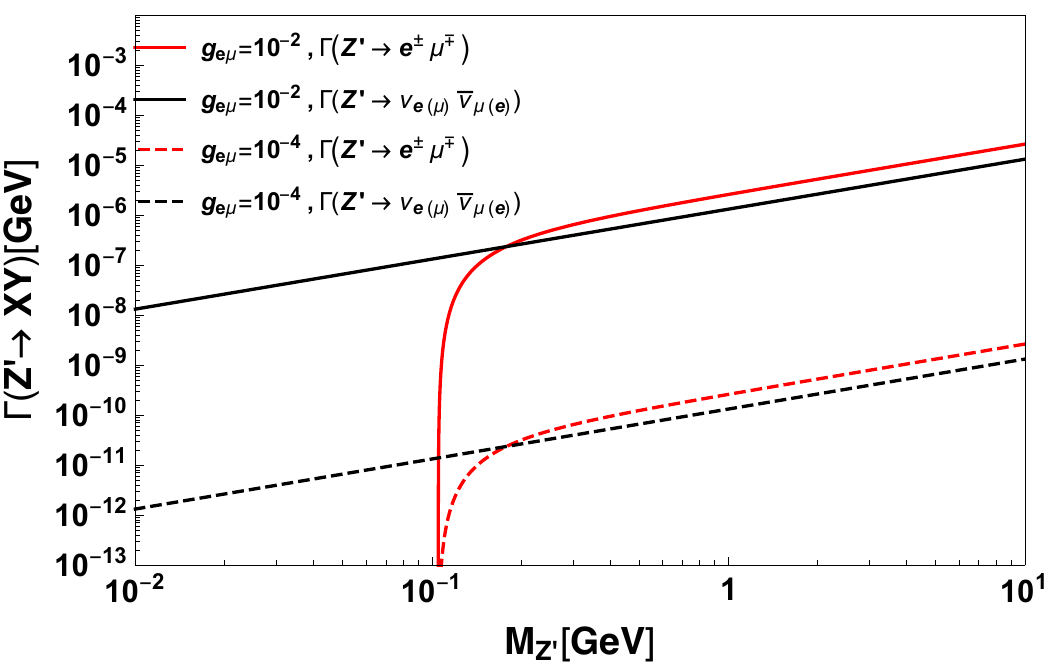}
    \caption{The decay width of $Z'$ correlated with charged and neutral leptons with $g_{e\mu} =10^{-2} (\text{solid}), 10^{-4} (\text{dashed})$. The red lines label the $\Gamma (Z' \to e^\pm \mu^\mp)$ and the black lines label the $\Gamma(Z' \to \nu_{e(\mu)} \bar{\nu}_{\mu(e)}) $.  }
    \label{fig:LFVdecay}
\end{figure}

Next, the decay width of the $Z'$ can be straightforwardly calculated from the above interaction, considering the limit where $m_e \rightarrow 0$,
\bea
\Gamma(Z' \to e^{\pm} \mu^{\mp}) &=& \frac{g^2_{e\mu} (M^2_{Z'} -m^2_\mu )^2 (2 M^2_{Z'} + m^2_\mu)}{24 \pi M_{Z'}^5} \label{eq:Zemu},\\
\Gamma(Z' \to \nu_{e(\mu)} \bar{\nu}_{\mu(e)}) &=& \frac{g^2_{e\mu}M_{Z'}}{24\pi} \label{eq:Znuemue},
\eea
Fig.~\ref{fig:LFVdecay} depicts the decay widths of $Z'$ across a mass range from 0.01 GeV to 10 GeV, with $g_{e\mu} =10^{-2}$ (solid line) and $10^{-4}$ (dashed line). It becomes apparent that the two decay channels become comparable when $M_{Z'}$ is significantly larger than $m_\mu + m_e$. Furthermore, for large $M_{Z'}$, the decay width into neutrinos is half that of the charged lepton decay,
\bea
2 \,\mathcal{B}(Z' \to \nu_{e(\mu)} \bar{\nu}_{\mu(e)}) =
\mathcal{B} (Z' \to e^{\pm} \mu^{\mp}) ,
\eea
because the neutrinos only have left-handed spinors.

\subsection{An ultraviolet (UV) model with maximal LFV $Z'$}

The preceding section examines a simplified LFV $Z'$ and explores its new gauge interaction. This prompts the inquiry into how such a $Z'$ could be generated within a UV framework. In our UV model, we establish the existence of a model comprising three scalar doublets and an additional real singlet, with specific charge assignments outlined in Tab.~\ref{tab:gaugecharge} as extended by Ref.~\cite{Foot:1994vd}.

\begin{table}[!h]
    \centering
    \begin{tabular}{ccccccccc}
    \hline
    \hline
      Group  & $\ell_{1L}$ & $\ell_{2L}$  &$e_{1R}$&$e_{2R}$ & $\phi_{1}$ & $\phi_2$ & $\phi_3$ & $\phi_4$   \\
      \hline 
      $ SU(3)_c $   & $ \textbf{1} $ &  $\textbf{1} $  &
       $\textbf{1} $ & $\textbf{1} $ & $ \textbf{1} $ &  $ \textbf{1} $ & $ \textbf{1}$ &  $\textbf{1} $ \\
      $ SU(2)_L  $ & $ \textbf{2} $ & $ \textbf{2} $  & $ \textbf{1} $ & $ \textbf{1} $ & $\textbf{2} $ &  $\textbf{2} $ &  $ \textbf{2} $ & $ \textbf{1} $ \\
      $ U(1)_Y $  & -1 &  -1  & -2& -2 & 1& 1&  1 &  0 \\
       $U(1)' $   & 2a &  -2a  & 2a& -2a & -4a & 4a&  0 & $ 1 $ \\
       \hline
       \hline
    \end{tabular}
    \caption{Particle contents and charges in the $U(1)'$ model with $\ell_{1L,2L},e_{1R,2R}$ and four scalars $\phi_{1,2,3,4}$.}
    \label{tab:gaugecharge}
\end{table}

The electromagnetic neutral components of three scalar doublets and one singlet are given in the following,
\bea 
\phi_1 &\supset& v_1 + \left(   \rho_1 + i \eta_1 \right) /\sqrt{2} ,\, \phi_2 \supset v_2 + \left( \rho_2 + i \eta_2\right)/\sqrt{2} ,\,\non\\
\phi_3 &\supset& v_3 +\left( \rho_3 + i \eta_3\right)/\sqrt{2},\,
\phi_4 = v_4 +   \rho_4  /\sqrt{2},
\eea
where we will propose the discrete symmetry $1 \leftrightarrow 2$ ($\ell_{1L} \to \ell_{2L}, e_{1R} \to e_{2R}, \phi_1 \to \phi_2$), thereby enforcing $v_1 = v_2 \equiv v_0$. In contrast to Ref.~\cite{Foot:1994vd}, we introduced an additional singlet scalar $\phi_4$, which is charged solely under $U(1)'$. The possible scalar potential is given in Appendix~\ref{app:A}.

On the other side, the different generation of leptons have exactly opposite charges under $U(1)'$ in Table~\ref{tab:gaugecharge} , which ensures the anomaly free condition. Therefore, the Yukawa interaction for leptons under the above charge assignments would be expressed as follows~\cite{Foot:1994vd},
\bea
\mathcal{L}^{\ell}_{\text{Yukawa}} &=& \mathcal{Y}_1 \bar{\ell}_{1L} e_{1R} \phi_3 + \mathcal{Y}_1 \bar{\ell}_{2L} e_{2R} \phi_3 + \mathcal{Y}_3 \bar{\ell}_{3L} e_{3R} \phi_3 \nonumber \\ 
&+& \mathcal{Y}_{12} (\bar{\ell}_{1L} e_{2R} \phi_2 + \bar{\ell}_{2L} e_{1R} \phi_1) + \rm{h.c},
\eea
where the Yukawa sector obeys the discrete symmetry $1 \leftrightarrow 2$. The scalar $\phi_4$ does not participate in Yukawa interactions due to its specific charge assignment. Instead, it contributes solely to the mass generation of the $Z'$ boson through the process of symmetry breaking. As detailed in Appendix~\ref{app:A}, the mass eigenstate $H$ emerges from the diagonalization of mass matrices, particularly when the flavor state $\rho_3$ aligns with the physical state $H$, facilitated by setting $\lambda_4 \to 0$. The SM-like Higgs ($H$) exhibits Yukawa couplings with leptons as follows,
\bea
\mathcal{L}^{\ell}_{H} = \frac{m_\tau}{\sqrt{2} v_3} \bar{\tau} \tau H  + \frac{m_\mu + m_e}{2\sqrt{2} v_3} (\bar{\mu}\mu + \bar{e}e) H, 
\label{eq:H-e-mu-simple}
\eea
where the off-diagonal Yukawa terms vanish. This cancellation occurs because the terms $(\bar{\ell}_{1L} e_{2R})$ and $(\bar{\ell}_{2L} e_{1R})$ eliminate the off-diagonal contributions:
\bea
&& \bar{\ell}_{1L} e_{2R} \phi_2 + \bar{\ell}_{2L} e_{1R} \phi_1 + \text{h.c} \non\\ &=& \frac{1}{2\sqrt{2}} ( \bar{e}_L e_R \mathcal{Y}_{12} +\bar{e}_L \mu_R \mathcal{Y}_{12} - \bar{\mu}_L e_R \mathcal{Y}_{12} - \bar{\mu}_L \mu_R \mathcal{Y}_{12}) \rho_1 \non\\
&+& \frac{1}{2\sqrt{2}} ( \bar{e}_L e_R \mathcal{Y}_{12} -\bar{e}_L \mu_R \mathcal{Y}_{12} + \bar{\mu}_L e_R \mathcal{Y}_{12} - \bar{\mu}_L \mu_R \mathcal{Y}_{12}) \rho_2 \non\\
&+& \frac{\sqrt{2}}{2} ( \bar{e}_L e_R \mathcal{Y}_{1}  + \bar{\mu}_L \mu_R \mathcal{Y}_{1}  ) \rho_3 +  \text{h.c},
\eea
Applying the linear transformation from $\rho_1, \rho_2 $ to the neutral states $S_{1},S_{2}$,
\bea \label{eq:HpHmtor1r2}
 S_{1} = \frac{\rho_1 + \rho_2}{\sqrt{2}},  S_2 = \frac{\rho_1 - \rho_2}{\sqrt{2}}, 
\eea 
results in the elimination of off-diagonal Yukawa-type interactions involving $S_1$. All off-diagonal interactions are transferred to the additional scalar $S_2$, which we note could be significantly heavier due to large number of parameters from scalar potential and thus not of primary interest in this context. The Yukawa couplings from the Lagrangian above display a distinctive feature wherein the couplings for $\mu$ and $e$ are identical. In Ref.~\cite{Foot:1994vd}, the model presents a unique aspect where the Yukawa couplings between  $H-\bar{\mu}\mu$ and $H-\bar{\tau}\tau$ are equal. However, the measurement of Higgs decay to muon and tau leptons have already ruled out this possibility~\cite{ATLAS:2022vkf,CMS:2022dwd}. Hence, our attention is directed exclusively towards the mixing between $\mu$ and $e$, given the current relatively low precision in measurements of $H-\bar{\mu}\mu$ and $H-\bar{e}e$. Latest measurements of the branching ratios for $\mathcal{B} (H \to \mu^+ \mu^-) < 2.6\times 10^{-4}$ by ATLAS~\cite{ATLAS:2022vkf} and $\mathcal{B} (H \to e^+ e^- ) < 3\times 10^{-4}$ by CMS~\cite{CMS:2022urr} suggest that setting the Yukawa couplings for both electrons and muons to $\mathcal{Y}_e \sim \mathcal{Y}_\mu < \mathcal{O}(10^{-4})$ would not impose stringent constraints on the model.

Conversely, following the transformation in Eq.~(\ref{eq:HpHmtor1r2}), the CP-even squared mass matrices result in $S_1, \rho_3$ being non-diagonal when $\lambda_4 \neq 0$, indicating that $S_1, \rho_3$ are not mass eigenstates. Instead, after diagonalization (as shown in Eqs.~(\ref{eq:l4neq0mass} and \ref{eq: mixH})), the mass eigenstates are $S'_1,H$ respectively. The transformation from flavor states ($\rho_1,\rho_2, \rho_3,\rho_4$) to mass states ($S'_1,S_2 ,H,S_4$) leads to the new SM Yukawa interaction between $H$ and leptons, represented as follows (where $S_2$ and $S_4$ are already the mass eigenstates):
\bea
\mathcal{L}^{\ell}_{H} &=& \frac{m_\tau}{\sqrt{2} v_3} \bar{\tau} \tau H + \left[\frac{m_{e}+m_{\mu}}{2\sqrt{2}v_{3}}+\frac{\lambda_{4}^2}{(\lambda_{1}+\lambda_{4})^2}\frac{v_{0}}{v_{3}} \frac{m_{\mu}-m_{e}}{2\sqrt{2}v_{3}}\right] \bar{\mu}\mu H \nonumber\\ &&+ \left[\frac{m_{e}+m_{\mu}}{2\sqrt{2}v_{3}}+\frac{\lambda_{4}^2}{(\lambda_{1}+\lambda_{4})^2}\frac{v_{0}}{v_{3}} \frac{m_{e}-m_{\mu}}{2\sqrt{2}v_{3}}\right] \bar{e}e H.
\eea
Once again, the off-diagonal interaction terms vanish for the same reasons as previously mentioned. It is noteworthy that the Yukawa couplings between $H$ and $\mu,e$ are no longer identical. However, directly converting the formal expression back to the SM $\mu$ and $e$ Yukawa couplings would require setting $v_0 = v_3$, which would affect the Yukawa couplings for other fermions, such as the $\tau$ Yukawa coupling. Instead, simplifying this scenario involves setting $\lambda_4 = 0$ to reduce one parameter of the model. In this case, the Yukawa couplings for $\mu$ and $e$ become nearly equivalent and revert to the formal condition in Eq.~\eqref{eq:H-e-mu-simple}.

From the $U(1)'$ covariant derivative ($D_\mu = \partial_\mu  + i g' Q_{U(1)'} Z^{'}_\mu$), the gauge interaction between $Z'$ and lepton flavor states can be easily expressed as follows,
\bea
\mathcal{L}^{Z'}_{\text{gauge}} = g_{e\mu} ( \bar{\ell}_{1L} \gamma^\mu \ell_{1L} - \bar{\ell}_{2L} \gamma^\mu \ell_{2L} +   \bar{e}_{1R} \gamma^\mu e_{1R} -  \bar{e}_{2R} \gamma^\mu e_{2R}) \,,
\eea
where $g_{e\mu} =g'\, 2a$ is the product of charge and coupling under the $U(1)'$. According to the $U(1)'$ charge assignments, $\phi_3$ would not couple with $U(1)'$ gauge boson and $\phi_4$ only connects with $Z'$. After electroweak symmetry breaking associated with $U(1)'$, the masses of the SM $Z$ boson and $Z'$ are determined from the covariant derivative,
\bea
|D_\mu \phi_1|^2 + |D_\mu \phi_2|^2 + |D_\mu \phi_3|^2 &\supset & \frac{1}{2} M^2_{Z} Z^\mu Z_\mu, \non\\
|D_\mu \phi_1|^2 + |D_\mu \phi_2|^2 + |D_\mu \phi_4|^2 &\supset & \frac{1}{2} M^2_{Z'} Z'^\mu Z'_\mu, \non \\
M^2_{Z} = \frac{1}{4} (g^2+ g^2_Y) (v^2_3 + 2\,v^2_0),\,\, M^2_{Z'} &\sim& g^2_{e\mu} (2 \,v^2_0 + v^2_4) \sim  g^2_{e\mu}\,v^2_{Z'},
\eea
where $v_0 < v_3$. By taking $(v^2_3 + 2 v^2_0) \approx v^2_{\text{SM}}$, this ensures that $M_Z$ remains unchanged while allowing $M_{Z'}$ to be a free parameter through varying the value of $v_{Z'}$. In comparison to Ref.~\cite{Foot:1994vd}, their model features $M_{Z'}$ strongly constrained by $M_{Z}$, resulting in no available region due to both Higgs-Yukawa and $M_{Z}$ constraints. 

\section{LFV limits}\label{sec:3}

In this section, our attention turns to the phenomenology of low mass $Z'$, where we conduct a thorough analysis to constrain $g_{e\mu}$ and $M_{Z'}$. Initially, we consider the anomalous magnetic moment ($g-2$) of the muon/electron and the constraints provided by electron-positron colliders such as the Belle-II experiment and LEP measurements. Then, we consider the LFV decays ($\mu$ decays) since the majority interaction between $Z'$ with leptons are $\mu$ and $e$. Thirdly, inverse $\mu$ decay and Muonium-anti Muonium oscillation are taking into account. Finally, the neutrino trident process would be discussed and we will present the difference between SM process and distinctive signal from the new $Z'$ contribution.   

\subsection{$(g - 2)_\mu$ and $(g - 2)_e$}
According to the window observable theory \cite{Ce:2022kxy, Colangelo:2022vok} and the SM prediction \cite{Acaroglu:2023cza}, the current disagreement concerning the $\Delta a_\mu$ designations has emerged as follows~\cite{Muong-2:2023cdq, Armando:2023zwz, Acaroglu:2023cza, Borsanyi:2020mff}: 
\bea\label{eq:amu}
\Delta a^{\rm{window}}_\mu =  (1.81 \pm 0.47) \times 10^{-9},
\eea
On the other hand, the electron magnetic dipole moment~\cite{Giudice:2012ms} with its experiment measurement through Rb in 2022~\cite{Fan:2022eto}, the discrepancy with SM prediction can be expressed as follows:
\bea \label{eq:ae}
\Delta a^{\text{Rb}}_e = 34(16) \times 10^{-14} ,
\eea
whereas the Cs atoms measurement method search has obtained a lower bound~\cite{Parker:2018vye},
\bea
\Delta a^{\text{Cs}}_e = - 102(26)   \times 10^{-14} ,
\eea
\begin{figure}
    \centering
\tikzset{every picture/.style={line width=0.75pt}} %set default line width to 0.75pt        
\begin{tikzpicture}[x=0.75pt,y=0.75pt,yscale=-1,xscale=1]
%uncomment if require: \path (0,683); %set diagram left start at 0, and has height of 683

%Shape: Wave [id:dp8423433343747229] 
\draw   (223.37,342.97) .. controls (220.74,341.77) and (218.23,340.61) .. (218.24,339.25) .. controls (218.26,337.89) and (220.79,336.71) .. (223.44,335.48) .. controls (226.1,334.24) and (228.63,333.06) .. (228.64,331.7) .. controls (228.65,330.34) and (226.14,329.19) .. (223.51,327.98) .. controls (220.88,326.77) and (218.37,325.61) .. (218.38,324.25) .. controls (218.39,322.89) and (220.92,321.71) .. (223.58,320.48) .. controls (226.24,319.24) and (228.76,318.06) .. (228.78,316.7) .. controls (228.79,315.34) and (226.28,314.19) .. (223.65,312.98) .. controls (221.01,311.77) and (218.5,310.61) .. (218.52,309.25) .. controls (218.53,307.9) and (221.06,306.71) .. (223.71,305.48) .. controls (226.37,304.24) and (228.9,303.06) .. (228.91,301.7) .. controls (228.93,300.34) and (226.42,299.19) .. (223.78,297.98) .. controls (221.36,296.86) and (219.04,295.79) .. (218.7,294.57) ;
%Shape: Wave [id:dp33989878997555345] 
\draw   (251.85,386.56) .. controls (250.5,386.61) and (249.45,389.14) .. (248.35,391.8) .. controls (247.25,394.46) and (246.2,396.99) .. (244.84,397.03) .. controls (243.49,397.08) and (242.2,394.63) .. (240.85,392.05) .. controls (239.51,389.47) and (238.22,387.02) .. (236.86,387.07) .. controls (235.51,387.11) and (234.46,389.64) .. (233.36,392.3) .. controls (232.26,394.96) and (231.21,397.49) .. (229.85,397.53) .. controls (228.5,397.58) and (227.21,395.13) .. (225.86,392.55) .. controls (224.51,389.98) and (223.23,387.52) .. (221.87,387.57) .. controls (220.51,387.62) and (219.46,390.15) .. (218.37,392.8) .. controls (217.27,395.46) and (216.22,397.99) .. (214.86,398.04) .. controls (213.51,398.08) and (212.22,395.63) .. (210.87,393.05) .. controls (209.52,390.48) and (208.24,388.03) .. (206.88,388.07) .. controls (205.52,388.12) and (204.47,390.65) .. (203.37,393.31) .. controls (202.28,395.96) and (201.23,398.49) .. (199.87,398.54) .. controls (198.51,398.58) and (197.23,396.13) .. (195.88,393.56) .. controls (194.53,390.98) and (193.24,388.53) .. (191.89,388.57) .. controls (191.21,388.6) and (190.61,389.24) .. (190.04,390.21) ;
%Straight Lines [id:da21213904183352428] 
\draw    (172.34,415.65) -- (196.35,380.75) ;
\draw [shift={(184.35,398.2)}, rotate = 124.53] [fill={rgb, 255:red, 0; green, 0; blue, 0 }  ][line width=0.08]  [draw opacity=0] (8.93,-4.29) -- (0,0) -- (8.93,4.29) -- cycle    ;
%Straight Lines [id:da04442186133355064] 
\draw    (196.35,380.75) -- (222.19,342.26) ;
\draw [shift={(212.06,357.35)}, rotate = 123.87] [fill={rgb, 255:red, 0; green, 0; blue, 0 }  ][line width=0.08]  [draw opacity=0] (8.93,-4.29) -- (0,0) -- (8.93,4.29) -- cycle    ;
%Straight Lines [id:da1302607574799759] 
\draw    (222.19,342.26) -- (253.14,386.91) ;
\draw [shift={(240.51,368.69)}, rotate = 235.28] [fill={rgb, 255:red, 0; green, 0; blue, 0 }  ][line width=0.08]  [draw opacity=0] (8.93,-4.29) -- (0,0) -- (8.93,4.29) -- cycle    ;
%Straight Lines [id:da40094662650607127] 
\draw    (253.14,386.91) -- (272.34,415.78) ;
\draw [shift={(265.51,405.51)}, rotate = 236.37] [fill={rgb, 255:red, 0; green, 0; blue, 0 }  ][line width=0.08]  [draw opacity=0] (8.93,-4.29) -- (0,0) -- (8.93,4.29) -- cycle    ;
%Shape: Wave [id:dp5253387106679405] 
\draw   (394.37,342.97) .. controls (391.74,341.77) and (389.23,340.61) .. (389.24,339.25) .. controls (389.26,337.89) and (391.79,336.71) .. (394.44,335.48) .. controls (397.1,334.24) and (399.63,333.06) .. (399.64,331.7) .. controls (399.65,330.34) and (397.14,329.19) .. (394.51,327.98) .. controls (391.88,326.77) and (389.37,325.61) .. (389.38,324.25) .. controls (389.39,322.89) and (391.92,321.71) .. (394.58,320.48) .. controls (397.24,319.24) and (399.76,318.06) .. (399.78,316.7) .. controls (399.79,315.34) and (397.28,314.19) .. (394.65,312.98) .. controls (392.01,311.77) and (389.5,310.61) .. (389.52,309.25) .. controls (389.53,307.9) and (392.06,306.71) .. (394.71,305.48) .. controls (397.37,304.24) and (399.9,303.06) .. (399.91,301.7) .. controls (399.93,300.34) and (397.42,299.19) .. (394.78,297.98) .. controls (392.36,296.86) and (390.04,295.79) .. (389.7,294.57) ;
%Shape: Wave [id:dp6456694623818346] 
\draw   (422.85,386.56) .. controls (421.5,386.61) and (420.45,389.14) .. (419.35,391.8) .. controls (418.25,394.46) and (417.2,396.99) .. (415.84,397.03) .. controls (414.49,397.08) and (413.2,394.63) .. (411.85,392.05) .. controls (410.51,389.47) and (409.22,387.02) .. (407.86,387.07) .. controls (406.51,387.11) and (405.46,389.64) .. (404.36,392.3) .. controls (403.26,394.96) and (402.21,397.49) .. (400.85,397.53) .. controls (399.5,397.58) and (398.21,395.13) .. (396.86,392.55) .. controls (395.51,389.98) and (394.23,387.52) .. (392.87,387.57) .. controls (391.51,387.62) and (390.46,390.15) .. (389.37,392.8) .. controls (388.27,395.46) and (387.22,397.99) .. (385.86,398.04) .. controls (384.51,398.08) and (383.22,395.63) .. (381.87,393.05) .. controls (380.52,390.48) and (379.24,388.03) .. (377.88,388.07) .. controls (376.52,388.12) and (375.47,390.65) .. (374.37,393.31) .. controls (373.28,395.96) and (372.23,398.49) .. (370.87,398.54) .. controls (369.51,398.58) and (368.23,396.13) .. (366.88,393.56) .. controls (365.53,390.98) and (364.24,388.53) .. (362.89,388.57) .. controls (362.21,388.6) and (361.61,389.24) .. (361.04,390.21) ;
%Straight Lines [id:da8504806815588826] 
\draw    (343.34,415.65) -- (367.35,380.75) ;
\draw [shift={(355.35,398.2)}, rotate = 124.53] [fill={rgb, 255:red, 0; green, 0; blue, 0 }  ][line width=0.08]  [draw opacity=0] (8.93,-4.29) -- (0,0) -- (8.93,4.29) -- cycle    ;
%Straight Lines [id:da18645048136814302] 
\draw    (367.35,380.75) -- (393.19,342.26) ;
\draw [shift={(383.06,357.35)}, rotate = 123.87] [fill={rgb, 255:red, 0; green, 0; blue, 0 }  ][line width=0.08]  [draw opacity=0] (8.93,-4.29) -- (0,0) -- (8.93,4.29) -- cycle    ;
%Straight Lines [id:da4160057539819566] 
\draw    (393.19,342.26) -- (424.14,386.91) ;
\draw [shift={(411.51,368.69)}, rotate = 235.28] [fill={rgb, 255:red, 0; green, 0; blue, 0 }  ][line width=0.08]  [draw opacity=0] (8.93,-4.29) -- (0,0) -- (8.93,4.29) -- cycle    ;
%Straight Lines [id:da46899445907988346] 
\draw    (424.14,386.91) -- (443.34,415.78) ;
\draw [shift={(436.51,405.51)}, rotate = 236.37] [fill={rgb, 255:red, 0; green, 0; blue, 0 }  ][line width=0.08]  [draw opacity=0] (8.93,-4.29) -- (0,0) -- (8.93,4.29) -- cycle    ;

% Text Node
\draw (219,273) node [anchor=north west][inner sep=0.75pt]   [align=left] {$\displaystyle \gamma $};
% Text Node
\draw (212,398) node [anchor=north west][inner sep=0.75pt]   [align=left] {$\displaystyle Z^{'}$};
% Text Node
\draw (159,389.17) node [anchor=north west][inner sep=0.75pt]   [align=left] {$\displaystyle e^{-}$};
% Text Node
\draw (272,391.17) node [anchor=north west][inner sep=0.75pt]   [align=left] {$\displaystyle e^{-}$};
% Text Node
\draw (183,348) node [anchor=north west][inner sep=0.75pt]   [align=left] {$\displaystyle \mu ^{-}$};
% Text Node
\draw (241,345) node [anchor=north west][inner sep=0.75pt]   [align=left] {$\displaystyle \mu ^{-}$};
% Text Node
\draw (390,273) node [anchor=north west][inner sep=0.75pt]   [align=left] {$\displaystyle \gamma $};
% Text Node
\draw (383,398) node [anchor=north west][inner sep=0.75pt]   [align=left] {$\displaystyle Z^{'}$};
% Text Node
\draw (348,351.17) node [anchor=north west][inner sep=0.75pt]   [align=left] {$\displaystyle e^{-}$};
% Text Node
\draw (415,350.17) node [anchor=north west][inner sep=0.75pt]   [align=left] {$\displaystyle e^{-}$};
% Text Node
\draw (327,386) node [anchor=north west][inner sep=0.75pt]   [align=left] {$\displaystyle \mu ^{-}$};
% Text Node
\draw (440,386) node [anchor=north west][inner sep=0.75pt]   [align=left] {$\displaystyle \mu ^{-}$};
\end{tikzpicture}
    \caption{The representative Feynman diagrams for $(g-2)_{e/\mu} $ contribution via the LFV $Z'$ interaction. }
    \label{fig:g2Zp}
\end{figure}
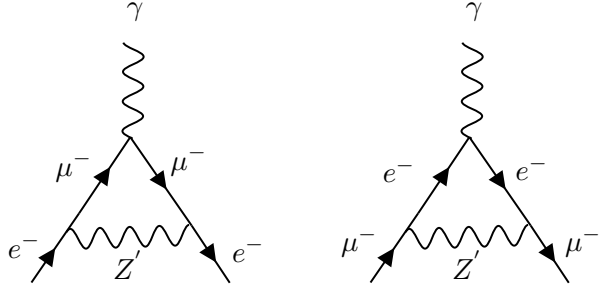
The LFV coupling ($g_{ij}$) would thus contribute to ($g-2$) with the Fig.~\ref{fig:g2Zp} and its contribution calculated through package-X~\cite{Patel:2015tea} via the following expression, 

\bea \label{eq:damu}
\Delta a_e &=& \frac{g_{e\mu}^2 m_{e}m_{\mu}\left(-m_{\mu}^{6}+4M_{Z'}^6-3m_{\mu}^2M_{Z'}^4-6m_{\mu}^2M_{Z'}^4\log\left(\frac{M_{Z'}^2}{m_{\mu}^2}\right)\right)}{16\pi^2 M_{Z'}^2(M_{Z'}^2-m_{\mu}^2)^{3}},\\
\Delta a_\mu  &=& \frac{g^2_{e\mu}}{16\pi^2}\left(-\frac{(2 M^2_{Z'} -m^2_\mu)^2}{m^2_\mu M^2_{Z'}} - \frac{2 M^2_{Z'}(3 m^2_\mu -2 M^2_{Z'})\log\left(\frac{ M^2_{Z'}}{ |M^2_{Z'}- m^2_\mu|}\right) }{m^4_\mu} \right).
\eea
We apply the 2$\sigma$ bounds based on the recent measurements of Eqs.~(\ref{eq:amu} and \ref{eq:ae}) to separately illustrate $\Delta a_\mu$ and $\Delta a_e$.

\subsection{Collider constraints}

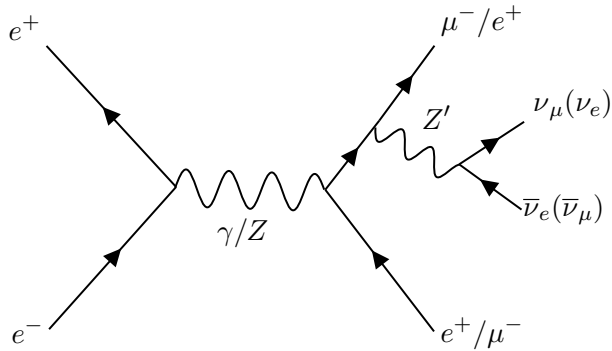
\begin{figure}[h]
    \centering
\tikzset{every picture/.style={line width=0.75pt}} %set default line width to 0.75pt        

\begin{tikzpicture}[x=0.75pt,y=0.75pt,yscale=-1,xscale=1]
%uncomment if require: \path (0,1705); %set diagram left start at 0, and has height of 1705

%Shape: Wave [id:dp9725509206012799] 
\draw   (218.23,1170.32) .. controls (220.03,1165.59) and (221.75,1161.07) .. (223.71,1161.14) .. controls (225.66,1161.21) and (227.32,1165.84) .. (229.05,1170.7) .. controls (230.79,1175.56) and (232.44,1180.19) .. (234.4,1180.26) .. controls (236.36,1180.33) and (238.08,1175.82) .. (239.87,1171.08) .. controls (241.67,1166.34) and (243.39,1161.83) .. (245.35,1161.89) .. controls (247.31,1161.96) and (248.96,1166.59) .. (250.7,1171.45) .. controls (252.43,1176.31) and (254.09,1180.94) .. (256.04,1181.01) .. controls (258,1181.08) and (259.72,1176.57) .. (261.52,1171.83) .. controls (263.31,1167.09) and (265.03,1162.58) .. (266.99,1162.65) .. controls (268.95,1162.71) and (270.6,1167.34) .. (272.34,1172.2) .. controls (274.07,1177.07) and (275.73,1181.69) .. (277.68,1181.76) .. controls (279.64,1181.83) and (281.36,1177.32) .. (283.16,1172.58) .. controls (284.95,1167.84) and (286.67,1163.33) .. (288.63,1163.4) .. controls (290.37,1163.46) and (291.87,1167.12) .. (293.4,1171.35) ;
%Straight Lines [id:da7119448609728173] 
\draw    (153.34,1098.79) -- (218.7,1169.97) ;
\draw [shift={(181.63,1129.59)}, rotate = 47.44] [fill={rgb, 255:red, 0; green, 0; blue, 0 }  ][line width=0.08]  [draw opacity=0] (8.93,-4.29) -- (0,0) -- (8.93,4.29) -- cycle    ;
%Straight Lines [id:da168583930070074] 
\draw    (218.7,1169.97) -- (154.64,1241.78) ;
\draw [shift={(191,1201.02)}, rotate = 131.74] [fill={rgb, 255:red, 0; green, 0; blue, 0 }  ][line width=0.08]  [draw opacity=0] (8.93,-4.29) -- (0,0) -- (8.93,4.29) -- cycle    ;
%Straight Lines [id:da4035470322598006] 
\draw    (293.93,1171.37) -- (321.15,1136.51) ;
\draw [shift={(310.62,1150)}, rotate = 127.97] [fill={rgb, 255:red, 0; green, 0; blue, 0 }  ][line width=0.08]  [draw opacity=0] (8.93,-4.29) -- (0,0) -- (8.93,4.29) -- cycle    ;
%Straight Lines [id:da7268234443654159] 
\draw    (293.93,1171.37) -- (348.78,1242.89) ;
\draw [shift={(317.4,1201.97)}, rotate = 52.51] [fill={rgb, 255:red, 0; green, 0; blue, 0 }  ][line width=0.08]  [draw opacity=0] (8.93,-4.29) -- (0,0) -- (8.93,4.29) -- cycle    ;
%Shape: Wave [id:dp43419163008441486] 
\draw   (319.22,1139.96) .. controls (319.16,1143.23) and (319.17,1146.26) .. (320.45,1147.04) .. controls (321.79,1147.85) and (324.15,1145.98) .. (326.63,1144.02) .. controls (329.11,1142.06) and (331.47,1140.2) .. (332.81,1141.01) .. controls (334.15,1141.82) and (334.1,1145.08) .. (334.04,1148.5) .. controls (333.97,1151.92) and (333.92,1155.18) .. (335.26,1155.99) .. controls (336.6,1156.8) and (338.96,1154.94) .. (341.44,1152.98) .. controls (343.92,1151.01) and (346.28,1149.15) .. (347.62,1149.96) .. controls (348.96,1150.77) and (348.91,1154.03) .. (348.85,1157.45) .. controls (348.78,1160.88) and (348.73,1164.14) .. (350.07,1164.95) .. controls (351.41,1165.76) and (353.77,1163.89) .. (356.25,1161.93) .. controls (358.32,1160.29) and (360.31,1158.72) .. (361.69,1158.72) ;
%Straight Lines [id:da6142895892364789] 
\draw    (321.15,1136.51) -- (349,1098.67) ;
\draw [shift={(338.04,1113.56)}, rotate = 126.36] [fill={rgb, 255:red, 0; green, 0; blue, 0 }  ][line width=0.08]  [draw opacity=0] (8.93,-4.29) -- (0,0) -- (8.93,4.29) -- cycle    ;
%Straight Lines [id:da4713867686564662] 
\draw    (361.46,1158.56) -- (392.91,1181.41) ;
\draw [shift={(371.93,1166.17)}, rotate = 35.99] [fill={rgb, 255:red, 0; green, 0; blue, 0 }  ][line width=0.08]  [draw opacity=0] (8.93,-4.29) -- (0,0) -- (8.93,4.29) -- cycle    ;
%Straight Lines [id:da615522443785338] 
\draw    (361.46,1158.56) -- (394.21,1137.09) ;
\draw [shift={(382.02,1145.08)}, rotate = 146.75] [fill={rgb, 255:red, 0; green, 0; blue, 0 }  ][line width=0.08]  [draw opacity=0] (8.93,-4.29) -- (0,0) -- (8.93,4.29) -- cycle    ;

% Text Node
\draw (134.11,1234.75) node [anchor=north west][inner sep=0.75pt]   [align=left] {$\displaystyle e^{-}$};
% Text Node
\draw (132.68,1081.64) node [anchor=north west][inner sep=0.75pt]   [align=left] {$\displaystyle e^{+}$};
% Text Node
\draw (418.05,1155) node [anchor=north west][inner sep=0.75pt]   [align=left] {$ $};
% Text Node
\draw (350.53,1234.05) node [anchor=north west][inner sep=0.75pt]   [align=left] {$\displaystyle e^{+} /\mu ^{-}$};
% Text Node
\draw (237.92,1184.45) node [anchor=north west][inner sep=0.75pt]   [align=left] {$\displaystyle \gamma /Z$};
% Text Node
\draw (341.2,1127.24) node [anchor=north west][inner sep=0.75pt]   [align=left] {$\displaystyle Z'$};
% Text Node
\draw (392.24,1172.2) node [anchor=north west][inner sep=0.75pt]   [align=left] {$\displaystyle \overline{\nu }$$\displaystyle _{e}(\overline{\nu }_{\mu })$};
% Text Node
\draw (397.33,1119.3) node [anchor=north west][inner sep=0.75pt]   [align=left] {$\displaystyle \nu $$\displaystyle _{\mu }( \nu _{e})$};
% Text Node
\draw (350.63,1076.6) node [anchor=north west][inner sep=0.75pt]   [align=left] {$\displaystyle \mu ^{-} /e^{+}$};

\end{tikzpicture}
    \caption{Typical Feynman diagrams for $e^+e^- \to e^\pm \mu^\mp + \slashed{E}$. }
    \label{fig:belle2miss}
\end{figure}

For an electron-positron collider, particularly the Belle-II experiment,  the search for LFV $Z'$ bosons is linked with $e\mu + \slashed{E}$ final states. We use their constraints on the product of the cross-section $\sigma(e^+e^- \to \mu^\pm e^\pm + \slashed{E})$ (see Fig.~\ref{fig:belle2miss}) and the detector efficiency $\epsilon$ to set the upper bound~\cite{Belle-II:2019qfb}. We used the product ($\epsilon \times \sigma$) to constrain the coupling in the mass region up to approximately 8 GeV. The numerical cross-section for the $Z'$ is directly proportional to $g^2_{e\mu}$ and can be computed by solving the relation between the existing product limit and their selection of $M_{Z'}$ values.

%If $\tau_{Z'}$ is large, $\sigma \times 1$, if  $\tau_{Z'}$ is small, $\sigma \times \mathcal{B}(Z'\to \bar{\nu}_e \nu_\mu)$.

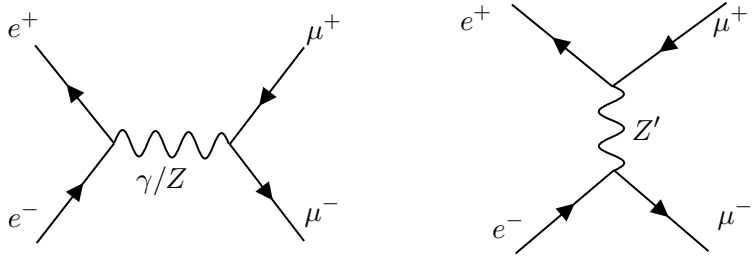
\begin{figure}[!t]
    \centering
\tikzset{every picture/.style={line width=0.75pt}} %set default line width to 0.75pt        

\begin{tikzpicture}[x=0.75pt,y=0.75pt,yscale=-1,xscale=1]
%uncomment if require: \path (0,1242); %set diagram left start at 0, and has height of 1242

%Shape: Wave [id:dp21030505825084733] 
\draw   (208.83,642.54) .. controls (208.83,642.54) and (208.83,642.54) .. (208.83,642.54) .. controls (210.22,639.23) and (211.54,636.07) .. (213.04,636.12) .. controls (214.55,636.17) and (215.83,639.41) .. (217.16,642.81) .. controls (218.49,646.21) and (219.77,649.44) .. (221.27,649.49) .. controls (222.78,649.54) and (224.1,646.38) .. (225.48,643.07) .. controls (226.86,639.75) and (228.18,636.6) .. (229.69,636.65) .. controls (231.2,636.69) and (232.47,639.93) .. (233.8,643.33) .. controls (235.14,646.73) and (236.41,649.97) .. (237.92,650.02) .. controls (239.42,650.06) and (240.74,646.91) .. (242.13,643.59) .. controls (243.51,640.28) and (244.83,637.13) .. (246.34,637.17) .. controls (247.84,637.22) and (249.12,640.46) .. (250.45,643.86) .. controls (251.78,647.26) and (253.06,650.49) .. (254.56,650.54) .. controls (256.07,650.59) and (257.39,647.43) .. (258.77,644.12) .. controls (260.15,640.81) and (261.48,637.65) .. (262.98,637.7) .. controls (264.32,637.74) and (265.48,640.3) .. (266.65,643.26) ;
%Straight Lines [id:da6777911445561717] 
\draw    (168.91,593.39) -- (208.78,642.84) ;
\draw [shift={(184.77,613.05)}, rotate = 51.12] [fill={rgb, 255:red, 0; green, 0; blue, 0 }  ][line width=0.08]  [draw opacity=0] (8.93,-4.29) -- (0,0) -- (8.93,4.29) -- cycle    ;
%Straight Lines [id:da6540792354437119] 
\draw    (208.78,642.84) -- (169.71,693.16) ;
\draw [shift={(193.23,662.86)}, rotate = 127.83] [fill={rgb, 255:red, 0; green, 0; blue, 0 }  ][line width=0.08]  [draw opacity=0] (8.93,-4.29) -- (0,0) -- (8.93,4.29) -- cycle    ;
%Straight Lines [id:da13477756595148493] 
\draw    (267.06,643.27) -- (304.57,594.39) ;
\draw [shift={(281.86,623.99)}, rotate = 307.5] [fill={rgb, 255:red, 0; green, 0; blue, 0 }  ][line width=0.08]  [draw opacity=0] (8.93,-4.29) -- (0,0) -- (8.93,4.29) -- cycle    ;
%Straight Lines [id:da23831377644458118] 
\draw    (267.06,643.27) -- (304.57,692.16) ;
\draw [shift={(288.86,671.68)}, rotate = 232.5] [fill={rgb, 255:red, 0; green, 0; blue, 0 }  ][line width=0.08]  [draw opacity=0] (8.93,-4.29) -- (0,0) -- (8.93,4.29) -- cycle    ;
%Shape: Wave [id:dp2600294115684275] 
\draw   (459.48,613.68) .. controls (462.74,615.04) and (465.84,616.35) .. (465.85,617.93) .. controls (465.85,619.5) and (462.76,620.91) .. (459.52,622.38) .. controls (456.27,623.85) and (453.19,625.26) .. (453.19,626.83) .. controls (453.2,628.41) and (456.3,629.72) .. (459.56,631.08) .. controls (462.81,632.45) and (465.91,633.76) .. (465.92,635.33) .. controls (465.93,636.91) and (462.84,638.32) .. (459.59,639.79) .. controls (456.35,641.26) and (453.26,642.67) .. (453.26,644.24) .. controls (453.27,645.82) and (456.37,647.12) .. (459.63,648.49) .. controls (462.89,649.86) and (465.99,651.17) .. (465.99,652.74) .. controls (466,654.17) and (463.47,655.45) .. (460.59,656.78) ;
%Straight Lines [id:da031049996611033692] 
\draw    (409.35,572.78) -- (460.01,614.12) ;
\draw [shift={(429.64,589.34)}, rotate = 39.22] [fill={rgb, 255:red, 0; green, 0; blue, 0 }  ][line width=0.08]  [draw opacity=0] (8.93,-4.29) -- (0,0) -- (8.93,4.29) -- cycle    ;
%Straight Lines [id:da5243041744443079] 
\draw    (460.85,656.42) -- (411.2,698.49) ;
\draw [shift={(440.99,673.25)}, rotate = 139.72] [fill={rgb, 255:red, 0; green, 0; blue, 0 }  ][line width=0.08]  [draw opacity=0] (8.93,-4.29) -- (0,0) -- (8.93,4.29) -- cycle    ;
%Straight Lines [id:da2508154542809117] 
\draw    (460.01,614.12) -- (517.61,571.89) ;
\draw [shift={(483.57,596.85)}, rotate = 323.75] [fill={rgb, 255:red, 0; green, 0; blue, 0 }  ][line width=0.08]  [draw opacity=0] (8.93,-4.29) -- (0,0) -- (8.93,4.29) -- cycle    ;
%Straight Lines [id:da535353627993882] 
\draw    (460.85,656.42) -- (508.51,697.29) ;
\draw [shift={(488.48,680.11)}, rotate = 220.62] [fill={rgb, 255:red, 0; green, 0; blue, 0 }  ][line width=0.08]  [draw opacity=0] (8.93,-4.29) -- (0,0) -- (8.93,4.29) -- cycle    ;

% Text Node
\draw (153.63,671.09) node [anchor=north west][inner sep=0.75pt]   [align=left] {$\displaystyle e^{-}$};
% Text Node
\draw (153.53,575.32) node [anchor=north west][inner sep=0.75pt]   [align=left] {$\displaystyle e^{+}$};
% Text Node
\draw (304.05,576.32) node [anchor=north west][inner sep=0.75pt]   [align=left] {$\displaystyle \mu ^{+}$};
% Text Node
\draw (303.25,668.1) node [anchor=north west][inner sep=0.75pt]   [align=left] {$\displaystyle \mu ^{-}$};
% Text Node
\draw (397.98,676.58) node [anchor=north west][inner sep=0.75pt]   [align=left] {$\displaystyle e^{-}$};
% Text Node
\draw (381.76,571.47) node [anchor=north west][inner sep=0.75pt]   [align=left] {$\displaystyle e^{+}$};
% Text Node
\draw (509.26,572.4) node [anchor=north west][inner sep=0.75pt]   [align=left] {$\displaystyle \mu ^{+}$};
% Text Node
\draw (511.56,669.51) node [anchor=north west][inner sep=0.75pt]   [align=left] {$\displaystyle \mu ^{-}$};
% Text Node
\draw (218.36,652.81) node [anchor=north west][inner sep=0.75pt]   [align=left] {$\displaystyle \gamma /Z$};
% Text Node
\draw (468.36,627.81) node [anchor=north west][inner sep=0.75pt]   [align=left] {$\displaystyle Z'$};
\end{tikzpicture}
    \caption{Typical Feynman diagrams for $e^+e^- \to \mu^{+} \mu^{-}$. Left panel: SM contribution. Right panel: $Z'$ contribution.}
    \label{fig:collider1}
\end{figure}

In the process $e^+e^- \to \mu^+ \mu^-$~\cite{DELPHI:2000ztm}, the LFV interaction occurs via a $t$-channel exchange mediated by $M_{Z'}$. As reported in the aforementioned reference, LEP conducted measurements of this process at $\sqrt{s} = 189$ GeV. The background from the $s$-channel in the SM (mediated by $H$ bosons) is suppressed by $m_\mu m_e/M^2_{H}$ due to small Yukawa couplings, resulting in significant interference contributions between new physics and the SM background associated only with s-channel ($\gamma$ and Z) diagrams ($\mathcal{M}_{\text{total}} =\mathcal{M}_{Z'(t)} + \mathcal{M}_{\gamma(s)} + \mathcal{M}_{Z(s)}$) as depicted in Fig.~\ref{fig:collider1}. 

The cross-section of the process $e^+ e^- \to \mu^+ \mu^-$expands the terms to $\mathcal{O} (s^{-1})$ as follows:
\bea
\sigma&=& \frac{e^4 \left(1+24 s_w^4\right)}{768 \pi   s_w^4 \left(1-s_w^2\right)^2 s }+\frac{e^2 g^2_{e\mu} \left(1+4 s_w^2\right) \left(2 \log
  (s/M_{Z'}^2) -3\right)}{64 \pi  s_w^2 \left(1-s_w^2\right) s}+\frac{g^4_{e\mu}}{4 \pi  M_{Z'}^2},
\eea
The squared amplitude of the new physics process is proportional to $1/(t-M_{Z'}^{2})^{2}$, where the variable $t$ depends on the scattering angle $\theta$. Upon integrating over $\theta$, the cross-section becomes proportional to the $1/(t-M_{Z'}^{2})$ term. Consequently, this factor reaches its maximum value of $1/M^2_{Z'}$ when \(t\) approaches zero. Furthermore, the contribution from the term $q^{\mu}q^{\nu}/M_{Z'}^{2}$ is proportional to $m_{\mu}^{4}/(M_{Z'}^{4}s)$, which is negligible compared to the $1/M^2_{Z'}$ contribution associated with the term $g^{\mu\nu}$ in the region of interest for $M_{Z'}$ (below 10 GeV). In summary, the factor $1/M^2_{Z'}$ in the aforementioned equation arises from the t-channel in the context of new physics.

We conducted an analysis of the LEP cross-sections ~\cite{DELPHI:2000ztm} at $\sqrt{s} = 189 \,$GeV, corresponding to an integrated luminosity of $155.21 \, \text{pb}^{-1}$. This analysis encompasses the range of \(\cos \theta\) from $-0.97$ to $0.97$, partitioned into 10 bins, in order to establish constraints on the $Z'$ boson. This enables us to perform a $\chi^2$ test based on the differential cross-section of the process (denoted as $\text{N}^j_{\text{signal}} = \int d\sigma /d \cos\theta \times \mathcal{L}$),
\bea
\chi^2 =\sum^{10}_{j=1} \frac{\left(\text{N}^j_{\text{experiment}} - (\text{N}^j_{\text{background}} + \text{N}^j_{\text{signal}}) \right)^2}{(\sigma^j_{\text{error}})^2} .
\eea
where $\text{N}^i_{\text{experiment}}$ represents the number of events measured by the LEP, and $\text{N}^i_{\text{background}}$ represents the difference between the theoretical predictions and the measured events in each bin. The two independent parameters, $g_{e\mu}$ and $M_{Z'}$, can then be constrained using the 10 bins with $\chi^2 \leq 5.991$ (95\% CL).

\begin{figure}[t]
\centering

\tikzset{every picture/.style={line width=0.75pt}} %set default line width to 0.75pt        

\begin{tikzpicture}[x=0.75pt,y=0.75pt,yscale=-1,xscale=1]
%uncomment if require: \path (0,3034); %set diagram left start at 0, and has height of 3034

%Straight Lines [id:da5380349644206269] 
\draw    (125.6,152.8) -- (175.6,152.2) -- (215.6,151.2) ;
\draw [shift={(155.6,152.44)}, rotate = 179.31] [fill={rgb, 255:red, 0; green, 0; blue, 0 }  ][line width=0.08]  [draw opacity=0] (8.93,-4.29) -- (0,0) -- (8.93,4.29) -- cycle    ;
\draw [shift={(200.6,151.58)}, rotate = 178.57] [fill={rgb, 255:red, 0; green, 0; blue, 0 }  ][line width=0.08]  [draw opacity=0] (8.93,-4.29) -- (0,0) -- (8.93,4.29) -- cycle    ;
%Shape: Wave [id:dp8719248853648139] 
\draw   (214.28,151.68) .. controls (214.16,151.8) and (214.04,151.91) .. (213.93,152.02) .. controls (211.1,154.77) and (208.42,157.39) .. (209,158.62) .. controls (209.58,159.84) and (213.26,159.34) .. (217.14,158.8) .. controls (221.01,158.26) and (224.69,157.76) .. (225.28,158.99) .. controls (225.86,160.21) and (223.17,162.83) .. (220.34,165.58) .. controls (217.52,168.33) and (214.83,170.95) .. (215.41,172.18) .. controls (215.99,173.4) and (219.68,172.9) .. (223.55,172.36) .. controls (227.42,171.82) and (231.11,171.32) .. (231.69,172.55) .. controls (232.27,173.77) and (229.58,176.39) .. (226.75,179.14) .. controls (223.93,181.89) and (221.24,184.51) .. (221.82,185.74) .. controls (222.4,186.96) and (226.09,186.46) .. (229.96,185.92) .. controls (233.83,185.38) and (237.52,184.88) .. (238.1,186.11) .. controls (238.68,187.33) and (235.99,189.95) .. (233.17,192.7) .. controls (230.34,195.45) and (227.65,198.07) .. (228.23,199.3) .. controls (228.81,200.52) and (232.5,200.02) .. (236.37,199.48) .. controls (239.5,199.05) and (242.51,198.64) .. (243.85,199.14) ;
%Straight Lines [id:da12423278992302877] 
\draw    (267.6,118.8) -- (215.6,151.2) ;
\draw [shift={(247.12,131.56)}, rotate = 148.07] [fill={rgb, 255:red, 0; green, 0; blue, 0 }  ][line width=0.08]  [draw opacity=0] (8.93,-4.29) -- (0,0) -- (8.93,4.29) -- cycle    ;
%Shape: Wave [id:dp902519824612556] 
\draw   (172.28,152.68) .. controls (172.16,152.8) and (172.04,152.91) .. (171.93,153.02) .. controls (169.1,155.77) and (166.42,158.39) .. (167,159.62) .. controls (167.58,160.84) and (171.26,160.34) .. (175.14,159.8) .. controls (179.01,159.26) and (182.69,158.76) .. (183.28,159.99) .. controls (183.86,161.21) and (181.17,163.83) .. (178.34,166.58) .. controls (175.52,169.33) and (172.83,171.95) .. (173.41,173.18) .. controls (173.99,174.4) and (177.68,173.9) .. (181.55,173.36) .. controls (185.42,172.82) and (189.11,172.32) .. (189.69,173.55) .. controls (190.27,174.77) and (187.58,177.39) .. (184.75,180.14) .. controls (181.93,182.89) and (179.24,185.51) .. (179.82,186.74) .. controls (180.4,187.96) and (184.09,187.46) .. (187.96,186.92) .. controls (191.83,186.38) and (195.52,185.88) .. (196.1,187.11) .. controls (196.68,188.33) and (193.99,190.95) .. (191.17,193.7) .. controls (188.34,196.45) and (185.65,199.07) .. (186.23,200.3) .. controls (186.81,201.52) and (190.5,201.02) .. (194.37,200.48) .. controls (197.5,200.05) and (200.51,199.64) .. (201.85,200.14) ;
%Straight Lines [id:da9522927854754611] 
\draw    (337.6,152.8) -- (387.6,152.2) -- (427.6,151.2) ;
\draw [shift={(367.6,152.44)}, rotate = 179.31] [fill={rgb, 255:red, 0; green, 0; blue, 0 }  ][line width=0.08]  [draw opacity=0] (8.93,-4.29) -- (0,0) -- (8.93,4.29) -- cycle    ;
\draw [shift={(412.6,151.58)}, rotate = 178.57] [fill={rgb, 255:red, 0; green, 0; blue, 0 }  ][line width=0.08]  [draw opacity=0] (8.93,-4.29) -- (0,0) -- (8.93,4.29) -- cycle    ;
%Shape: Wave [id:dp8626142015496403] 
\draw   (426.28,151.68) .. controls (426.16,151.8) and (426.04,151.91) .. (425.93,152.02) .. controls (423.1,154.77) and (420.42,157.39) .. (421,158.62) .. controls (421.58,159.84) and (425.26,159.34) .. (429.14,158.8) .. controls (433.01,158.26) and (436.69,157.76) .. (437.28,158.99) .. controls (437.86,160.21) and (435.17,162.83) .. (432.34,165.58) .. controls (429.52,168.33) and (426.83,170.95) .. (427.41,172.18) .. controls (427.99,173.4) and (431.68,172.9) .. (435.55,172.36) .. controls (439.42,171.82) and (443.11,171.32) .. (443.69,172.55) .. controls (444.27,173.77) and (441.58,176.39) .. (438.75,179.14) .. controls (435.93,181.89) and (433.24,184.51) .. (433.82,185.74) .. controls (434.4,186.96) and (438.09,186.46) .. (441.96,185.92) .. controls (445.83,185.38) and (449.52,184.88) .. (450.1,186.11) .. controls (450.68,187.33) and (447.99,189.95) .. (445.17,192.7) .. controls (442.34,195.45) and (439.65,198.07) .. (440.23,199.3) .. controls (440.81,200.52) and (444.5,200.02) .. (448.37,199.48) .. controls (451.5,199.05) and (454.51,198.64) .. (455.85,199.14) ;
%Straight Lines [id:da7970345095411902] 
\draw    (479.6,118.8) -- (427.6,151.2) ;
\draw [shift={(459.12,131.56)}, rotate = 148.07] [fill={rgb, 255:red, 0; green, 0; blue, 0 }  ][line width=0.08]  [draw opacity=0] (8.93,-4.29) -- (0,0) -- (8.93,4.29) -- cycle    ;
%Shape: Wave [id:dp5709169778367518] 
\draw   (384.28,152.68) .. controls (384.16,152.8) and (384.04,152.91) .. (383.93,153.02) .. controls (381.1,155.77) and (378.42,158.39) .. (379,159.62) .. controls (379.58,160.84) and (383.26,160.34) .. (387.14,159.8) .. controls (391.01,159.26) and (394.69,158.76) .. (395.28,159.99) .. controls (395.86,161.21) and (393.17,163.83) .. (390.34,166.58) .. controls (387.52,169.33) and (384.83,171.95) .. (385.41,173.18) .. controls (385.99,174.4) and (389.68,173.9) .. (393.55,173.36) .. controls (397.42,172.82) and (401.11,172.32) .. (401.69,173.55) .. controls (402.27,174.77) and (399.58,177.39) .. (396.75,180.14) .. controls (393.93,182.89) and (391.24,185.51) .. (391.82,186.74) .. controls (392.4,187.96) and (396.09,187.46) .. (399.96,186.92) .. controls (403.83,186.38) and (407.52,185.88) .. (408.1,187.11) .. controls (408.68,188.33) and (405.99,190.95) .. (403.17,193.7) .. controls (400.34,196.45) and (397.65,199.07) .. (398.23,200.3) .. controls (398.81,201.52) and (402.5,201.02) .. (406.37,200.48) .. controls (409.5,200.05) and (412.51,199.64) .. (413.85,200.14) ;

% Text Node
\draw (98,146) node [anchor=north west][inner sep=0.75pt]   [align=left] {$\displaystyle \mu ^{-}$};
% Text Node
\draw (269,107) node [anchor=north west][inner sep=0.75pt]   [align=left] {$\displaystyle e^{-}$};
% Text Node
\draw (187,201) node [anchor=north west][inner sep=0.75pt]   [align=left] {$\displaystyle Z^{'}$};
% Text Node
\draw (236,209) node [anchor=north west][inner sep=0.75pt]   [align=left] {$\displaystyle \gamma $};
% Text Node
\draw (310,146) node [anchor=north west][inner sep=0.75pt]   [align=left] {$\displaystyle \mu ^{-}$};
% Text Node
\draw (483,107) node [anchor=north west][inner sep=0.75pt]   [align=left] {$\displaystyle e^{-}$};
% Text Node
\draw (449,199) node [anchor=north west][inner sep=0.75pt]   [align=left] {$\displaystyle Z^{'}$};
% Text Node
\draw (394,211) node [anchor=north west][inner sep=0.75pt]   [align=left] {$\displaystyle \gamma $};
% Text Node
\draw (396,126) node [anchor=north west][inner sep=0.75pt]   [align=left] {$\displaystyle \mu ^{-}$};
% Text Node
\draw (187,127) node [anchor=north west][inner sep=0.75pt]   [align=left] {$\displaystyle e^{-}$};
\end{tikzpicture}
\caption{The representative Feynman diagrams for 3-body decay of the muon which $Z'$ is lighter than $m_\mu$.}
\label{fig:LFV3bodydecay}
\end{figure}
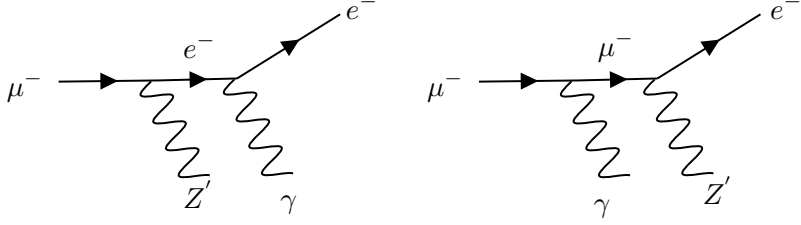
\subsection{Muon decay limits}
The occurrence of LFV decays, characterized as $\ell_i \to \ell_j + X$, imposes constraints on the underlying interactions. In the context of muon decays, such as $\mu \to e \gamma$, $\mu \to 3e$ and $\mu \to e + X$ (A summarized overview is provided in~\cite{Renga:2019mpg}), these processes define significant boundaries for new physics. Notably, the $\mu \to e \gamma$ decay has been scrutinized extensively at the MEG experiment~\cite{MEG:2011naj,MEG:2013oxv,MEG:2016leq}, while $\mu \to 3e$ has been explored by various experiments~\cite{SINDRUM:1987nra, Mu3e:2020gyw, Perrevoort:2023qhn}. In previous searches, flavor-violating interactions associated with muon decay involving massless Goldstone bosons, also known as familons, have been thoroughly investigated. These studies have addressed not only the phenomenological constraints~\cite{Anselm:1985bp,Andreev:2006wh} but also the cosmological consequences and implications of such interactions~\cite{berezhiani1991cosmology,sakharov1994horizontal}. Furthermore, the TWIST collaboration has explored the decay process $\mu \to e + X$ involving a non-massless $X$~\cite{TWIST:2014ymv}. The absence of kinetic mixing, such as $Z-Z'$ or $A-Z'$ interactions, precludes the occurrence of $\mu \to 3e$ due to the lack of diagonal coupling vertices between $Z'$ and charged leptons. Consequently, the constraints on LFV decays primarily focus on two-body processes, where the final decay channel involves $\mu \to e Z'$ and subsequent $Z'$ decays exclusively to neutrinos with a branching ratio of 100\%. The TWIST experiment utilized muon stopped events to establish an upper limit for $\mathcal{B} (\mu \to e + X)$ of approximately $\mathcal{O}(10^{-5})$ within the mass region of $X$, ranging from 13 MeV to 80 MeV~\cite{TWIST:2014ymv}. In addition, the PSI experiment with a HPGe detector,  testing masses ranging from 100 MeV to 150 MeV, obtained an upper limit of $\mathcal{B} (\mu^+ \to e^+ +X) < 5.7 \times 10^{-4}$~\cite{Bilger:1998rp}. Therefore, we will employ the current limit established by the TWIST collaboration to assess the constraint on the $Z'$. 

The decay width of $\mu \to e + Z'$ can be expressed as follows under the approximation of neglecting $m_e$,
\bea
\Gamma (\mu \to e + Z') = 
 \frac{g_{e \mu}^2(m_{\mu}^2-M_{Z'}^2)^2(m_{\mu}^2+2M^{2}_{Z'})}{16 \pi m^3_\mu M^2_{Z'}},
\eea
with its branching ratio $\left(\tau_\mu = 2.2\times 10^{-6} s \sim \Gamma_\mu = 3\times 10^{-19} \text{GeV} \right)$~\cite{Workman:2022ynf},
\bea
\mathcal{B} (\mu \to e + Z') =  \Gamma (\mu \to e + Z') / \Gamma_\mu,
\eea

However, as reported by Ref.~\cite{Renga:2019mpg}, the Crystal Box collaboration~\cite{Goldman:1987hy} has showcased the capability to examine the inclusive $\mu \to e + \gamma + X$ decay process with $\mu \to e + \gamma$. They have established an upper bound for the branching ratio $\mathcal{B}(\mu \to e + \gamma + X) < 1.3 \times 10^{-9}$~\cite{ Goldman:1987hy}. The three-body decay, $\mu \to e + \gamma +X$, can be expressed as follows,
\bea
\Gamma ( \mu \to e + \gamma + Z') \simeq \frac{\alpha_{\text{EM}} g_{e\mu}^2 m_{\mu}^{5}}{128 \pi^2 E_{\gamma}^{2} M_{Z'}^{2}}.
\eea
where $E_\gamma$ correspond to minimum energy cut to prevent infrared divergence (typical energy is picked for 38 MeV in~\cite{Goldman:1987hy}). For  $M_{Z'}$ reduces to $m_e$, the decay process is predominantly governed by $m_\mu$, leading to tighter constraints on its coupling. Conversely, as the mass of $Z'$ approaches $m_\mu$, the situation is reversed. This is due to the enhancement of $\sim 1/M^2_{Z'}$ (arising from the longitudinal polarization mode of $Z'$), which has been extensively discussed for many channels based on the Goldstone Boson Equivalence Theorem (e.g., $\tau \to \mu Z'$, $t \to b W^+$~\cite{Heeck:2016xkh,CHANOWITZ1985379,Peskin:2017emn,Hill:2023dym}). Meanwhile, an additional $\alpha_{\text{EM}}$ factor and the three body phase space factor further suppresses the new physics contribution compared to the $\mu \to e + Z'$ process. Thus, the three-body channel would then be the most stringent limit compares with the two-body channel.

\subsection{Inverse $\mu$ decay}
\begin{figure}[!h]
    \centering

\tikzset{every picture/.style={line width=0.75pt}} %set default line width to 0.75pt        

\begin{tikzpicture}[x=0.75pt,y=0.75pt,yscale=-1,xscale=1]
%uncomment if require: \path (0,1919); %set diagram left start at 0, and has height of 1919

%Shape: Wave [id:dp4398160107941409] 
\draw   (141.53,1481.37) .. controls (143.5,1483.05) and (145.37,1484.65) .. (145.37,1486.56) .. controls (145.38,1488.47) and (143.52,1490.15) .. (141.56,1491.91) .. controls (139.61,1493.67) and (137.75,1495.35) .. (137.76,1497.26) .. controls (137.76,1499.16) and (139.63,1500.77) .. (141.6,1502.45) .. controls (143.56,1504.12) and (145.44,1505.73) .. (145.44,1507.63) .. controls (145.45,1509.54) and (143.59,1511.22) .. (141.63,1512.98) .. controls (139.68,1514.74) and (137.82,1516.42) .. (137.83,1518.33) .. controls (137.83,1520.24) and (139.7,1521.84) .. (141.67,1523.52) .. controls (143.63,1525.2) and (145.51,1526.8) .. (145.51,1528.71) .. controls (145.52,1530.61) and (143.66,1532.3) .. (141.7,1534.06) .. controls (139.75,1535.81) and (137.89,1537.5) .. (137.89,1539.4) .. controls (137.9,1540.03) and (138.1,1540.63) .. (138.45,1541.21) ;
%Straight Lines [id:da1122446460886597] 
\draw    (71.87,1479.79) -- (140.42,1480.66) ;
\draw [shift={(111.14,1480.29)}, rotate = 180.72] [fill={rgb, 255:red, 0; green, 0; blue, 0 }  ][line width=0.08]  [draw opacity=0] (8.93,-4.29) -- (0,0) -- (8.93,4.29) -- cycle    ;
%Straight Lines [id:da5593809892860259] 
\draw    (138.38,1541.54) -- (71.89,1541.07) ;
\draw [shift={(111.64,1541.35)}, rotate = 180.41] [fill={rgb, 255:red, 0; green, 0; blue, 0 }  ][line width=0.08]  [draw opacity=0] (8.93,-4.29) -- (0,0) -- (8.93,4.29) -- cycle    ;
%Straight Lines [id:da660878981022929] 
\draw    (140.42,1480.66) -- (222.97,1479.79) ;
\draw [shift={(186.69,1480.17)}, rotate = 179.4] [fill={rgb, 255:red, 0; green, 0; blue, 0 }  ][line width=0.08]  [draw opacity=0] (8.93,-4.29) -- (0,0) -- (8.93,4.29) -- cycle    ;
%Straight Lines [id:da9560834264586711] 
\draw    (138.38,1541.54) -- (222.05,1541.07) ;
\draw [shift={(185.21,1541.28)}, rotate = 179.67] [fill={rgb, 255:red, 0; green, 0; blue, 0 }  ][line width=0.08]  [draw opacity=0] (8.93,-4.29) -- (0,0) -- (8.93,4.29) -- cycle    ;
%Shape: Wave [id:dp197785007718969] 
\draw   (394.53,1482.37) .. controls (396.5,1484.05) and (398.37,1485.65) .. (398.37,1487.56) .. controls (398.38,1489.47) and (396.52,1491.15) .. (394.56,1492.91) .. controls (392.61,1494.67) and (390.75,1496.35) .. (390.76,1498.26) .. controls (390.76,1500.16) and (392.63,1501.77) .. (394.6,1503.45) .. controls (396.56,1505.12) and (398.44,1506.73) .. (398.44,1508.63) .. controls (398.45,1510.54) and (396.59,1512.22) .. (394.63,1513.98) .. controls (392.68,1515.74) and (390.82,1517.42) .. (390.83,1519.33) .. controls (390.83,1521.24) and (392.7,1522.84) .. (394.67,1524.52) .. controls (396.63,1526.2) and (398.51,1527.8) .. (398.51,1529.71) .. controls (398.52,1531.61) and (396.66,1533.3) .. (394.7,1535.06) .. controls (392.75,1536.81) and (390.89,1538.5) .. (390.89,1540.4) .. controls (390.9,1541.03) and (391.1,1541.63) .. (391.45,1542.21) ;
%Straight Lines [id:da8724231800271282] 
\draw    (324.87,1480.79) -- (393.42,1481.66) ;
\draw [shift={(352.64,1481.14)}, rotate = 0.72] [fill={rgb, 255:red, 0; green, 0; blue, 0 }  ][line width=0.08]  [draw opacity=0] (8.93,-4.29) -- (0,0) -- (8.93,4.29) -- cycle    ;
%Straight Lines [id:da0851175133233516] 
\draw    (391.38,1542.54) -- (324.89,1542.07) ;
\draw [shift={(364.64,1542.35)}, rotate = 180.41] [fill={rgb, 255:red, 0; green, 0; blue, 0 }  ][line width=0.08]  [draw opacity=0] (8.93,-4.29) -- (0,0) -- (8.93,4.29) -- cycle    ;
%Straight Lines [id:da15269082139159418] 
\draw    (393.42,1481.66) -- (475.97,1480.79) ;
\draw [shift={(428.19,1481.29)}, rotate = 359.4] [fill={rgb, 255:red, 0; green, 0; blue, 0 }  ][line width=0.08]  [draw opacity=0] (8.93,-4.29) -- (0,0) -- (8.93,4.29) -- cycle    ;
%Straight Lines [id:da5510249457389504] 
\draw    (391.38,1542.54) -- (475.05,1542.07) ;
\draw [shift={(438.21,1542.28)}, rotate = 179.67] [fill={rgb, 255:red, 0; green, 0; blue, 0 }  ][line width=0.08]  [draw opacity=0] (8.93,-4.29) -- (0,0) -- (8.93,4.29) -- cycle    ;

% Text Node
\draw (46.89,1471.42) node [anchor=north west][inner sep=0.75pt]   [align=left] {$\displaystyle \nu _{\mu }$};
% Text Node
\draw (52.93,1531.79) node [anchor=north west][inner sep=0.75pt]   [align=left] {$\displaystyle e^{-}$};
% Text Node
\draw (148.67,1501.75) node [anchor=north west][inner sep=0.75pt]   [align=left] {$\displaystyle W^{+}$};
% Text Node
\draw (224.28,1465.54) node [anchor=north west][inner sep=0.75pt]   [align=left] {$\displaystyle \mu ^{-}$};
% Text Node
\draw (225.66,1536.42) node [anchor=north west][inner sep=0.75pt]   [align=left] {$\displaystyle \nu _{e}$};
% Text Node
\draw (303.89,1472.42) node [anchor=north west][inner sep=0.75pt]   [align=left] {$\displaystyle \overline{\nu }_{\mu }$};
% Text Node
\draw (305.93,1530.79) node [anchor=north west][inner sep=0.75pt]   [align=left] {$\displaystyle e^{-}$};
% Text Node
\draw (404.67,1501.75) node [anchor=north west][inner sep=0.75pt]   [align=left] {$\displaystyle Z'$};
% Text Node
\draw (476.66,1528.42) node [anchor=north west][inner sep=0.75pt]   [align=left] {$\displaystyle \mu ^{-}$};
% Text Node
\draw (478.89,1471.42) node [anchor=north west][inner sep=0.75pt]   [align=left] {$\displaystyle \overline{\nu }_{e}$};

\end{tikzpicture}
    \caption{Feynman diagrams for inverse $\mu$ decay with the SM electroweak interaction and the LFV $Z'$ interaction.}
    \label{fig:inversemuon}
\end{figure}
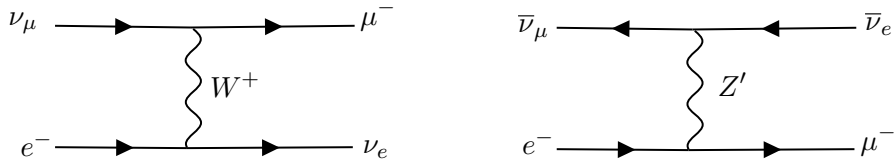
In Ref.~\cite{Amsterdam-CERN-Hamburg-Moscow-Rome:1980qbb, NuTeV:2001bgq}, CHARM and NuTeV collaborations tested the inverse $\mu$ decay $\nu_\mu e^- \to \mu^- \nu_e$ and $\bar{\nu}_\mu e^- \to \mu^- \bar{\nu}_e$, established the upper limit for the ratio of the process as Fig.~\ref{fig:inversemuon} (where there is no SM contribution for the second channel),
\bea \label{eq:inverseratio}
R_{\nu e}=\frac{\sigma(\bar{\nu}_\mu e^- \to \mu^- \bar{\nu}_e)}{\sigma(\nu_\mu e^- \to \mu^- \nu_e)} < \bigg\{
\begin{array}{c}

     0.09\, ( 90\%\, \text{C.L.)~\cite{Amsterdam-CERN-Hamburg-Moscow-Rome:1980qbb} }  \\
    \, 0.017\, ( 90\%\, \text{C.L.)~\cite{NuTeV:2001bgq} }
\end{array}
,
\eea

In fact, there is an additional $u$-channel diagram from $Z'$ contribution supply for the $\nu_\mu e^- \to \mu^- \nu_e$  despite the SM ($W$) diagram. However, compared to the SM contribution, the $Z'$ contribution can be neglected. Therefore, the resulting ratio can be simplified as follows:
\bea \label{eq:inverseRa}
\frac{\sigma(\bar{\nu}_\mu e^- \to \mu^- \bar{\nu}_e)}{\sigma(\nu_\mu e^- \to \mu^- \nu_e)} \simeq  
\frac{g^4_{e\mu}}{2 \,G_{F}^{2}\,M^2_{Z'}s},(M_{Z'} <\sqrt{s}),
\eea
\bea \label{eq:inverseRb}
\frac{\sigma(\bar{\nu}_\mu e^- \to \mu^- \bar{\nu}_e)}{\sigma(\nu_\mu e^- \to \mu^- \nu_e)} \simeq  
\frac{g^4_{e\mu}}{3 \,G_{F}^{2}\,M^4_{Z'}},(M_{Z'} >\sqrt{s}).
\eea
where $\sqrt{s} \approx  \sqrt{2 E_{\nu_\mu}m_e} $. In this study, the limit was established by the newest NuTeV search.

\subsection{Muonium-to-antimuonium oscillation ($\rm{Mu}-\overline{\rm{Mu}}$) }
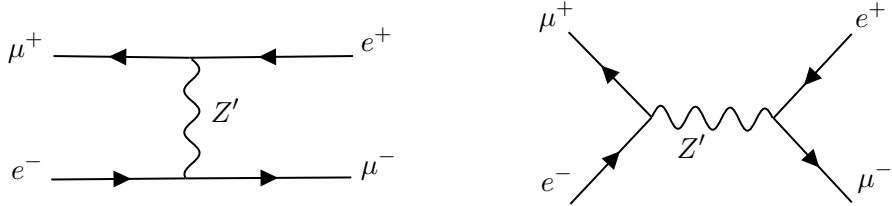
\begin{figure}[!h]
    \centering

\tikzset{every picture/.style={line width=0.75pt}} %set default line width to 0.75pt        

\begin{tikzpicture}[x=0.75pt,y=0.75pt,yscale=-1,xscale=1]
%uncomment if require: \path (0,2018); %set diagram left start at 0, and has height of 2018

%Shape: Wave [id:dp912821218987275] 
\draw   (188.53,1665.16) .. controls (190.5,1666.84) and (192.37,1668.44) .. (192.37,1670.35) .. controls (192.38,1672.26) and (190.52,1673.94) .. (188.56,1675.7) .. controls (186.61,1677.46) and (184.75,1679.14) .. (184.76,1681.05) .. controls (184.76,1682.95) and (186.63,1684.56) .. (188.6,1686.24) .. controls (190.56,1687.91) and (192.44,1689.52) .. (192.44,1691.42) .. controls (192.45,1693.33) and (190.59,1695.01) .. (188.63,1696.77) .. controls (186.68,1698.53) and (184.82,1700.21) .. (184.83,1702.12) .. controls (184.83,1704.03) and (186.7,1705.63) .. (188.67,1707.31) .. controls (190.63,1708.99) and (192.51,1710.59) .. (192.51,1712.5) .. controls (192.52,1714.4) and (190.66,1716.09) .. (188.7,1717.85) .. controls (186.75,1719.6) and (184.89,1721.29) .. (184.89,1723.19) .. controls (184.9,1723.83) and (185.1,1724.42) .. (185.45,1725) ;
%Straight Lines [id:da6493256682270944] 
\draw    (118.87,1663.58) -- (187.42,1664.45) ;
\draw [shift={(146.64,1663.93)}, rotate = 0.72] [fill={rgb, 255:red, 0; green, 0; blue, 0 }  ][line width=0.08]  [draw opacity=0] (8.93,-4.29) -- (0,0) -- (8.93,4.29) -- cycle    ;
%Straight Lines [id:da8364154934775649] 
\draw    (185.38,1725.34) -- (117.74,1725.9) ;
\draw [shift={(158.06,1725.56)}, rotate = 179.52] [fill={rgb, 255:red, 0; green, 0; blue, 0 }  ][line width=0.08]  [draw opacity=0] (8.93,-4.29) -- (0,0) -- (8.93,4.29) -- cycle    ;
%Straight Lines [id:da6914615325466917] 
\draw    (187.42,1664.45) -- (269.97,1663.58) ;
\draw [shift={(222.19,1664.08)}, rotate = 359.4] [fill={rgb, 255:red, 0; green, 0; blue, 0 }  ][line width=0.08]  [draw opacity=0] (8.93,-4.29) -- (0,0) -- (8.93,4.29) -- cycle    ;
%Straight Lines [id:da9765407636099361] 
\draw    (185.38,1725.34) -- (269.05,1724.86) ;
\draw [shift={(232.21,1725.07)}, rotate = 179.67] [fill={rgb, 255:red, 0; green, 0; blue, 0 }  ][line width=0.08]  [draw opacity=0] (8.93,-4.29) -- (0,0) -- (8.93,4.29) -- cycle    ;
%Shape: Wave [id:dp6947298606742727] 
\draw   (420.69,1694.26) .. controls (422.14,1691.38) and (423.54,1688.64) .. (425.12,1688.68) .. controls (426.71,1688.72) and (428.05,1691.54) .. (429.45,1694.49) .. controls (430.85,1697.45) and (432.19,1700.26) .. (433.78,1700.3) .. controls (435.37,1700.34) and (436.76,1697.6) .. (438.21,1694.72) .. controls (439.67,1691.84) and (441.06,1689.1) .. (442.64,1689.14) .. controls (444.23,1689.18) and (445.57,1691.99) .. (446.97,1694.95) .. controls (448.37,1697.91) and (449.72,1700.72) .. (451.3,1700.76) .. controls (452.89,1700.8) and (454.28,1698.06) .. (455.73,1695.18) .. controls (457.19,1692.3) and (458.58,1689.56) .. (460.16,1689.6) .. controls (461.75,1689.64) and (463.09,1692.45) .. (464.49,1695.41) .. controls (465.9,1698.37) and (467.24,1701.18) .. (468.82,1701.22) .. controls (470.41,1701.26) and (471.8,1698.52) .. (473.25,1695.64) .. controls (474.71,1692.76) and (476.1,1690.01) .. (477.68,1690.06) .. controls (479.09,1690.09) and (480.31,1692.32) .. (481.55,1694.89) ;
%Straight Lines [id:da8961358494998716] 
\draw    (378.67,1651.52) -- (420.63,1694.52) ;
\draw [shift={(395.11,1668.37)}, rotate = 45.7] [fill={rgb, 255:red, 0; green, 0; blue, 0 }  ][line width=0.08]  [draw opacity=0] (8.93,-4.29) -- (0,0) -- (8.93,4.29) -- cycle    ;
%Straight Lines [id:da6643680460574012] 
\draw    (420.63,1694.52) -- (379.51,1738.28) ;
\draw [shift={(404.52,1711.67)}, rotate = 133.22] [fill={rgb, 255:red, 0; green, 0; blue, 0 }  ][line width=0.08]  [draw opacity=0] (8.93,-4.29) -- (0,0) -- (8.93,4.29) -- cycle    ;
%Straight Lines [id:da6279865376511562] 
\draw    (481.98,1694.9) -- (521.46,1652.39) ;
\draw [shift={(497.29,1678.41)}, rotate = 312.88] [fill={rgb, 255:red, 0; green, 0; blue, 0 }  ][line width=0.08]  [draw opacity=0] (8.93,-4.29) -- (0,0) -- (8.93,4.29) -- cycle    ;
%Straight Lines [id:da24244368952996587] 
\draw    (481.98,1694.9) -- (521.46,1737.42) ;
\draw [shift={(505.12,1719.82)}, rotate = 227.12] [fill={rgb, 255:red, 0; green, 0; blue, 0 }  ][line width=0.08]  [draw opacity=0] (8.93,-4.29) -- (0,0) -- (8.93,4.29) -- cycle    ;

% Text Node
\draw (95.93,1711.58) node [anchor=north west][inner sep=0.75pt]   [align=left] {$\displaystyle e^{-}$};
% Text Node
\draw (272.28,1647.33) node [anchor=north west][inner sep=0.75pt]   [align=left] {$\displaystyle e^{+}$};
% Text Node
\draw (196.67,1683.55) node [anchor=north west][inner sep=0.75pt]   [align=left] {$\displaystyle Z'$};
% Text Node
\draw (272.28,1709.33) node [anchor=north west][inner sep=0.75pt]   [align=left] {$\displaystyle \mu ^{-}$};
% Text Node
\draw (94.28,1650.33) node [anchor=north west][inner sep=0.75pt]   [align=left] {$\displaystyle \mu ^{+}$};
% Text Node
\draw (363.19,1717.53) node [anchor=north west][inner sep=0.75pt]   [align=left] {$\displaystyle e^{-}$};
% Text Node
\draw (362.08,1634.23) node [anchor=north west][inner sep=0.75pt]   [align=left] {$\displaystyle \mu ^{+}$};
% Text Node
\draw (521.62,1635.1) node [anchor=north west][inner sep=0.75pt]   [align=left] {$\displaystyle e^{+}$};
% Text Node
\draw (522.78,1715.93) node [anchor=north west][inner sep=0.75pt]   [align=left] {$\displaystyle \mu ^{-}$};
% Text Node
\draw (431.77,1701.63) node [anchor=north west][inner sep=0.75pt]   [align=left] {$\displaystyle Z'$};

\end{tikzpicture}
    \caption{Feynman diagrams for muonium (Mu) bound state.}
    \label{fig:muonium}
\end{figure}

One specific observable, Muonium (Mu), forming a bound state of $\mu^+e^-$, would be relevant for exploring new physics, as it could induce Muonium-anti Muonium transitions due to the mediation of the $Z'$ between two bound states (e.g., $\mu^+ + e^- \to \mu^- + e^+ $ in Fig.~\ref{fig:muonium}.)~\cite{Kriewald:2022erk, Fukuyama:2021iyw}. 

This transition probability has been tested by the PSI experiment~\cite{Willmann:1998gd}, and its expression could be labeled as follows (only vector current and left/right-handed vertices are the same result in the constraint only proportional to $g^2_{e\mu}/M^2_{Z'}$),
\bea
P &=& \frac{2.57\times 10^{-5}}{\text{G}^2_\text{F}} \left( |c_{0,0}|^2 |- G_3 + G_{123}|^2 + |c_{1,0}|^2 | G_3 + G_{123}|^2\right) < 8.3 \times10^{-11},
\eea
with parameters,
\bea
G_1 &=& G_2 = \sqrt{2}\frac{g^2_{e\mu}}{8M^2_{Z'}},\, G_3 = \sqrt{2} \frac{2g^2_{e\mu}}{8M^2_{Z'}}, G_{123} =\frac{G_1 +G_2 -1/2\, G_3}{\sqrt{1+X^2}}.
\eea
In the above equations $|c_{0,0}|^2=0.32$ and $|c_{1,0}|^2=0.18$ represent the population states of the Mu and $X = 6.31 \times B / \rm{Tesla}$~\cite{Fukuyama:2021iyw}, with $B = 0.1\,\,\rm{Tesla}$~\cite{Willmann:1998gd}. Recently, the proposed Muonium-to-Antimuonium Conversion Experiment (MACE) Collaboration at CSNS aims to enhance transition sensitivity to 
 $\mathcal{O} (10^{-14})$~\cite{Bai:2022sxq,Zhao:2023plv}. This improvement leverages new intense and slow muon sources, along with an optimized detector design. In this study, we utilize the existing PSI measurements to establish the $Z'$ limit and display the proposed MACE expectations.

\subsection{Neutrino trident production with maximal LFV $Z'$}

\begin{figure}[!h]
    \centering
\tikzset{every picture/.style={line width=0.75pt}} %set default line width to 0.75pt        

\begin{tikzpicture}[x=0.75pt,y=0.75pt,yscale=-1,xscale=1]
%uncomment if require: \path (0,3034); %set diagram left start at 0, and has height of 3034

%Shape: Wave [id:dp6810793744081621] 
\draw   (146.54,1905.44) .. controls (149.02,1906.2) and (151.39,1906.93) .. (151.39,1907.81) .. controls (151.4,1908.69) and (149.04,1909.47) .. (146.57,1910.29) .. controls (144.1,1911.11) and (141.74,1911.89) .. (141.74,1912.77) .. controls (141.75,1913.65) and (144.11,1914.38) .. (146.6,1915.14) .. controls (149.08,1915.9) and (151.45,1916.63) .. (151.45,1917.51) .. controls (151.46,1918.39) and (149.1,1919.17) .. (146.63,1919.99) .. controls (144.15,1920.81) and (141.8,1921.6) .. (141.8,1922.47) .. controls (141.81,1923.35) and (144.17,1924.08) .. (146.66,1924.85) .. controls (149.14,1925.61) and (151.5,1926.34) .. (151.51,1927.22) .. controls (151.51,1928.09) and (149.16,1928.88) .. (146.68,1929.7) .. controls (144.21,1930.52) and (141.85,1931.3) .. (141.86,1932.18) .. controls (141.87,1933.06) and (144.23,1933.79) .. (146.71,1934.55) .. controls (149.2,1935.31) and (151.56,1936.04) .. (151.57,1936.92) .. controls (151.57,1937.8) and (149.22,1938.58) .. (146.74,1939.4) .. controls (145.61,1939.78) and (144.49,1940.15) .. (143.63,1940.52) ;
%Straight Lines [id:da38245049642818163] 
\draw    (44.5,1905.64) -- (146.88,1905.32) ;
\draw [shift={(100.69,1905.46)}, rotate = 179.82] [fill={rgb, 255:red, 0; green, 0; blue, 0 }  ][line width=0.08]  [draw opacity=0] (8.93,-4.29) -- (0,0) -- (8.93,4.29) -- cycle    ;
%Straight Lines [id:da8415469607022967] 
\draw    (142.64,1940.81) -- (143.33,1981.45) ;
\draw [shift={(143.07,1966.13)}, rotate = 269.02] [fill={rgb, 255:red, 0; green, 0; blue, 0 }  ][line width=0.08]  [draw opacity=0] (8.93,-4.29) -- (0,0) -- (8.93,4.29) -- cycle    ;
%Straight Lines [id:da18500924463015256] 
\draw    (146.88,1905.32) -- (249.24,1905.18) ;
\draw [shift={(203.06,1905.25)}, rotate = 179.92] [fill={rgb, 255:red, 0; green, 0; blue, 0 }  ][line width=0.08]  [draw opacity=0] (8.93,-4.29) -- (0,0) -- (8.93,4.29) -- cycle    ;
%Straight Lines [id:da2718482766886384] 
\draw    (142.64,1940.81) -- (247.24,1941.14) ;
\draw [shift={(188.44,1940.96)}, rotate = 0.18] [fill={rgb, 255:red, 0; green, 0; blue, 0 }  ][line width=0.08]  [draw opacity=0] (8.93,-4.29) -- (0,0) -- (8.93,4.29) -- cycle    ;
%Straight Lines [id:da5200873882428273] 
\draw    (143.33,1981.45) -- (248.38,1981.03) ;
\draw [shift={(200.86,1981.22)}, rotate = 179.77] [fill={rgb, 255:red, 0; green, 0; blue, 0 }  ][line width=0.08]  [draw opacity=0] (8.93,-4.29) -- (0,0) -- (8.93,4.29) -- cycle    ;
%Shape: Wave [id:dp24323480030386413] 
\draw   (145.2,1981.68) .. controls (145.2,1981.68) and (145.2,1981.68) .. (145.2,1981.68) .. controls (147.68,1982.44) and (150.05,1983.17) .. (150.05,1984.05) .. controls (150.06,1984.92) and (147.7,1985.71) .. (145.23,1986.53) .. controls (142.75,1987.35) and (140.4,1988.13) .. (140.4,1989.01) .. controls (140.41,1989.89) and (142.77,1990.62) .. (145.26,1991.38) .. controls (147.74,1992.14) and (150.11,1992.87) .. (150.11,1993.75) .. controls (150.12,1994.63) and (147.76,1995.41) .. (145.29,1996.23) .. controls (142.81,1997.05) and (140.46,1997.83) .. (140.46,1998.71) .. controls (140.47,1999.59) and (142.83,2000.32) .. (145.32,2001.08) .. controls (147.8,2001.84) and (150.16,2002.57) .. (150.17,2003.45) .. controls (150.17,2004.33) and (147.82,2005.11) .. (145.34,2005.93) .. controls (142.87,2006.75) and (140.51,2007.54) .. (140.52,2008.42) .. controls (140.52,2009.29) and (142.89,2010.02) .. (145.37,2010.79) .. controls (147.86,2011.55) and (150.22,2012.28) .. (150.23,2013.16) .. controls (150.23,2014.03) and (147.88,2014.82) .. (145.4,2015.64) .. controls (144.27,2016.01) and (143.15,2016.38) .. (142.28,2016.76) ;
%Straight Lines [id:da9048215997689105] 
\draw    (44.5,2018.04) -- (146.87,2017.73)(44.51,2021.04) -- (146.88,2020.73) ;
%Straight Lines [id:da3753680466343595] 
\draw    (146.87,2017.73) -- (249.24,2017.59)(146.88,2020.73) -- (249.24,2020.59) ;
%Shape: Wave [id:dp45289507358399594] 
\draw   (474.53,1905.44) .. controls (477.01,1906.2) and (479.37,1906.93) .. (479.38,1907.81) .. controls (479.38,1908.69) and (477.03,1909.47) .. (474.55,1910.29) .. controls (472.08,1911.11) and (469.72,1911.89) .. (469.73,1912.77) .. controls (469.73,1913.65) and (472.1,1914.38) .. (474.58,1915.14) .. controls (477.07,1915.9) and (479.43,1916.63) .. (479.44,1917.51) .. controls (479.44,1918.39) and (477.09,1919.17) .. (474.61,1919.99) .. controls (472.14,1920.81) and (469.78,1921.6) .. (469.79,1922.47) .. controls (469.79,1923.35) and (472.16,1924.08) .. (474.64,1924.85) .. controls (477.12,1925.61) and (479.49,1926.34) .. (479.49,1927.22) .. controls (479.5,1928.09) and (477.14,1928.88) .. (474.67,1929.7) .. controls (472.19,1930.52) and (469.84,1931.3) .. (469.84,1932.18) .. controls (469.85,1933.06) and (472.21,1933.79) .. (474.7,1934.55) .. controls (477.18,1935.31) and (479.55,1936.04) .. (479.55,1936.92) .. controls (479.56,1937.8) and (477.2,1938.58) .. (474.73,1939.4) .. controls (473.59,1939.78) and (472.48,1940.15) .. (471.61,1940.52) ;
%Straight Lines [id:da20016659158853123] 
\draw    (372.49,1905.64) -- (474.86,1905.32) ;
\draw [shift={(428.67,1905.46)}, rotate = 179.82] [fill={rgb, 255:red, 0; green, 0; blue, 0 }  ][line width=0.08]  [draw opacity=0] (8.93,-4.29) -- (0,0) -- (8.93,4.29) -- cycle    ;
%Straight Lines [id:da2847335445649015] 
\draw    (470.62,1940.81) -- (471.32,1981.45) ;
\draw [shift={(470.86,1954.63)}, rotate = 89.02] [fill={rgb, 255:red, 0; green, 0; blue, 0 }  ][line width=0.08]  [draw opacity=0] (8.93,-4.29) -- (0,0) -- (8.93,4.29) -- cycle    ;
%Straight Lines [id:da2872644405500169] 
\draw    (474.86,1905.32) -- (577.23,1905.18) ;
\draw [shift={(531.04,1905.25)}, rotate = 179.92] [fill={rgb, 255:red, 0; green, 0; blue, 0 }  ][line width=0.08]  [draw opacity=0] (8.93,-4.29) -- (0,0) -- (8.93,4.29) -- cycle    ;
%Straight Lines [id:da05228001117283276] 
\draw    (470.62,1940.81) -- (575.22,1941.14) ;
\draw [shift={(527.92,1940.99)}, rotate = 180.18] [fill={rgb, 255:red, 0; green, 0; blue, 0 }  ][line width=0.08]  [draw opacity=0] (8.93,-4.29) -- (0,0) -- (8.93,4.29) -- cycle    ;
%Straight Lines [id:da9820738199658854] 
\draw    (471.32,1981.45) -- (576.37,1981.03) ;
\draw [shift={(517.34,1981.27)}, rotate = 359.77] [fill={rgb, 255:red, 0; green, 0; blue, 0 }  ][line width=0.08]  [draw opacity=0] (8.93,-4.29) -- (0,0) -- (8.93,4.29) -- cycle    ;
%Shape: Wave [id:dp7251310924069799] 
\draw   (473.18,1981.68) .. controls (475.67,1982.44) and (478.03,1983.17) .. (478.04,1984.05) .. controls (478.04,1984.92) and (475.69,1985.71) .. (473.21,1986.53) .. controls (470.74,1987.35) and (468.38,1988.13) .. (468.39,1989.01) .. controls (468.39,1989.89) and (470.76,1990.62) .. (473.24,1991.38) .. controls (475.73,1992.14) and (478.09,1992.87) .. (478.1,1993.75) .. controls (478.1,1994.63) and (475.75,1995.41) .. (473.27,1996.23) .. controls (470.8,1997.05) and (468.44,1997.83) .. (468.45,1998.71) .. controls (468.45,1999.59) and (470.82,2000.32) .. (473.3,2001.08) .. controls (475.78,2001.84) and (478.15,2002.57) .. (478.15,2003.45) .. controls (478.16,2004.33) and (475.8,2005.11) .. (473.33,2005.93) .. controls (470.85,2006.75) and (468.5,2007.54) .. (468.5,2008.42) .. controls (468.51,2009.29) and (470.87,2010.02) .. (473.36,2010.79) .. controls (475.84,2011.55) and (478.21,2012.28) .. (478.21,2013.16) .. controls (478.22,2014.03) and (475.86,2014.82) .. (473.39,2015.64) .. controls (472.25,2016.01) and (471.14,2016.38) .. (470.27,2016.76) ;
%Shape: Wave [id:dp4155984141795297] 
\draw   (144.53,2064.44) .. controls (147.01,2065.2) and (149.37,2065.93) .. (149.38,2066.81) .. controls (149.38,2067.69) and (147.03,2068.47) .. (144.55,2069.29) .. controls (142.08,2070.11) and (139.72,2070.89) .. (139.73,2071.77) .. controls (139.73,2072.65) and (142.1,2073.38) .. (144.58,2074.14) .. controls (147.07,2074.9) and (149.43,2075.63) .. (149.44,2076.51) .. controls (149.44,2077.39) and (147.09,2078.17) .. (144.61,2078.99) .. controls (142.14,2079.81) and (139.78,2080.6) .. (139.79,2081.47) .. controls (139.79,2082.35) and (142.16,2083.08) .. (144.64,2083.85) .. controls (147.12,2084.61) and (149.49,2085.34) .. (149.49,2086.22) .. controls (149.5,2087.09) and (147.14,2087.88) .. (144.67,2088.7) .. controls (142.19,2089.52) and (139.84,2090.3) .. (139.84,2091.18) .. controls (139.85,2092.06) and (142.21,2092.79) .. (144.7,2093.55) .. controls (147.18,2094.31) and (149.55,2095.04) .. (149.55,2095.92) .. controls (149.56,2096.8) and (147.2,2097.58) .. (144.73,2098.4) .. controls (143.59,2098.78) and (142.48,2099.15) .. (141.61,2099.52) ;
%Straight Lines [id:da9927444398965708] 
\draw    (42.49,2064.64) -- (144.86,2064.32) ;
\draw [shift={(98.67,2064.46)}, rotate = 179.82] [fill={rgb, 255:red, 0; green, 0; blue, 0 }  ][line width=0.08]  [draw opacity=0] (8.93,-4.29) -- (0,0) -- (8.93,4.29) -- cycle    ;
%Straight Lines [id:da44962827001323213] 
\draw    (140.62,2099.81) -- (141.32,2140.45) ;
\draw [shift={(140.86,2113.63)}, rotate = 89.02] [fill={rgb, 255:red, 0; green, 0; blue, 0 }  ][line width=0.08]  [draw opacity=0] (8.93,-4.29) -- (0,0) -- (8.93,4.29) -- cycle    ;
%Straight Lines [id:da3791700335889393] 
\draw    (144.86,2064.32) -- (247.23,2064.18) ;
\draw [shift={(201.04,2064.25)}, rotate = 179.92] [fill={rgb, 255:red, 0; green, 0; blue, 0 }  ][line width=0.08]  [draw opacity=0] (8.93,-4.29) -- (0,0) -- (8.93,4.29) -- cycle    ;
%Straight Lines [id:da38231354523338446] 
\draw    (140.62,2099.81) -- (245.22,2100.14) ;
\draw [shift={(197.92,2099.99)}, rotate = 180.18] [fill={rgb, 255:red, 0; green, 0; blue, 0 }  ][line width=0.08]  [draw opacity=0] (8.93,-4.29) -- (0,0) -- (8.93,4.29) -- cycle    ;
%Straight Lines [id:da023905238203448365] 
\draw    (141.32,2140.45) -- (246.37,2140.03) ;
\draw [shift={(187.34,2140.27)}, rotate = 359.77] [fill={rgb, 255:red, 0; green, 0; blue, 0 }  ][line width=0.08]  [draw opacity=0] (8.93,-4.29) -- (0,0) -- (8.93,4.29) -- cycle    ;
%Shape: Wave [id:dp9421128776837834] 
\draw   (143.18,2140.68) .. controls (143.18,2140.68) and (143.18,2140.68) .. (143.18,2140.68) .. controls (145.67,2141.44) and (148.03,2142.17) .. (148.04,2143.05) .. controls (148.04,2143.92) and (145.69,2144.71) .. (143.21,2145.53) .. controls (140.74,2146.35) and (138.38,2147.13) .. (138.39,2148.01) .. controls (138.39,2148.89) and (140.76,2149.62) .. (143.24,2150.38) .. controls (145.73,2151.14) and (148.09,2151.87) .. (148.1,2152.75) .. controls (148.1,2153.63) and (145.75,2154.41) .. (143.27,2155.23) .. controls (140.8,2156.05) and (138.44,2156.83) .. (138.45,2157.71) .. controls (138.45,2158.59) and (140.82,2159.32) .. (143.3,2160.08) .. controls (145.78,2160.84) and (148.15,2161.57) .. (148.15,2162.45) .. controls (148.16,2163.33) and (145.8,2164.11) .. (143.33,2164.93) .. controls (140.85,2165.75) and (138.5,2166.54) .. (138.5,2167.42) .. controls (138.51,2168.29) and (140.87,2169.02) .. (143.36,2169.79) .. controls (145.84,2170.55) and (148.21,2171.28) .. (148.21,2172.16) .. controls (148.22,2173.03) and (145.86,2173.82) .. (143.39,2174.64) .. controls (142.25,2175.01) and (141.14,2175.38) .. (140.27,2175.76) ;
%Straight Lines [id:da3791794684919624] 
\draw    (42.48,2177.04) -- (144.86,2176.73)(42.49,2180.04) -- (144.87,2179.73) ;
%Straight Lines [id:da06564687011718784] 
\draw    (144.86,2176.73) -- (247.22,2176.59)(144.86,2179.73) -- (247.23,2179.59) ;
%Shape: Wave [id:dp5771540530314572] 
\draw   (472.51,2064.44) .. controls (474.99,2065.2) and (477.36,2065.93) .. (477.36,2066.81) .. controls (477.37,2067.69) and (475.01,2068.47) .. (472.54,2069.29) .. controls (470.06,2070.11) and (467.71,2070.89) .. (467.71,2071.77) .. controls (467.72,2072.65) and (470.08,2073.38) .. (472.57,2074.14) .. controls (475.05,2074.9) and (477.42,2075.63) .. (477.42,2076.51) .. controls (477.43,2077.39) and (475.07,2078.17) .. (472.6,2078.99) .. controls (470.12,2079.81) and (467.77,2080.6) .. (467.77,2081.47) .. controls (467.78,2082.35) and (470.14,2083.08) .. (472.63,2083.85) .. controls (475.11,2084.61) and (477.47,2085.34) .. (477.48,2086.22) .. controls (477.48,2087.09) and (475.13,2087.88) .. (472.65,2088.7) .. controls (470.18,2089.52) and (467.82,2090.3) .. (467.83,2091.18) .. controls (467.83,2092.06) and (470.2,2092.79) .. (472.68,2093.55) .. controls (475.17,2094.31) and (477.53,2095.04) .. (477.54,2095.92) .. controls (477.54,2096.8) and (475.19,2097.58) .. (472.71,2098.4) .. controls (471.57,2098.78) and (470.46,2099.15) .. (469.59,2099.52) ;
%Straight Lines [id:da5615741691940231] 
\draw    (370.47,2064.64) -- (472.85,2064.32) ;
\draw [shift={(426.66,2064.46)}, rotate = 179.82] [fill={rgb, 255:red, 0; green, 0; blue, 0 }  ][line width=0.08]  [draw opacity=0] (8.93,-4.29) -- (0,0) -- (8.93,4.29) -- cycle    ;
%Straight Lines [id:da3074646630046556] 
\draw    (468.61,2099.81) -- (469.3,2140.45) ;
\draw [shift={(468.84,2113.63)}, rotate = 89.02] [fill={rgb, 255:red, 0; green, 0; blue, 0 }  ][line width=0.08]  [draw opacity=0] (8.93,-4.29) -- (0,0) -- (8.93,4.29) -- cycle    ;
%Straight Lines [id:da013390327369870336] 
\draw    (472.85,2064.32) -- (575.21,2064.18) ;
\draw [shift={(529.03,2064.25)}, rotate = 179.92] [fill={rgb, 255:red, 0; green, 0; blue, 0 }  ][line width=0.08]  [draw opacity=0] (8.93,-4.29) -- (0,0) -- (8.93,4.29) -- cycle    ;
%Straight Lines [id:da41361939194617314] 
\draw    (468.61,2099.81) -- (573.21,2100.14) ;
\draw [shift={(525.91,2099.99)}, rotate = 180.18] [fill={rgb, 255:red, 0; green, 0; blue, 0 }  ][line width=0.08]  [draw opacity=0] (8.93,-4.29) -- (0,0) -- (8.93,4.29) -- cycle    ;
%Straight Lines [id:da8525071545660965] 
\draw    (469.3,2140.45) -- (574.35,2140.03) ;
\draw [shift={(515.33,2140.27)}, rotate = 359.77] [fill={rgb, 255:red, 0; green, 0; blue, 0 }  ][line width=0.08]  [draw opacity=0] (8.93,-4.29) -- (0,0) -- (8.93,4.29) -- cycle    ;
%Shape: Wave [id:dp2165576376410756] 
\draw   (471.17,2140.68) .. controls (473.65,2141.44) and (476.02,2142.17) .. (476.02,2143.05) .. controls (476.03,2143.92) and (473.67,2144.71) .. (471.2,2145.53) .. controls (468.72,2146.35) and (466.37,2147.13) .. (466.37,2148.01) .. controls (466.38,2148.89) and (468.74,2149.62) .. (471.23,2150.38) .. controls (473.71,2151.14) and (476.07,2151.87) .. (476.08,2152.75) .. controls (476.09,2153.63) and (473.73,2154.41) .. (471.26,2155.23) .. controls (468.78,2156.05) and (466.43,2156.83) .. (466.43,2157.71) .. controls (466.44,2158.59) and (468.8,2159.32) .. (471.28,2160.08) .. controls (473.77,2160.84) and (476.13,2161.57) .. (476.14,2162.45) .. controls (476.14,2163.33) and (473.79,2164.11) .. (471.31,2164.93) .. controls (468.84,2165.75) and (466.48,2166.54) .. (466.49,2167.42) .. controls (466.49,2168.29) and (468.86,2169.02) .. (471.34,2169.79) .. controls (473.83,2170.55) and (476.19,2171.28) .. (476.2,2172.16) .. controls (476.2,2173.03) and (473.84,2173.82) .. (471.37,2174.64) .. controls (470.23,2175.01) and (469.12,2175.38) .. (468.25,2175.76) ;
%Straight Lines [id:da2302685365228424] 
\draw    (369.5,2018.04) -- (471.87,2017.73)(369.51,2021.04) -- (471.88,2020.73) ;
%Straight Lines [id:da13725675111988778] 
\draw    (471.87,2017.73) -- (574.24,2017.59)(471.88,2020.73) -- (574.24,2020.59) ;
%Straight Lines [id:da7089149147143458] 
\draw    (370.48,2177.04) -- (472.86,2176.73)(370.49,2180.04) -- (472.87,2179.73) ;
%Straight Lines [id:da9062659520856858] 
\draw    (472.86,2176.73) -- (575.22,2176.59)(472.86,2179.73) -- (575.23,2179.59) ;

% Text Node
\draw (22.08,1897.27) node [anchor=north west][inner sep=0.75pt]   [align=left] {$\displaystyle \nu _{\mu }$};
% Text Node
\draw (150.07,1950.61) node [anchor=north west][inner sep=0.75pt]   [align=left] {$\displaystyle e^{-}$};
% Text Node
\draw (156.22,1912.47) node [anchor=north west][inner sep=0.75pt]   [align=left] {$\displaystyle Z'$};
% Text Node
\draw (246.7,1927.96) node [anchor=north west][inner sep=0.75pt]   [align=left] {$\displaystyle \mu ^{+}$};
% Text Node
\draw (251.85,1897.5) node [anchor=north west][inner sep=0.75pt]   [align=left] {$\displaystyle \nu _{e}$};
% Text Node
\draw (248.34,1968.46) node [anchor=north west][inner sep=0.75pt]   [align=left] {$\displaystyle e^{-}$};
% Text Node
\draw (22.02,2011.75) node [anchor=north west][inner sep=0.75pt]   [align=left] {$\displaystyle N$};
% Text Node
\draw (252.49,2011.75) node [anchor=north west][inner sep=0.75pt]   [align=left] {$\displaystyle N$};
% Text Node
\draw (350.07,1897.27) node [anchor=north west][inner sep=0.75pt]   [align=left] {$\displaystyle \nu _{\mu }$};
% Text Node
\draw (478.05,1950.61) node [anchor=north west][inner sep=0.75pt]   [align=left] {$\displaystyle \mu ^{+}$};
% Text Node
\draw (484.21,1912.47) node [anchor=north west][inner sep=0.75pt]   [align=left] {$\displaystyle Z'$};
% Text Node
\draw (574.69,1927.96) node [anchor=north west][inner sep=0.75pt]   [align=left] {$\displaystyle e^{-}$};
% Text Node
\draw (579.84,1896.5) node [anchor=north west][inner sep=0.75pt]   [align=left] {$\displaystyle \nu _{e}$};
% Text Node
\draw (576.32,1968.46) node [anchor=north west][inner sep=0.75pt]   [align=left] {$\displaystyle \mu ^{+}$};
% Text Node
\draw (351,2011.75) node [anchor=north west][inner sep=0.75pt]   [align=left] {$\displaystyle N$};
% Text Node
\draw (581.48,2009.75) node [anchor=north west][inner sep=0.75pt]   [align=left] {$\displaystyle N$};
% Text Node
\draw (154.92,1991.67) node [anchor=north west][inner sep=0.75pt]   [align=left] {$\displaystyle \gamma $};
% Text Node
\draw (482.92,1989.67) node [anchor=north west][inner sep=0.75pt]   [align=left] {$\displaystyle \gamma $};
% Text Node
\draw (20.07,2056.27) node [anchor=north west][inner sep=0.75pt]   [align=left] {$\displaystyle \nu _{\mu }$};
% Text Node
\draw (148.05,2109.61) node [anchor=north west][inner sep=0.75pt]   [align=left] {$\displaystyle \mu ^{+} /e^{+}$};
% Text Node
\draw (151.21,2070.47) node [anchor=north west][inner sep=0.75pt]   [align=left] {$\displaystyle Z$};
% Text Node
\draw (245.69,2086.96) node [anchor=north west][inner sep=0.75pt]   [align=left] {$\displaystyle \mu ^{-} /e^{-}$};
% Text Node
\draw (249.84,2054.5) node [anchor=north west][inner sep=0.75pt]   [align=left] {$\displaystyle \nu _{\mu }$};
% Text Node
\draw (21,2169.75) node [anchor=north west][inner sep=0.75pt]   [align=left] {$\displaystyle N$};
% Text Node
\draw (249.48,2169.75) node [anchor=north west][inner sep=0.75pt]   [align=left] {$\displaystyle N$};
% Text Node
\draw (348.05,2054.27) node [anchor=north west][inner sep=0.75pt]   [align=left] {$\displaystyle \nu _{\mu }$};
% Text Node
\draw (482.19,2071.47) node [anchor=north west][inner sep=0.75pt]   [align=left] {$\displaystyle W^{+}$};
% Text Node
\draw (348.98,2168.75) node [anchor=north west][inner sep=0.75pt]   [align=left] {$\displaystyle N$};
% Text Node
\draw (580.46,2167.75) node [anchor=north west][inner sep=0.75pt]   [align=left] {$\displaystyle N$};
% Text Node
\draw (149.91,2150.67) node [anchor=north west][inner sep=0.75pt]   [align=left] {$\displaystyle \gamma $};
% Text Node
\draw (479.91,2148.67) node [anchor=north west][inner sep=0.75pt]   [align=left] {$\displaystyle \gamma $};
% Text Node
\draw (245.05,2127.61) node [anchor=north west][inner sep=0.75pt]   [align=left] {$\displaystyle \mu ^{+} /e^{+}$};
% Text Node
\draw (577.69,2054.96) node [anchor=north west][inner sep=0.75pt]   [align=left] {$\displaystyle \mu ^{-}$};
% Text Node
\draw (575.84,2085.5) node [anchor=north west][inner sep=0.75pt]   [align=left] {$\displaystyle \nu _{\mu } /\nu _{e}$};
% Text Node
\draw (475.05,2110.61) node [anchor=north west][inner sep=0.75pt]   [align=left] {$\displaystyle \mu ^{+} /e^{+}$};
% Text Node
\draw (573.05,2127.61) node [anchor=north west][inner sep=0.75pt]   [align=left] {$\displaystyle \mu ^{+} /e^{+}$};
\end{tikzpicture}
\caption{Top panels: Feynman diagrams illustrating the neutrino trident process ($\nu_\mu + N$) featuring off-diagonal lepton flavor violation (LFV) interactions. Lower panels:  The muon neutrino trident process mediated by the SM electroweak interactions through $W/Z$ exchange. In the latter, photon couplings with negatively charged leptons diagrams are omitted.    }
    \label{fig:tridiag}
\end{figure}
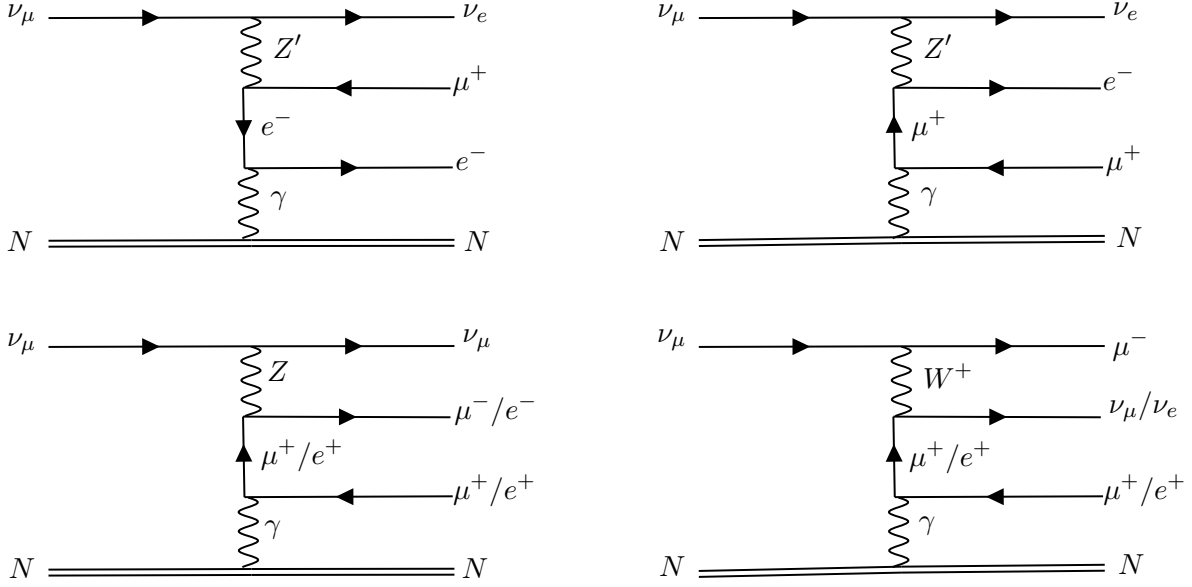

Neutrino trident production~\cite{Fujikawa:1971nx,Czyz:1964zz,Koike:1971tu,Koike:1971vg,Brown:1972vne}, where a neutrino collides with  the electric field of a nucleus and produces a pair of charged leptons, is a rare process involve neutral and charged lepton electroweak interactions. This phenomenon has been investigated by several experimental groups, as listed in Table~\ref{tab:trident}. It is worth noting that the proposed SHiP experiment~\cite{Pastore:2762117,DiCrescenzo:2023czg}  will be sensitive to LFV in future searches. Additionally, these processes probe new types of interactions mediated by particles beyond the SM, such as the LFV $Z'$ boson. Previous experiments have concentrated on utilizing produced $\mu^+\mu^-$ pairs to constrain the couplings of $Z'$ with leptons. These searches involving $Z'$ consider both off-diagonal couplings between different flavors and diagonal couplings with the same flavor. The main challenge lies in distinguishing the new physics signal from significant SM backgrounds, where final states could include $e^+e^-\nu,\mu^+\mu^-\nu$, or $e^+ \mu^-\nu$, as shown in the lower figures of Fig.~\ref{fig:tridiag}. However, the incoming $\nu_\mu$ beam also provides sensitivity to $\mu^+ e^-\nu$, a final state not produced by SM. For instance, the DUNE experiment~\cite{Altmannshofer:2019zhy} distinguishes between incoming $\nu_\mu$ and $\bar{\nu}_\mu$ fluxes, facilitating the isolation of final lepton states. Hence, it offers a clear signal for searches involving the $Z'\mu^+e^-$ vertex, as shown in the upper two figures of Fig.~\ref{fig:tridiag}. 

\begin{table}[!h]
    \centering
    \begin{tabular}{lll}
    \hline
    \hline
    Experiment & $E^{\rm{mean}}_{\nu_\mu}$ (GeV) & Target \\
    \hline
      CHARM-II~\cite{CHARM-II:1990dvf, Conrad:1997ne}   & $\sim 24 $  & silicon (Z=14,A=28) \\

       CCFR~\cite{Sakumoto:1990py, CCFR:1991lpl, King:1991gs, Spentzouris:1998pf}  &  160 & iron (Z=26,A=56) \\
    
       NuTeV~\cite{NuTeV:1999wlw, NuTeV:2001bgq} & 140 & iron (Z=26,A=56)\\

       DUNE~\cite{Magill:2016hgc, Ballett:2018uuc, Altmannshofer:2019zhy} & $\thicksim$ 3 & argon (Z=18,A=40)\\
    
       SHiP(proposed)~\cite{Pastore:2762117, DiCrescenzo:2023czg} & $\thicksim 40$ & Tungsten (Z=74,A=184) \\
       \hline
       \hline
    \end{tabular}
    \caption{
Previous and proposed experiments on neutrino tridents along with their respective setup configurations. }
    \label{tab:trident}
\end{table}

The amplitude for the distinctive clean signal channel ($\nu_\mu (k_{1}) + \gamma(q) \to \nu_e (k_{2}) + \mu^+ (p_1) +  e^-(p_2) $) based on EPA method~\cite{vonWeizsacker:1934nji,Williams:1934ad} will be expressed as follows\footnote{The EPA method is valid for dimuon channels when the neutrino energy exceeds 20 GeV~\cite{Ballett:2018uuc}. For mixed flavor states, its shows a factor of two enhancement against the full 2-to-4 calculations. To account for this, we adjust the cross-section enhancement applying a factor of one-third for the mean neutrino energy ($\Bar{E}_{\nu}\sim 3$ GeV).}~\cite{Belusevic:1987cw, Vysotsky:2002ix, Altmannshofer:2014pba},
\bea
\mathcal{M}_1 &=& g_{e\mu} \,e\, \bar{u}(p2) \bigg( \epsilon^{\mu}\gamma_\mu. \frac{g^{\alpha \beta}-\frac{k^{\alpha}k^{\beta}}{M_{Z'}^2}}{k^2- M^2_{Z'}} \frac{\gamma.(-p_1 +q) + m_\mu }{(-p_1 + q)^2 - m^2_\mu} \gamma^\alpha   \\ 
&&+ \epsilon^{\mu}\gamma_\mu. \frac{g^{\alpha \beta}-\frac{k^{\alpha}k^{\beta}}{M_{Z'}^2}}{k^2- M^2_{Z'}}\frac{\gamma.(p_2 -q) + m_e }{(p_2 -q)^2 - m^2_e} \gamma^\alpha \bigg) v(p_1), \nonumber \\
\mathcal{M}_2 & =& g_{e\mu} \bar{u}(k_2) \gamma^\beta u(k_1) , \,\\
d \sigma^{\rm{Z'}}_{\nu_\mu \gamma} &\approx& \frac{1}{2 s} \frac{1}{2} \frac{d t}{2 s }\frac{dl}{2 \pi} v \frac{d\Omega}{4 \pi} |\mathcal{M}_{\nu_\mu \gamma \to \nu_e \mu^+ e^-}|^2.
\eea
In the above expression $k=k_1-k_2$, and the velocity $v$ is defined as $v= \frac{l-m^2_\mu}{l+m^2_\mu}$ where $m_e \to 0$. Here, $l$ represents the invariant momentum squared of the $\mu$ and $e$, given by $l = (P_{\mu^+} + P_{e^-})^2$. The resulting amplitude is formulated to exclude terms involving $m_\mu$, as they are deemed negligible compared to other contributions. Subsequently, we reformulate the momentum products using new Mandelstam variables~\cite{Vysotsky:2002ix} (refer to the Appendix \ref{app:B} for details). After integrating out $t$ and $l$, we obtain the following expression when $m_\mu \ll M_{Z'} \ll \sqrt{s}$, 
\bea\label{eq:largezp}
\sigma^{Z'}_{\nu_\mu \gamma} \simeq \frac{1}{M_{Z'}^2} \frac{g_{e\mu}^4 \alpha_{\text{EM}}}{2\pi^{2}}\log\left(\frac{M_{Z'}^2}{m_{\mu}^2}\right),
\eea
while for $M_{Z'} \ll  m_{\mu} \ll \sqrt{s}$, the cross-section can be simplified as follows,
\bea \label{eq:smallzp}
\sigma^{Z'}_{\nu_\mu \gamma} \simeq \frac{1}{m_{\mu}^2} \frac{3\,g_{e\mu}^4 \alpha_{\text{EM}}}{10\,\pi^{2}}\log\left(\frac{m_{\mu}^2}{M_{Z'}^2}\right),
\eea

Thus, the total cross-section for the neutrino trident production with distinctive $Z'$ signals can be expressed as follows, 
\bea
\sigma (\nu_\mu  \rm{N} \to \nu_e \mu^+  e^-  \rm{N} ) &\approx & \frac{Z^2\alpha_{\text{EM}}}{\pi} \int^{s_{\text{Max}}}_{s_{\text{Min}}} \frac{1}{s} ds \,\int^{q^2_{\text{Max}}}_{q^2_{\text{Min}}} dq^2\, \frac{1}{q^2}\, F^2(A,q^2)\, \,\sigma^{\rm{Z'}}_{\nu_\mu \gamma} ,  \\
q^2_{\text{Max}} & \thicksim & \left(\frac{m_\pi}{A^{1/3}}\right)^2 ,\, q^2_{\text{Min}} = \left(\frac{s}{2E_\nu}\right)^2,\, \non \\ 
s_{\text{Max}} &=& 2 E_\nu \, q_{\text{Max}},\, s_{\text{Min}} =  m^2_\mu,
\eea
where $Z,A$ are the relevant atomic number and atomic mass of the material in the target and we adopted the electromagnetic nucleus form factor as in Ref.~\cite{Altmannshofer:2019zhy},
\bea
F(A,q^2)= \text{exp} \bigg \{ \frac{-\left(1.3 \,\text{fm}\, A^{1/3} \right)^2 q^2}{10 } \bigg\},
\eea 
We used the full $\sigma^{Z'}_{\nu \gamma}$ results to evaluate the expected $Z'$ signals. The number of coherent events will be obtained by the following form~\cite{Altmannshofer:2019zhy},
\bea
N_{\text{trident}} &=& F_{\nu_\mu} \Bar{\sigma}(\nu_\mu N\to \nu_e \mu^+ e^- N) \frac{M_\text{detector}}{M_{\text{target atom}}} N_{\text{POT}},\\
\Bar{\sigma}&=& \int \sigma{(E_\nu}) f(E_\nu) d E_\nu,
\eea
where $F_{\nu_\mu}$ is the integrated $\nu_\mu$ flux correspond to the incoming $\nu_\mu$ energy beam, and $N_{\text{POT}}$ is the number of protons on target. Here, $f(E_{\nu})$ represents the normalized energy spectrum of the neutrino~\cite{Altmannshofer:2019zhy}. For the DUNE experiment, the values are given by $F_{\nu_\mu} = 1.04 \times 10^{-3}$ m$^{-2}$ POT$^{-1}$ and $N_{\text{POT}} = 1.1\times 10^{21}$ POT per year. The ratio $M_\text{detector}/M_{\text{target atom}}$ is correlated with DUNE's argon liquid detector (LArTPC) mass (147 tonnes) and atomic mass number (A = 40). Although the approach outlined above is free of SM background, the experiment faces the additional challenge of distinguishing between $\mu^+ e^-$ and $\mu^- e^+$ final states. A recent study in DUNE suggests that the charge identification efficiency for muons and antimuons can reach up to 95\%~\cite{Denton:2024glz,ICAL:2015stm}. Considering the mixed flavor states from the SM background, particularly $e^+ \mu^- \nu$ as described in Table 2 of \cite{Altmannshofer:2019zhy}, the total number of background events is significantly reduced, leaving approximately 5\% after signal selection. Starting with an estimated 4000 background events for $e^+ \mu^-$ states over 3 years, the expected upper limit of the signal can be determined using the inverse chi-squared function at a given confidence level~\cite{ParticleDataGroup:2024cfk}. In our scenario, the upper limit of the signal is approximately 31 events at 95\% confidence level (C.L.). In this study, our focus lies on the DUNE experiment in general, with the anticipation that estimations for other experiments would be similar.

\begin{figure}[!t]
    \centering
    \includegraphics[width=0.8\textwidth]{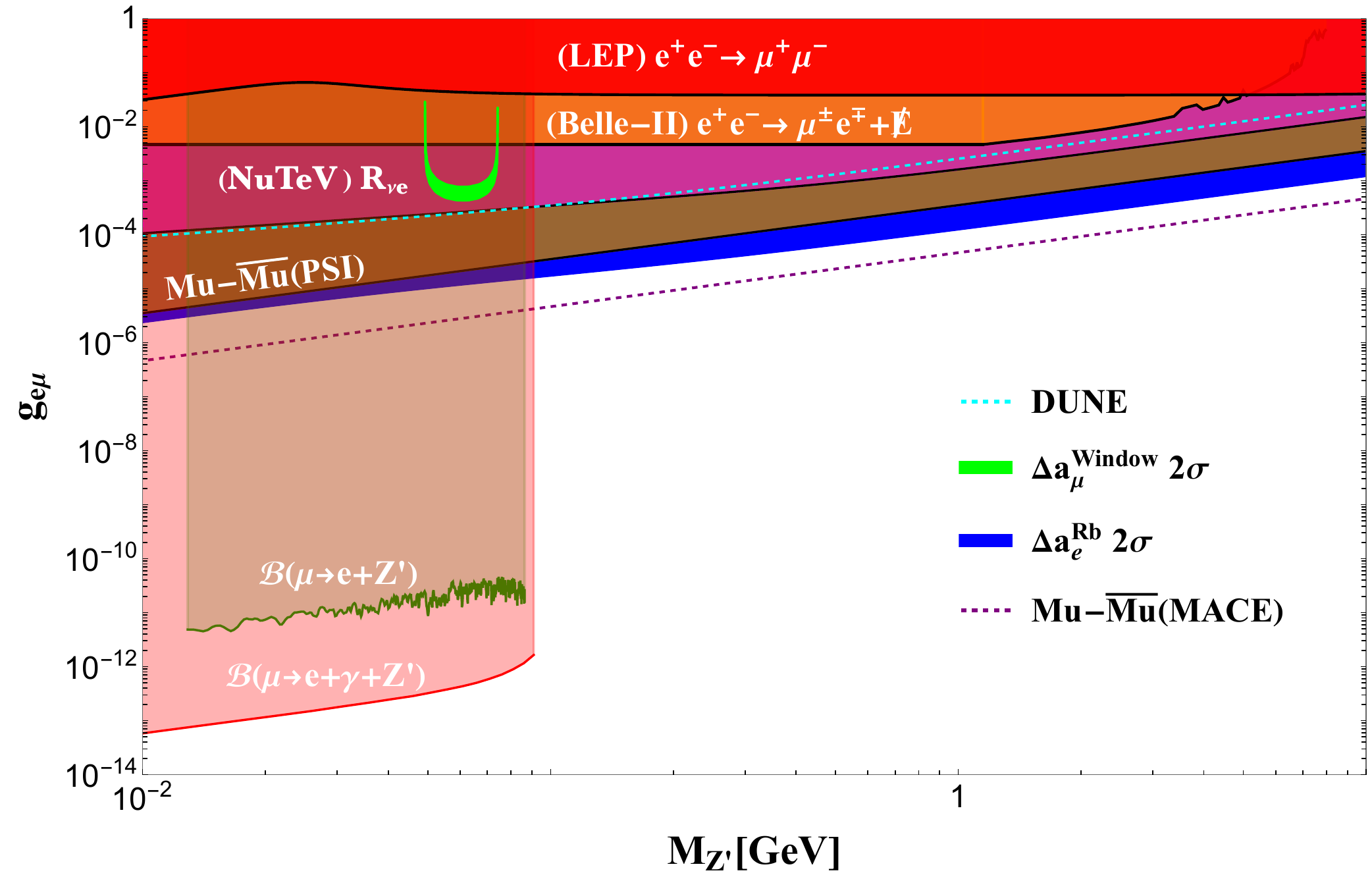}
    \caption{The exclusion limit in the ($M_{Z'}$, $g_{e\mu}$) parameter space under various constraints. The blue region denotes $\Delta a_e (2 \sigma)$. The green band indicates the $\Delta a_\mu (2 \sigma)$. The exclusion region for Mu-$\overline{\rm{Mu}}$ oscillation by PSI is represented by the brown area~\cite{Willmann:1998gd}. The proposed sensitivity expectation at MACE is indicated by the dashed purple line~\cite{Bai:2022sxq}, while the cyan dashed line illustrates the exclusion limits for DUNE.~\cite{Altmannshofer:2019zhy}. The top red region is exclusion from the LEP ($e^+ e^- \to \mu^+ \mu^-$) at $\sqrt{s} = 189$ GeV~\cite{DELPHI:2000ztm} and the orange region is Belle-II exclusion from ($e^+ e^- \to \mu^\pm e^\mp + \slashed{E}$)~\cite{Belle-II:2019qfb}. The inverse $\mu$ decay mode from the NuTeV search, providing the ratio $R_{\nu e}$, Eq.~(\ref{eq:inverseratio}), is labeled in magenta region~\cite{NuTeV:2001bgq}. The dark green line labels the exclusion region from $\mathcal{B} (\mu \to e + X)$~\cite{TWIST:2014ymv}. The most stringent constraints for $g_{ e \mu}$ below $m_\mu$ originate from $\mathcal{B}(\mu \to e + \gamma + X)$ (red)~\cite{Goldman:1987hy}.   }
    \label{fig:LFVresult}
\end{figure}

\section{Results}\label{sec:4}

To verify the bounds for $g_{e\mu}$ and $M_{Z'}$, we adopted the Lagrangian generated by the Feynrules~\cite{Alloul:2013bka} and using the Madgraph5~\cite{Alwall:2011uj} to calculate the LFV three-body decay and cross section results to compare with the measurements. The decay widths of $Z'$ and analytical evaluations are obtained by using the FeynArt~\cite{Hahn:2000kx} and FeynCalc~\cite{MERTIG1991345, Shtabovenko:2016sxi, Shtabovenko:2020gxv}. 
Since our focus is within the mass range of $0.01$ GeV to $10$ GeV, we do not consider the constraints from BBN and cosmology.

The constraints of $g_{e\mu}$ versus $M_{Z'}$ are shown in Fig.~\ref{fig:LFVresult}. For the collider approach, the LEP measurement of $\sigma(e^+ e^- \to \mu^+ \mu^-)$~\cite{DELPHI:2000ztm} has labeled the top red region, while the orange region is excluded by Belle-II ($e^+ e^- \to \mu^\pm e^\mp + \slashed{E}$)~\cite{Belle-II:2019qfb}. For the $e^+ e^- \to \mu^+ \mu^-$ process, the exclusion bound remains flat for $M_{Z'} > m_\mu$, primarily due to the dominant effect of the $1/s$ term. However, as $M_{Z'}$ decreases further, the enhancement of $1/M^2_{Z'}$ dominates, results in a reduction of the cross-section, leading to tighter limits as $M_{Z'}$ approaches $m_e$. This tightening effect is attributed to the $1/(t-M^2_{Z'})$ from new physics diagram. 

The magenta and brown areas represent exclusion regions for inverse $\mu$ decay ratio $R_{\nu e}$ and Mu-$\overline{\rm{Mu}}$ oscillation, respectively. From the simplified analytical formulas Eqs.~(\ref{eq:inverseRa} and~\ref{eq:inverseRb}), the constraints are proportional to $ g^4_{e\mu}/M^2_{Z'}$ with an additional $1/s$ suppression for $M_{Z'} < \sqrt{s}$ and to $ g^4_{e\mu}/M^4_{Z'}$ for $M_{Z'} > \sqrt{s}$. The limit from the brown region is proportional to $ g^2_{e\mu}/M^2_{Z'}$, leading to a linear enhancement of the bounds as $M_{Z'} \to m_e$. From the plot, the inverse muon decay is weaker than the Mu-$\overline{\rm{Mu}}$ oscillation by PSI. 

In the context of the neutrino trident process, the DUNE experiment estimates the significance with the dashed cyan line for a mean neutrino energy of $E_{\nu_\mu} \sim  3$ GeV. The exclusion region derived from trident events is slightly more stringent than those obtained from collider searches and remains comparable to the limits from past PSI Mu-$\overline{\rm{Mu}}$ oscillation as $M_{Z'}$ approaches 10 GeV, as indicated by the plot. From Eqs.~(\ref{eq:largezp} and \ref{eq:smallzp}), the cross-sections exhibit dependencies proportional to $g^4_{e\mu} \log\left(1/M^2_{Z'}\right)$ and $g^4_{e\mu}/M^2_{Z'} \log\left(M^2_{Z'}\right)$, respectively.  As $M_{Z'}$ increases from below $m_\mu$ to around 10 GeV, the constraint curve transitions smoothly, reflecting the enhanced cross-section when $M_{Z'} < m_\mu$ due to the logarithmic term $\log\left(1/M^2_{Z'}\right)$, which strengthens the coupling limit $g_{e\mu}$. Conversely, the trident cross-section is suppressed as $M_{Z'} $ exceeds $m_\mu$ owing to the inverse-square dependence $1/M_{Z'}^2$, resulting in a weaker constraint on the coupling.

The green band represents the 2$\sigma$ allowed range of $\Delta a_\mu$. It is observed that the $Z'$ itself, with $e-\mu$ couplings, can only resolve the $\Delta a_\mu$ discrepancy within a narrow mass region ($0.04\, \text{GeV}  < M_{Z'} < 0.08 \,\text{GeV}$), which is already excluded by above existing experimental constraints. Conversely, the $\Delta a_e$ provides stringent limits for $m_\mu < M_{Z'}$.

Based on the plot, it is apparent that the most stringent constraint arises from the $\mathcal{B}(\mu \to e + \gamma + Z')$ process when $M_{Z'} < m_\mu$. This limitation stems from the $M_{Z'}$ enhancement in the denominator, an extra factor of $\alpha_{\text{EM}}$, and the suppression of the three-body phase space, which includes an additional factor related to the density of states when integrating the final momenta of the particles, $\prod\limits^{3}_{i=1} \frac{d^3\vec{p}_i}{(2\pi)^{3}}$. From the aforementioned experimental constraints, $\Delta a_e$ plays a major role for $M_{Z'} > m_\mu$. However, regarding the proposed Mu-$\overline{\rm{Mu}}$ oscillation at MACE~\cite{Bai:2022sxq}, the limit on $Z'$ is expected to be stricter compared to the constraints imposed by $\Delta a_e$. This is represented by the dashed purple line in Fig.~\ref{fig:LFVresult}. Consequently, future measurements at MACE are anticipated to significantly enhance sensitivity towards detecting our maximal LFV $Z'$ scenario.

\section{Conclusions}\label{sec:5}

In this work, we investigated a novel $U(1)'$ gauge symmetry, focusing on the $Z'$ boson that couples with $e$ and $\mu$ through a specific charge configuration. We propose a UV model featuring four scalars—three doublets and one singlet—that contribute to the $M_{Z'}$ after symmetry breaking, without altering the SM $Z$ boson mass. The model employs a discrete symmetry $\ell_{1L} \to \ell_{2L},\, e_{1R} \to e_{2R},Z'_{\mu}\to - Z'_{\mu}$, leading to maximally off-diagonal interactions between $Z'$ and leptons, manifesting exclusively as lepton flavor-violating (LFV) processes. A notable aspect indicates that the Yukawa coupling between $\mu/e$ and the SM Higgs remains identical. Given recent measurements of the Higgs' leptonic decay branching ratios, we anticipate that future precision measurements will be sensitive to this model.

Next, we analyzed the existing experiments aimed at constraining the LFV coupling $g_{e\mu}$ of $Z'$. Specifically, the discrepancy between the SM prediction and the measured magnetic moment of the muon, $\Delta a_\mu$, cannot be explained by the $Z'$, as the parameter space where $Z'$ could account for this discrepancy has already been ruled out by other experimental measurements. We considered several LFV search limits, including those from Belle-II, LEP, LFV decays, inverse muon decay, Muonium-anti Muonium oscillation, and the newly proposed neutrino trident process ($\nu_\mu + \text{N} \to \nu_e + \mu^+ + e^- + \text{N}$). 

Neutrino trident processes are measured from the DUNE and some other similar experiments to test the LFV interaction. Most searches have focused on muon pair final states. However, the LFV interaction involve the maximal flavor violation can contribute a distinctive signal $\mu^+ e^-$ from $\nu_\mu$ beams without any SM background as the SM does not produce $\mu^+ e^-$ in the final state via $\nu_\mu$. We employed the neutrino trident production method to investigate the $Z'\mu^+ e^-$ vertex, utilizing its framework to incorporate the amplitude of a distinct, background-free signal through the DUNE experiment. In the appendix, we listed the kinematic variables for the 2-to-3 scattering process $\nu_\mu + \gamma \to \nu_e + \mu^+ + e^-$ and calculated the final cross-section for an initial $\nu_\mu$ collides with the nucleus. The resulting constraint on $Z'$ is stricter compared to those from colliders and LFV decays when $M_{Z'} > m_\mu$. This limit forms a quadratic curve with the couplings ranging from $\mathcal{O}(10^{-5})$ to $\mathcal{O}(10^{-2})$, though it does not surpass the constraint from $\Delta a_e$.  For $M_{Z'} < m_\mu$, the primary constraint comes from $\mathcal{B}(\mu \to e + \gamma + Z')$ due to the enhancement by the factor $1/M^2_{Z'}$ and the suppression both from the three-body phase space factor and QED coupling $\alpha_{\text{EM}}$. The Mu-$\overline{\rm{Mu}}$ oscillation searched at PSI exhibits a transition probability that differs notably from $\Delta a_e$. The proposed Muonium-to-Antimuonium Conversion Experiment (MACE), involving Mu-$\overline{\rm{Mu}}$ oscillations, is expected to surpass the sensitivity achieved in recent measurements of $\Delta a_e$, thereby enhancing validation of scenarios involving LFV $Z'$ and other LFV ($\mu$-$e$) interactions.

\appendix
\section{Scalar potential with neutral scalars}\label{app:A}

The scalar potential can be expressed as follows, incorporating the discrete symmetry ($\phi_1 \to \phi_2$)~\cite{Foot:1994vd},
\bea
V_{\phi_1,\phi_2,\phi_3,\phi_4} &=& \lambda_1 (\phi^{\dg}_3\phi_3 - v^2_3)^2 + \lambda_2 (\phi^{\dg}_1\phi_1+\phi^{\dg}_2\phi_2 - 2\,v^2_0)^2 \non \\ 
&+& \lambda_3 (\phi^{\dg}_1\phi_1 - \phi^{\dg}_2\phi_2)^2  +\lambda_4 (\phi^{\dg}_1\phi_1+\phi^{\dg}_2\phi_2 + \phi^{\dg}_3\phi_3 -v^2_3 -2\,v^2_0)^2 \non \\ 
&+& \lambda_5 [ \phi^{\dg}_1\phi_1 \phi^{\dg}_2\phi_2 - \phi^{\dg}_1\phi_2 \phi^{\dg}_2\phi_1 ] \non \\ 
&+& \lambda_6 [ \phi^{\dg}_3\phi_3 (\phi^{\dg}_1\phi_1+ \phi^{\dg}_2\phi_2) - \phi^{\dg}_3\phi_1 \phi^{\dg}_1\phi_3 - \phi^{\dg}_3\phi_2 \phi^{\dg}_2\phi_3 ] \non\\ 
&+& \lambda_7 [\phi^{\dg}_3\phi_3 (\phi^{\dg}_1\phi_1+ \phi^{\dg}_2\phi_2)- \phi^{\dg}_3\phi_1 \phi^{\dg}_3\phi_2 - \phi^{\dg}_1\phi_3 \phi^{\dg}_2\phi_3 ]
\non \\
&+& \frac{1}{2}\, \mu^2_{4} \phi^2_4 + \lambda_{44} \phi^4_4 ,
\eea
Based on the aforementioned reference and the alteration of our set up, one can redefine the CP-even scalar components, 
\bea\label{eq:lineartransform}
S_1= \frac{\rho_1 + \rho_2}{\sqrt{2}}, S_2= \frac{\rho_1 - \rho_2}{\sqrt{2}},
\eea
The CP even squared mass matrices can be expressed as follows ($v_0 \equiv v_1 = v_2$),
\bea 
M^2_{S_1}&=& 
 8 v^2_0 \lambda_2, \\
  M^2_{S_2} &=&  8 \lambda_3 v^2_0 + 2 v^2_3 \lambda_7, \label{eq: scalarsquaredhm} \\
  M^2_{H} &=& 4 v_3^2 \lambda_1 ,\\
M^2_{S_4} &=& \frac{1}{2} (\mu^2_4 + 8 v^2_4 \lambda_{44}) ,\label{eq: scalarsquaredh4}
\eea 
The above $M_H$ would be the SM Higgs mass, with the flavor states and mass states being identical ($\rho_1,\rho_2,\rho_3,\rho_4 \to S_1,S_2,H,S_4$) when we assume $\lambda_4 \to 0$. If $\lambda_4 \neq 0$, one could have non-diagonalized mass matrices,
\bea \label{eq:l4neq0mass}
M^2_{S_1,\rho_3} = \bigg( \begin{array}{cc}
 8 v_0^2 (\lambda_2 + \lambda_4)   & 4 \sqrt{2} v_0 v_3 \lambda_4 \\
  4 \sqrt{2} v_0 v_3 \lambda_4   &  4 v^2_3 (\lambda_1 + \lambda_4) 
\end{array}
\bigg),
\eea
On the other hand, $M^2_{S_2}$ and $M^2_{S_4}$ are automatically diagonalized as shown in Eqs.~(\ref{eq: scalarsquaredhm} and \ref{eq: scalarsquaredh4}), with  $S_2,S_4$ representing the mass states. The states ($S_1$ and $\rho_3$) are no longer directly equivalent to their mass eigenstates. Instead, the mass eigenstates ($S'_1$ and $H$) are introduced, and their relationship can be expressed through the following 2-by-2 mixing matrix, which rotates from the flavor states $S_1$ and $\rho_3$ to the mass states $S'_1$ and $H$),
\bea\label{eq: mixH}
\left(\begin{array}{cc}
  S_1     \\
   \rho_3   
\end{array}\right)
= 
\left(
\begin{array}{cc}
   \frac{\sqrt{2} v_0 v_3 \lambda_4 }{ v^2_3 \lambda_1 - 2 v^2_0 \lambda_2}   & 1 \\
    1 &  \frac{\sqrt{2} v_0 v_3 \lambda_4}{-v^3_3 \lambda_1 + 2 v^2_0 \lambda_2} 
\end{array} \right)\,
\left( \begin{array}{c}
    H \\
     S'_1
\end{array}
\right) + \mathcal{O} (\lambda^2_4),
\eea
Thus, the flavor states ($\rho_1,\rho_2,\rho_3,\rho_4$) transform into the physical mass states ($S'_1,S_2,H,S_4$) resembles the SM-like Higgs ($H$). 

In this context, we exclusively focus on the CP-even scalars due to their correlation with SM-like scalar Yukawa couplings. The CP-odd and charged components are not pertinent to our current discussion. The scalar potential introduces numerous parameters, resulting in a vast parameter space that could potentially include additional scalar states, which we do not investigate here.

\section{Kinematic variables for the neutrino trident process}\label{app:B}

The kinematic variables that transform the 4-momentum product into Mandelstam variables in the 2-to-3 neutrino trident production ($\nu_\mu (k_{1}) + \gamma(q) \to \nu_e (k_{2}) + \mu^+ (p_1) +  e^-(p_2) $) are expressed as follows:
\bea
 k_1^2 &=& k_2^2=0,\, k=k_1-k_2,\, k_1 + q = k_2+ p_1 +p_2,\\
\kappa_{1}&=&m_{\mu}^{2}-p_1^2,\,\kappa_{2}=-p_2^2, \\s &=& (k_1 + q)^2,\, l = (p_1 +p_2)^2, \, t=2 k.q=\kappa_{1}+\kappa_{2}=l-k^2-q^2,  \\
 \kappa_1 &\to& \frac{t}{2} \left(1+\frac{m^2_\mu}{l}+x(1-\frac{m^2_\mu}{l}) \right) , \\ \kappa_2 &\to & \frac{t}{2}\left(1-\frac{m^2_\mu}{l}-x (1-\frac{m^2_\mu}{l}) \right), \kappa_3 \to  l-t,\\
k_1.q &\to &\frac{s}{2},\, k_2.q\to \frac{s- t}{2}, \, p_1.q \to \frac{\kappa_1}{2},\, p_2.q \to \frac{\kappa_2}{2},  \\
 k_1.p_1&\to& \frac{l-\kappa_1}{2} + \frac{s-l}{4} \left(1+\frac{m^2_\mu}{l}+x(1 - \frac{m^2_\mu}{l})   \right),\\k_1.p_2 &\to& \frac{l-\kappa_2}{2} + \frac{s-l}{4} \left(1-\frac{m^2_\mu}{l}-x (1-\frac{m^2_\mu}{l})  \right), \\
  k_2.p_1&\to&  \frac{s-l}{4} \left(1 + \frac{m^2_\mu}{l}+x (1- \frac{m^2_\mu}{l} )  \right),\\
  k_2.p_2 &\to &  \frac{s-l}{4} \left(1- \frac{m^2_\mu}{l} - x(1- \frac{m^2_\mu}{l})  \right), \\
  p_1.p_2 &\to & \frac{l}{2} -m^2_\mu, \, k_1.k_2 \to -\frac{\kappa_3}{2}.
\eea
We employed the notations introduced in Ref.~\cite{Vysotsky:2002ix}, where $x=v \cos\theta_{qp_1} $, with $v$ representing the velocity of the muon in the center-of-mass (COM) frame of the $\mu^+(p_1)e^-(p_2)$ pair, and we take $q^2 \to 0$. In these expressions, the electron mass is neglected as it remains the lightest particle regardless of the value of $M_{Z'}$.

\acknowledgments
We thank Jinhui Guo and Yuxuan He for their discussion on the early stage of the work. The work of JL is supported by Natural Science Foundation of China under grant No. 12075005 and 12235001.

%\bibliographystyle{JHEP}
%\bibliography{refs}

\begin{thebibliography}{100}

\bibitem{Feinberg:1959ui}
G.~Feinberg, P.~Kabir and S.~Weinberg, \emph{{Transformation of muons into
  electrons}}, \href{https://doi.org/10.1103/PhysRevLett.3.527}{\emph{Phys.
  Rev. Lett.} {\bfseries 3} (1959) 527}.

\bibitem{Feinberg:1961zz}
G.~Feinberg and S.~Weinberg, \emph{{Law of Conservation of Muons}},
  \href{https://doi.org/10.1103/PhysRevLett.6.381}{\emph{Phys. Rev. Lett.}
  {\bfseries 6} (1961) 381}.

\bibitem{Cirigliano:2005ck}
V.~Cirigliano, B.~Grinstein, G.~Isidori and M.B.~Wise, \emph{{Minimal flavor
  violation in the lepton sector}},
  \href{https://doi.org/10.1016/j.nuclphysb.2005.08.037}{\emph{Nucl. Phys. B}
  {\bfseries 728} (2005) 121}
  [\href{https://arxiv.org/abs/hep-ph/0507001}{{\ttfamily hep-ph/0507001}}].

\bibitem{Calibbi:2017uvl}
L.~Calibbi and G.~Signorelli, \emph{{Charged Lepton Flavour Violation: An
  Experimental and Theoretical Introduction}},
  \href{https://doi.org/10.1393/ncr/i2018-10144-0}{\emph{Riv. Nuovo Cim.}
  {\bfseries 41} (2018) 71} [\href{https://arxiv.org/abs/1709.00294}{{\ttfamily
  1709.00294}}].

\bibitem{Bjorkeroth:2018dzu}
F.~Bj\"orkeroth, E.J.~Chun and S.F.~King, \emph{{Flavourful Axion
  Phenomenology}}, \href{https://doi.org/10.1007/JHEP08(2018)117}{\emph{JHEP}
  {\bfseries 08} (2018) 117}
  [\href{https://arxiv.org/abs/1806.00660}{{\ttfamily 1806.00660}}].

\bibitem{Cornella:2019uxs}
C.~Cornella, P.~Paradisi and O.~Sumensari, \emph{{Hunting for ALPs with Lepton
  Flavor Violation}},
  \href{https://doi.org/10.1007/JHEP01(2020)158}{\emph{JHEP} {\bfseries 01}
  (2020) 158} [\href{https://arxiv.org/abs/1911.06279}{{\ttfamily
  1911.06279}}].

\bibitem{Calibbi:2020jvd}
L.~Calibbi, D.~Redigolo, R.~Ziegler and J.~Zupan, \emph{{Looking forward to
  lepton-flavor-violating ALPs}},
  \href{https://doi.org/10.1007/JHEP09(2021)173}{\emph{JHEP} {\bfseries 09}
  (2021) 173} [\href{https://arxiv.org/abs/2006.04795}{{\ttfamily
  2006.04795}}].

\bibitem{Cheung:2021mol}
K.~Cheung, A.~Soffer, Z.S.~Wang and Y.-H.~Wu, \emph{{Probing charged lepton
  flavor violation with axion-like particles at Belle II}},
  \href{https://doi.org/10.1007/JHEP11(2021)218}{\emph{JHEP} {\bfseries 11}
  (2021) 218} [\href{https://arxiv.org/abs/2108.11094}{{\ttfamily
  2108.11094}}].

\bibitem{Calibbi:2024rcm}
L.~Calibbi, T.~Li, L.~Mukherjee and Y.~Yang, \emph{{Probing ALP Lepton Flavour
  Violation at $\mu$TRISTAN}},
  \href{https://arxiv.org/abs/2406.13234}{{\ttfamily 2406.13234}}.

\bibitem{Davoudiasl:2024vje}
H.~Davoudiasl, R.~Marcarelli and E.T.~Neil, \emph{{Flavor-violating ALPs,
  electron g-2, and the Electron-Ion Collider}},
  \href{https://doi.org/10.1103/PhysRevD.109.115013}{\emph{Phys. Rev. D}
  {\bfseries 109} (2024) 115013}
  [\href{https://arxiv.org/abs/2402.17821}{{\ttfamily 2402.17821}}].

\bibitem{Hisano:1995cp}
J.~Hisano, T.~Moroi, K.~Tobe and M.~Yamaguchi, \emph{{Lepton flavor violation
  via right-handed neutrino Yukawa couplings in supersymmetric standard
  model}}, \href{https://doi.org/10.1103/PhysRevD.53.2442}{\emph{Phys. Rev. D}
  {\bfseries 53} (1996) 2442}
  [\href{https://arxiv.org/abs/hep-ph/9510309}{{\ttfamily hep-ph/9510309}}].

\bibitem{Abada:2014kba}
A.~Abada, M.E.~Krauss, W.~Porod, F.~Staub, A.~Vicente and C.~Weiland,
  \emph{{Lepton flavor violation in low-scale seesaw models: SUSY and non-SUSY
  contributions}}, \href{https://doi.org/10.1007/JHEP11(2014)048}{\emph{JHEP}
  {\bfseries 11} (2014) 048} [\href{https://arxiv.org/abs/1408.0138}{{\ttfamily
  1408.0138}}].

\bibitem{Blanke:2007db}
M.~Blanke, A.J.~Buras, B.~Duling, A.~Poschenrieder and C.~Tarantino,
  \emph{{Charged Lepton Flavour Violation and (g-2)(mu) in the Littlest Higgs
  Model with T-Parity: A Clear Distinction from Supersymmetry}},
  \href{https://doi.org/10.1088/1126-6708/2007/05/013}{\emph{JHEP} {\bfseries
  05} (2007) 013} [\href{https://arxiv.org/abs/hep-ph/0702136}{{\ttfamily
  hep-ph/0702136}}].

\bibitem{Han:2011aq}
X.~Han, \emph{{The Lepton Flavor Violating Decays $Z\to l_i l_j$ in the
  Simplest little Higgs Model}},
  \href{https://doi.org/10.1142/S0217732312501581}{\emph{Mod. Phys. Lett. A}
  {\bfseries 27} (2012) 1250158}
  [\href{https://arxiv.org/abs/1104.3534}{{\ttfamily 1104.3534}}].

\bibitem{Ardu:2023yyw}
M.~Ardu, S.~Davidson and S.~Lavignac, \emph{{Distinguishing models with $\mu
  \to e$ observables}},
  \href{https://doi.org/10.1007/JHEP11(2023)101}{\emph{JHEP} {\bfseries 11}
  (2023) 101} [\href{https://arxiv.org/abs/2308.16897}{{\ttfamily
  2308.16897}}].

\bibitem{Altmannshofer:2023tsa}
W.~Altmannshofer, P.~Munbodh and T.~Oh, \emph{{Probing lepton flavor violation
  at Circular Electron-Positron Colliders}},
  \href{https://doi.org/10.1007/JHEP08(2023)026}{\emph{JHEP} {\bfseries 08}
  (2023) 026} [\href{https://arxiv.org/abs/2305.03869}{{\ttfamily
  2305.03869}}].

\bibitem{Altmannshofer:2022fvz}
W.~Altmannshofer, C.~Caillol, M.~Dam, S.~Xella and Y.~Zhang, \emph{{Charged
  Lepton Flavour Violation in Heavy Particle Decays}},  in \emph{{Snowmass
  2021}}, 5, 2022 [\href{https://arxiv.org/abs/2205.10576}{{\ttfamily
  2205.10576}}].

\bibitem{Huo:2024puy}
J.-P.~Huo, X.-X.~Dong, J.~Ma, S.-M.~Zhao, C.~Guo, H.-B.~Zhang et~al.,
  \emph{{Lepton flavor violating decays $Z\rightarrow l^{\pm}_{i}l^{\mp}_{j}$
  in the B-L Supersymmetric Standard Model}},
  \href{https://arxiv.org/abs/2406.03108}{{\ttfamily 2406.03108}}.

\bibitem{Toma:2013zsa}
T.~Toma and A.~Vicente, \emph{{Lepton Flavor Violation in the Scotogenic
  Model}}, \href{https://doi.org/10.1007/JHEP01(2014)160}{\emph{JHEP}
  {\bfseries 01} (2014) 160} [\href{https://arxiv.org/abs/1312.2840}{{\ttfamily
  1312.2840}}].

\bibitem{Tapender:2024ktc}
Tapender, S.~Verma and S.~Kumar, \emph{{On Lepton Flavor Violation and Dark
  Matter in Scotogenic model with Trimaximal Mixing}},
  \href{https://arxiv.org/abs/2402.16491}{{\ttfamily 2402.16491}}.

\bibitem{xiaogang:1991a}
X.-G.~He, G.C.~Joshi, H.~Lew and R.R.~Volkas, \emph{Simplest ${Z}^{'}$ model},
  \href{https://doi.org/10.1103/PhysRevD.44.2118}{\emph{Phys. Rev. D}
  {\bfseries 44} (1991) 2118}.

\bibitem{xiaogang:1991b}
X.G.~He, G.C.~Joshi, H.~Lew and R.R.~Volkas, \emph{New-${Z}^{'}$
  phenomenology}, \href{https://doi.org/10.1103/PhysRevD.43.R22}{\emph{Phys.
  Rev. D} {\bfseries 43} (1991) R22}.

\bibitem{Foot:1994vd}
R.~Foot, X.G.~He, H.~Lew and R.R.~Volkas, \emph{{Model for a light Z-prime
  boson}}, \href{https://doi.org/10.1103/PhysRevD.50.4571}{\emph{Phys. Rev. D}
  {\bfseries 50} (1994) 4571}
  [\href{https://arxiv.org/abs/hep-ph/9401250}{{\ttfamily hep-ph/9401250}}].

\bibitem{Iguro:2020rby}
S.~Iguro, Y.~Omura and M.~Takeuchi, \emph{{Probing $\mu\tau$ flavor-violating
  solutions for the muon $g-2$ anomaly at Belle II}},
  \href{https://doi.org/10.1007/JHEP09(2020)144}{\emph{JHEP} {\bfseries 09}
  (2020) 144} [\href{https://arxiv.org/abs/2002.12728}{{\ttfamily
  2002.12728}}].

\bibitem{AtzoriCorona:2022moj}
M.~Atzori~Corona, M.~Cadeddu, N.~Cargioli, F.~Dordei, C.~Giunti, Y.F.~Li
  et~al., \emph{{Probing light mediators and $(g-2)_{\mu}$ through detection of
  coherent elastic neutrino nucleus scattering at COHERENT}},
  \href{https://doi.org/10.1007/JHEP05(2022)109}{\emph{JHEP} {\bfseries 05}
  (2022) 109} [\href{https://arxiv.org/abs/2202.11002}{{\ttfamily
  2202.11002}}].

\bibitem{Crivellin:2023sig}
A.~Crivellin and S.~Iguro, \emph{{Accumulating Hints for Flavour Violating
  Higgses at the Electroweak Scale}},
  \href{https://arxiv.org/abs/2311.03430}{{\ttfamily 2311.03430}}.

\bibitem{Espinosa-Gomez:2023xrq}
D.~Espinosa-Gomez, F.~Ramirez-Zavaleta and E.S.~Tututi, \emph{{Decay of the
  $Z^\prime $ gauge boson with lepton flavor violation}},
  \href{https://doi.org/10.1140/epjc/s10052-023-12069-7}{\emph{Eur. Phys. J. C}
  {\bfseries 83} (2023) 909}
  [\href{https://arxiv.org/abs/2310.04509}{{\ttfamily 2310.04509}}].

\bibitem{Eguren:2024oov}
J.F.~Eguren, S.~Klingel, E.~Stamou, M.~Tabet and R.~Ziegler, \emph{{Flavor
  Phenomenology of Light Dark Vectors}},
  \href{https://arxiv.org/abs/2405.00108}{{\ttfamily 2405.00108}}.

\bibitem{marin:2024LFV}
M.~Marín, R.~Gaitán and R.~Martinez, \emph{Lepton flavor-violating higgs
  decays mediated by ultralight gauge boson},
  \href{https://arxiv.org/abs/2406.17040}{{\ttfamily 2406.17040}}.

\bibitem{Altmannshofer:2016brv}
W.~Altmannshofer, C.-Y.~Chen, P.S.~Bhupal~Dev and A.~Soni, \emph{{Lepton flavor
  violating Z' explanation of the muon anomalous magnetic moment}},
  \href{https://doi.org/10.1016/j.physletb.2016.09.046}{\emph{Phys. Lett. B}
  {\bfseries 762} (2016) 389}
  [\href{https://arxiv.org/abs/1607.06832}{{\ttfamily 1607.06832}}].

\bibitem{Kang:2019vng}
Z.~Kang and Y.~Shigekami, \emph{{$(g-2)_{\mu}$ versus flavor changing neutral
  current induced by the light $(B-L)_{\mu\tau}$ boson}},
  \href{https://doi.org/10.1007/JHEP11(2019)049}{\emph{JHEP} {\bfseries 11}
  (2019) 049} [\href{https://arxiv.org/abs/1905.11018}{{\ttfamily
  1905.11018}}].

\bibitem{Cheng:2021okr}
Y.~Cheng, X.-G.~He and J.~Sun, \emph{{Widening the $U(1)_{L_\mu - L_\tau} \,
  Z'$ mass range for resolving the muon $g-2$ anomaly}},
  \href{https://doi.org/10.1016/j.physletb.2022.136989}{\emph{Phys. Lett. B}
  {\bfseries 827} (2022) 136989}
  [\href{https://arxiv.org/abs/2112.09920}{{\ttfamily 2112.09920}}].

\bibitem{Eijima:2023yiw}
S.~Eijima, M.~Ibe and K.~Murai, \emph{{Muon $g-2$ and non-thermal leptogenesis
  in $ \textrm{U}{(1)}_{L_{\mu }-{L}_{\tau }} $ model}},
  \href{https://doi.org/10.1007/JHEP05(2023)010}{\emph{JHEP} {\bfseries 05}
  (2023) 010} [\href{https://arxiv.org/abs/2303.09751}{{\ttfamily
  2303.09751}}].

\bibitem{Buras:2021btx}
A.J.~Buras, A.~Crivellin, F.~Kirk, C.A.~Manzari and M.~Montull, \emph{{Global
  analysis of leptophilic Z' bosons}},
  \href{https://doi.org/10.1007/JHEP06(2021)068}{\emph{JHEP} {\bfseries 06}
  (2021) 068} [\href{https://arxiv.org/abs/2104.07680}{{\ttfamily
  2104.07680}}].

\bibitem{ATLAS:2018sky}
{\scshape ATLAS} collaboration, \emph{{A search for lepton-flavor-violating
  decays of the $Z$ boson into a $\tau$-lepton and a light lepton with the
  ATLAS detector}},
  \href{https://doi.org/10.1103/PhysRevD.98.092010}{\emph{Phys. Rev. D}
  {\bfseries 98} (2018) 092010}
  [\href{https://arxiv.org/abs/1804.09568}{{\ttfamily 1804.09568}}].

\bibitem{Araki:2022xqp}
T.~Araki, K.~Asai, H.~Otono, T.~Shimomura and Y.~Takubo, \emph{{Search for
  lepton flavor violating decay at FASER}},
  \href{https://doi.org/10.1007/JHEP01(2023)145}{\emph{JHEP} {\bfseries 01}
  (2023) 145} [\href{https://arxiv.org/abs/2210.12730}{{\ttfamily
  2210.12730}}].

\bibitem{ding2024study}
R.~Ding, J.~Li, M.~Lu, Z.~You, Z.~Wang and Q.~Li, \emph{{Study of charged
  Lepton Flavor Violation in electron muon interactions}},
  \href{https://arxiv.org/abs/2405.09417}{{\ttfamily 2405.09417}}.

\bibitem{CDF:2008zud}
{\scshape CDF} collaboration, \emph{{Search for Maximal Flavor Violating
  Scalars in Same-Charge Lepton Pairs in $p \bar{p}$ Collisions at $\sqrt{s}$ =
  1.96-TeV}}, \href{https://doi.org/10.1103/PhysRevLett.102.041801}{\emph{Phys.
  Rev. Lett.} {\bfseries 102} (2009) 041801}
  [\href{https://arxiv.org/abs/0809.4903}{{\ttfamily 0809.4903}}].

\bibitem{Foldenauer:2016rpi}
P.~Foldenauer and J.~Jaeckel, \emph{{Purely flavor-changing Z' bosons and where
  they might hide}}, \href{https://doi.org/10.1007/JHEP05(2017)010}{\emph{JHEP}
  {\bfseries 05} (2017) 010}
  [\href{https://arxiv.org/abs/1612.07789}{{\ttfamily 1612.07789}}].

\bibitem{Kriewald:2022erk}
J.~Kriewald, J.~Orloff, E.~Pinsard and A.M.~Teixeira, \emph{{Prospects for a
  flavour violating $Z^\prime $ explanation of $\Delta a_{\mu ,e}$}},
  \href{https://doi.org/10.1140/epjc/s10052-022-10776-1}{\emph{Eur. Phys. J. C}
  {\bfseries 82} (2022) 844}
  [\href{https://arxiv.org/abs/2204.13134}{{\ttfamily 2204.13134}}].

\bibitem{Bar-Shalom:2007xeu}
S.~Bar-Shalom and A.~Rajaraman, \emph{{Models and phenomenology of maximal
  flavor violation}},
  \href{https://doi.org/10.1103/PhysRevD.77.095011}{\emph{Phys. Rev. D}
  {\bfseries 77} (2008) 095011}
  [\href{https://arxiv.org/abs/0711.3193}{{\ttfamily 0711.3193}}].

\bibitem{CHARM-II:1990dvf}
{\scshape CHARM-II} collaboration, \emph{{First observation of neutrino trident
  production}}, \href{https://doi.org/10.1016/0370-2693(90)90146-W}{\emph{Phys.
  Lett. B} {\bfseries 245} (1990) 271}.

\bibitem{CCFR:1991lpl}
{\scshape CCFR} collaboration, \emph{{Neutrino Tridents and W Z Interference}},
  \href{https://doi.org/10.1103/PhysRevLett.66.3117}{\emph{Phys. Rev. Lett.}
  {\bfseries 66} (1991) 3117}.

\bibitem{NuTeV:1999wlw}
{\scshape NuTeV} collaboration, \emph{{Evidence for diffractive charm
  production in muon-neutrino Fe and anti-muon-neutrino Fe scattering at the
  Tevatron}}, \href{https://doi.org/10.1103/PhysRevD.61.092001}{\emph{Phys.
  Rev. D} {\bfseries 61} (2000) 092001}
  [\href{https://arxiv.org/abs/hep-ex/9909041}{{\ttfamily hep-ex/9909041}}].

\bibitem{Czyz:1964zz}
W.~Czyz, G.C.~Sheppey and J.D.~Walecka, \emph{{Neutrino production of lepton
  pairs through the point four-fermion interaction}},
  \href{https://doi.org/10.1007/BF02734586}{\emph{Nuovo Cim.} {\bfseries 34}
  (1964) 404}.

\bibitem{Fujikawa:1971nx}
K.~Fujikawa, \emph{{The self-coupling of weak lepton currents in high-energy
  neutrino and muon reactions}},
  \href{https://doi.org/10.1016/0003-4916(71)90244-2}{\emph{Annals Phys.}
  {\bfseries 68} (1971) 102}.

\bibitem{Francener:2024wul}
R.~Francener, V.P.~Goncalves and D.R.~Gratieri, \emph{{Neutrino trident
  scattering at the LHC energy regime}},
  \href{https://arxiv.org/abs/2406.13593}{{\ttfamily 2406.13593}}.

\bibitem{Altmannshofer:2024hqd}
W.~Altmannshofer, T.~M\"akel\"a, S.~Sarkar, S.~Trojanowski, K.~Xie and B.~Zhou,
  \emph{{Discovering neutrino tridents at the Large Hadron Collider}},
  \href{https://arxiv.org/abs/2406.16803}{{\ttfamily 2406.16803}}.

\bibitem{Ge:2017poy}
S.-F.~Ge, M.~Lindner and W.~Rodejohann, \emph{{Atmospheric Trident Production
  for Probing New Physics}},
  \href{https://doi.org/10.1016/j.physletb.2017.06.020}{\emph{Phys. Lett. B}
  {\bfseries 772} (2017) 164}
  [\href{https://arxiv.org/abs/1702.02617}{{\ttfamily 1702.02617}}].

\bibitem{Zhou:2019vxt}
B.~Zhou and J.F.~Beacom, \emph{{Neutrino-nucleus cross sections for W-boson and
  trident production}},
  \href{https://doi.org/10.1103/PhysRevD.101.036011}{\emph{Phys. Rev. D}
  {\bfseries 101} (2020) 036011}
  [\href{https://arxiv.org/abs/1910.08090}{{\ttfamily 1910.08090}}].

\bibitem{Zhou:2019frk}
B.~Zhou and J.F.~Beacom, \emph{{W-boson and trident production in
  TeV\textendash{}PeV neutrino observatories}},
  \href{https://doi.org/10.1103/PhysRevD.101.036010}{\emph{Phys. Rev. D}
  {\bfseries 101} (2020) 036010}
  [\href{https://arxiv.org/abs/1910.10720}{{\ttfamily 1910.10720}}].

\bibitem{Ballett:2018uuc}
P.~Ballett, M.~Hostert, S.~Pascoli, Y.F.~Perez-Gonzalez, Z.~Tabrizi and
  R.~Zukanovich~Funchal, \emph{{Neutrino Trident Scattering at Near
  Detectors}}, \href{https://doi.org/10.1007/JHEP01(2019)119}{\emph{JHEP}
  {\bfseries 01} (2019) 119}
  [\href{https://arxiv.org/abs/1807.10973}{{\ttfamily 1807.10973}}].

\bibitem{ATLAS:2022vkf}
{\scshape ATLAS} collaboration, \emph{{A detailed map of Higgs boson
  interactions by the ATLAS experiment ten years after the discovery}},
  \href{https://doi.org/10.1038/s41586-022-04893-w}{\emph{Nature} {\bfseries
  607} (2022) 52} [\href{https://arxiv.org/abs/2207.00092}{{\ttfamily
  2207.00092}}].

\bibitem{CMS:2022dwd}
{\scshape CMS} collaboration, \emph{{A portrait of the Higgs boson by the CMS
  experiment ten years after the discovery.}},
  \href{https://doi.org/10.1038/s41586-022-04892-x}{\emph{Nature} {\bfseries
  607} (2022) 60} [\href{https://arxiv.org/abs/2207.00043}{{\ttfamily
  2207.00043}}].

\bibitem{CMS:2022urr}
{\scshape CMS} collaboration, \emph{{Search for the Higgs boson decay to a pair
  of electrons in proton-proton collisions at $\sqrt{s} $ =13TeV}},
  \href{https://doi.org/10.1016/j.physletb.2023.137783}{\emph{Phys. Lett. B}
  {\bfseries 846} (2023) 137783}
  [\href{https://arxiv.org/abs/2208.00265}{{\ttfamily 2208.00265}}].

\bibitem{Ce:2022kxy}
M.~C\`e et~al., \emph{{Window observable for the hadronic vacuum polarization
  contribution to the $(g-2)_\mu$ from lattice QCD}},
  \href{https://doi.org/10.1103/PhysRevD.106.114502}{\emph{Phys. Rev. D}
  {\bfseries 106} (2022) 114502}
  [\href{https://arxiv.org/abs/2206.06582}{{\ttfamily 2206.06582}}].

\bibitem{Colangelo:2022vok}
G.~Colangelo, A.X.~El-Khadra, M.~Hoferichter, A.~Keshavarzi, C.~Lehner,
  P.~Stoffer et~al., \emph{{Data-driven evaluations of Euclidean windows to
  scrutinize hadronic vacuum polarization}},
  \href{https://doi.org/10.1016/j.physletb.2022.137313}{\emph{Phys. Lett. B}
  {\bfseries 833} (2022) 137313}
  [\href{https://arxiv.org/abs/2205.12963}{{\ttfamily 2205.12963}}].

\bibitem{Acaroglu:2023cza}
H.~Acaro\u{g}lu, M.~Blanke and M.~Tabet, \emph{{Opening the Higgs portal to
  lepton-flavoured dark matter}},
  \href{https://doi.org/10.1007/JHEP11(2023)079}{\emph{JHEP} {\bfseries 11}
  (2023) 079} [\href{https://arxiv.org/abs/2309.10700}{{\ttfamily
  2309.10700}}].

\bibitem{Muong-2:2023cdq}
{\scshape Muon g-2} collaboration, \emph{{Measurement of the Positive Muon
  Anomalous Magnetic Moment to 0.20~ppm}},
  \href{https://doi.org/10.1103/PhysRevLett.131.161802}{\emph{Phys. Rev. Lett.}
  {\bfseries 131} (2023) 161802}
  [\href{https://arxiv.org/abs/2308.06230}{{\ttfamily 2308.06230}}].

\bibitem{Armando:2023zwz}
G.~Armando, P.~Panci, J.~Weiss and R.~Ziegler, \emph{{Leptonic ALP portal to
  the dark sector}},
  \href{https://doi.org/10.1103/PhysRevD.109.055029}{\emph{Phys. Rev. D}
  {\bfseries 109} (2024) 055029}
  [\href{https://arxiv.org/abs/2310.05827}{{\ttfamily 2310.05827}}].

\bibitem{Borsanyi:2020mff}
S.~Borsanyi et~al., \emph{{Leading hadronic contribution to the muon magnetic
  moment from lattice QCD}},
  \href{https://doi.org/10.1038/s41586-021-03418-1}{\emph{Nature} {\bfseries
  593} (2021) 51} [\href{https://arxiv.org/abs/2002.12347}{{\ttfamily
  2002.12347}}].

\bibitem{Giudice:2012ms}
G.F.~Giudice, P.~Paradisi and M.~Passera, \emph{{Testing new physics with the
  electron g-2}}, \href{https://doi.org/10.1007/JHEP11(2012)113}{\emph{JHEP}
  {\bfseries 11} (2012) 113} [\href{https://arxiv.org/abs/1208.6583}{{\ttfamily
  1208.6583}}].

\bibitem{Fan:2022eto}
X.~Fan, T.G.~Myers, B.A.D.~Sukra and G.~Gabrielse, \emph{{Measurement of the
  Electron Magnetic Moment}},
  \href{https://doi.org/10.1103/PhysRevLett.130.071801}{\emph{Phys. Rev. Lett.}
  {\bfseries 130} (2023) 071801}
  [\href{https://arxiv.org/abs/2209.13084}{{\ttfamily 2209.13084}}].

\bibitem{Parker:2018vye}
R.H.~Parker, C.~Yu, W.~Zhong, B.~Estey and H.~M\"uller, \emph{{Measurement of
  the fine-structure constant as a test of the Standard Model}},
  \href{https://doi.org/10.1126/science.aap7706}{\emph{Science} {\bfseries 360}
  (2018) 191} [\href{https://arxiv.org/abs/1812.04130}{{\ttfamily
  1812.04130}}].

\bibitem{Patel:2015tea}
H.H.~Patel, \emph{{Package-X: A Mathematica package for the analytic
  calculation of one-loop integrals}},
  \href{https://doi.org/10.1016/j.cpc.2015.08.017}{\emph{Comput. Phys. Commun.}
  {\bfseries 197} (2015) 276}
  [\href{https://arxiv.org/abs/1503.01469}{{\ttfamily 1503.01469}}].

\bibitem{Belle-II:2019qfb}
{\scshape Belle-II} collaboration, \emph{{Search for an Invisibly Decaying
  $Z^{\prime}$ Boson at Belle II in $e^+ e^- \to \mu^+ \mu^- (e^{\pm}
  \mu^{\mp}) + \slashed{E}$ Final States}},
  \href{https://doi.org/10.1103/PhysRevLett.124.141801}{\emph{Phys. Rev. Lett.}
  {\bfseries 124} (2020) 141801}
  [\href{https://arxiv.org/abs/1912.11276}{{\ttfamily 1912.11276}}].

\bibitem{DELPHI:2000ztm}
{\scshape DELPHI} collaboration, \emph{{Measurement and interpretation of
  fermion-pair production at LEP energies of 183-GeV and 189-GeV}},
  \href{https://doi.org/10.1016/S0370-2693(00)00675-4}{\emph{Phys. Lett. B}
  {\bfseries 485} (2000) 45}
  [\href{https://arxiv.org/abs/hep-ex/0103025}{{\ttfamily hep-ex/0103025}}].

\bibitem{Renga:2019mpg}
F.~Renga, \emph{{Experimental searches for muon decays beyond the Standard
  Model}}, \href{https://doi.org/10.1016/j.revip.2019.100029}{\emph{Rev. Phys.}
  {\bfseries 4} (2019) 100029}
  [\href{https://arxiv.org/abs/1902.06291}{{\ttfamily 1902.06291}}].

\bibitem{MEG:2011naj}
{\scshape MEG} collaboration, \emph{{New limit on the lepton-flavour violating
  decay $\mu^{+} \to e^{+} \gamma$}},
  \href{https://doi.org/10.1103/PhysRevLett.107.171801}{\emph{Phys. Rev. Lett.}
  {\bfseries 107} (2011) 171801}
  [\href{https://arxiv.org/abs/1107.5547}{{\ttfamily 1107.5547}}].

\bibitem{MEG:2013oxv}
{\scshape MEG} collaboration, \emph{{New constraint on the existence of the
  $\mu^+ \to e^+\gamma$ decay}},
  \href{https://doi.org/10.1103/PhysRevLett.110.201801}{\emph{Phys. Rev. Lett.}
  {\bfseries 110} (2013) 201801}
  [\href{https://arxiv.org/abs/1303.0754}{{\ttfamily 1303.0754}}].

\bibitem{MEG:2016leq}
{\scshape MEG} collaboration, \emph{{Search for the lepton flavour violating
  decay $\mu ^+ \rightarrow \mathrm {e}^+ \gamma $ with the full dataset of the
  MEG experiment}},
  \href{https://doi.org/10.1140/epjc/s10052-016-4271-x}{\emph{Eur. Phys. J. C}
  {\bfseries 76} (2016) 434}
  [\href{https://arxiv.org/abs/1605.05081}{{\ttfamily 1605.05081}}].

\bibitem{SINDRUM:1987nra}
{\scshape SINDRUM} collaboration, \emph{{Search for the Decay $\mu^+ \to e^+
  e^+ e^-$}}, \href{https://doi.org/10.1016/0550-3213(88)90462-2}{\emph{Nucl.
  Phys. B} {\bfseries 299} (1988) 1}.

\bibitem{Mu3e:2020gyw}
{\scshape Mu3e} collaboration, \emph{{Technical design of the phase I Mu3e
  experiment}}, \href{https://doi.org/10.1016/j.nima.2021.165679}{\emph{Nucl.
  Instrum. Meth. A} {\bfseries 1014} (2021) 165679}
  [\href{https://arxiv.org/abs/2009.11690}{{\ttfamily 2009.11690}}].

\bibitem{Perrevoort:2023qhn}
{\scshape Mu3e} collaboration, \emph{{Searching for Charged Lepton Flavour
  Violation with Mu3e$^\dagger$}},
  \href{https://doi.org/10.3390/psf2023008030}{\emph{Phys. Sci. Forum}
  {\bfseries 8} (2023) 30} [\href{https://arxiv.org/abs/2308.11403}{{\ttfamily
  2308.11403}}].

\bibitem{Anselm:1985bp}
A.A.~Anselm, N.G.~Uraltsev and M.Y.~Khlopov, \emph{{$\mu \to e$ FAMILON
  DECAY}}, {\emph{Sov. J. Nucl. Phys.} {\bfseries 41} (1985) 1060}.

\bibitem{Andreev:2006wh}
V.A.~Andreev et~al., \emph{{Search for Two-Particle Muon Decay to Positron and
  Goldstone Massless Boson (FAMILON)}},
  \href{https://arxiv.org/abs/hep-ex/0612064}{{\ttfamily hep-ex/0612064}}.

\bibitem{berezhiani1991cosmology}
Z.~Berezhiani and M.Y.~Khlopov, \emph{Cosmology of spontaneously broken gauge
  family symmetry with axion solution of strong cp-problem}, {\emph{Zeitschrift
  f{\"u}r Physik C Particles and Fields} {\bfseries 49} (1991) 73}.

\bibitem{sakharov1994horizontal}
A.~Sakharov and M.Y.~Khlopov, \emph{Horizontal unification as the phenomenology
  of the theory of'everything'}, {\emph{Yadernaya Fizika} {\bfseries 57} (1994)
  690}.

\bibitem{TWIST:2014ymv}
{\scshape TWIST} collaboration, \emph{{Search for two body muon decay
  signals}}, \href{https://doi.org/10.1103/PhysRevD.91.052020}{\emph{Phys. Rev.
  D} {\bfseries 91} (2015) 052020}
  [\href{https://arxiv.org/abs/1409.0638}{{\ttfamily 1409.0638}}].

\bibitem{Bilger:1998rp}
R.~Bilger, K.~Foehl, H.~Clement, M.~Croni, A.~Erhardt, R.~Meier et~al.,
  \emph{{Search for exotic muon decays}},
  \href{https://doi.org/10.1016/S0370-2693(98)01507-X}{\emph{Phys. Lett. B}
  {\bfseries 446} (1999) 363}
  [\href{https://arxiv.org/abs/hep-ph/9811333}{{\ttfamily hep-ph/9811333}}].

\bibitem{Workman:2022ynf}
{\scshape Particle Data Group} collaboration, \emph{{Review of Particle
  Physics}}, \href{https://doi.org/10.1093/ptep/ptac097}{\emph{PTEP} {\bfseries
  2022} (2022) 083C01}.

\bibitem{Goldman:1987hy}
J.T.~Goldman et~al., \emph{{Light Boson Emission in the Decay of the $\mu^+$}},
  \href{https://doi.org/10.1103/PhysRevD.36.1543}{\emph{Phys. Rev. D}
  {\bfseries 36} (1987) 1543}.

\bibitem{Heeck:2016xkh}
J.~Heeck, \emph{{Lepton flavor violation with light vector bosons}},
  \href{https://doi.org/10.1016/j.physletb.2016.05.007}{\emph{Phys. Lett. B}
  {\bfseries 758} (2016) 101}
  [\href{https://arxiv.org/abs/1602.03810}{{\ttfamily 1602.03810}}].

\bibitem{CHANOWITZ1985379}
M.S.~Chanowitz and M.K.~Gaillard, \emph{The tev physics of strongly interacting
  w's and z's},
  \href{https://doi.org/https://doi.org/10.1016/0550-3213(85)90580-2}{\emph{Nuclear
  Physics B} {\bfseries 261} (1985) 379}.

\bibitem{Peskin:2017emn}
M.E.~Peskin, \emph{{Lectures on the Theory of the Weak Interaction}},  in
  \emph{{2016 European School of High-Energy Physics}}, pp.~1--70, 2017,
  \href{https://doi.org/10.23730/CYRSP-2017-005.1}{DOI}
  [\href{https://arxiv.org/abs/1708.09043}{{\ttfamily 1708.09043}}].

\bibitem{Hill:2023dym}
R.J.~Hill, R.~Plestid and J.~Zupan, \emph{{Searching for new physics at $\mu
  \to e$ facilities with $\mu^+$ and $\pi^+$ decays at rest}},
  \href{https://doi.org/10.1103/PhysRevD.109.035025}{\emph{Phys. Rev. D}
  {\bfseries 109} (2024) 035025}
  [\href{https://arxiv.org/abs/2310.00043}{{\ttfamily 2310.00043}}].

\bibitem{Amsterdam-CERN-Hamburg-Moscow-Rome:1980qbb}
{\scshape Amsterdam-CERN-Hamburg-Moscow-Rome} collaboration,
  \emph{{Experimental Study of Inverse Muon Decay}},
  \href{https://doi.org/10.1016/0370-2693(80)90127-6}{\emph{Phys. Lett. B}
  {\bfseries 93} (1980) 203}.

\bibitem{NuTeV:2001bgq}
{\scshape NuTeV} collaboration, \emph{{Search for the Lepton Number Violating
  Process $\bar{\nu}_{\mu} e^- \to \mu^- \bar{\nu}_e$}},
  \href{https://doi.org/10.1103/PhysRevLett.87.071803}{\emph{Phys. Rev. Lett.}
  {\bfseries 87} (2001) 071803}
  [\href{https://arxiv.org/abs/hep-ex/0104029}{{\ttfamily hep-ex/0104029}}].

\bibitem{Fukuyama:2021iyw}
T.~Fukuyama, Y.~Mimura and Y.~Uesaka, \emph{{Models of the muonium to
  antimuonium transition}},
  \href{https://doi.org/10.1103/PhysRevD.105.015026}{\emph{Phys. Rev. D}
  {\bfseries 105} (2022) 015026}
  [\href{https://arxiv.org/abs/2108.10736}{{\ttfamily 2108.10736}}].

\bibitem{Willmann:1998gd}
L.~Willmann et~al., \emph{{New bounds from searching for muonium to
  anti-muonium conversion}},
  \href{https://doi.org/10.1103/PhysRevLett.82.49}{\emph{Phys. Rev. Lett.}
  {\bfseries 82} (1999) 49}
  [\href{https://arxiv.org/abs/hep-ex/9807011}{{\ttfamily hep-ex/9807011}}].

\bibitem{Bai:2022sxq}
A.-Y.~Bai et~al., \emph{{Snowmass2021 Whitepaper: Muonium to antimuonium
  conversion}},  in \emph{{Snowmass 2021}}, 3, 2022
  [\href{https://arxiv.org/abs/2203.11406}{{\ttfamily 2203.11406}}].

\bibitem{Zhao:2023plv}
S.~Zhao and J.~Tang, \emph{{Optimization of muonium yield in perforated silica
  aerogel}}, \href{https://doi.org/10.1103/PhysRevD.109.072012}{\emph{Phys.
  Rev. D} {\bfseries 109} (2024) 072012}
  [\href{https://arxiv.org/abs/2401.00222}{{\ttfamily 2401.00222}}].

\bibitem{Koike:1971tu}
K.~Koike, M.~Konuma, K.~Kurata and K.~Sugano, \emph{{Neutrino production of
  lepton pairs. 1.}}, \href{https://doi.org/10.1143/PTP.46.1150}{\emph{Prog.
  Theor. Phys.} {\bfseries 46} (1971) 1150}.

\bibitem{Koike:1971vg}
K.~Koike, M.~Konuma, K.~Kurata and K.~Sugano, \emph{{Neutrino production of
  lepton pairs. 2.}}, \href{https://doi.org/10.1143/PTP.46.1799}{\emph{Prog.
  Theor. Phys.} {\bfseries 46} (1971) 1799}.

\bibitem{Brown:1972vne}
R.W.~Brown, R.H.~Hobbs, J.~Smith and N.~Stanko, \emph{{Intermediate boson. iii.
  virtual-boson effects in neutrino trident production}},
  \href{https://doi.org/10.1103/PhysRevD.6.3273}{\emph{Phys. Rev. D} {\bfseries
  6} (1972) 3273}.

\bibitem{Pastore:2762117}
{\scshape SHiP} collaboration, \emph{{Neutrino physics with the SHiP experiment
  at CERN}}, \href{https://doi.org/10.22323/1.364.0372}{\emph{PoS} {\bfseries
  EPS-HEP2019} (2020) 372}.

\bibitem{DiCrescenzo:2023czg}
A.~Di~Crescenzo, \emph{{Neutrino physics and dark matter search with SHiP at
  CERN}}, \href{https://doi.org/10.1142/S2010194523600029}{\emph{Int. J. Mod.
  Phys. Conf. Ser.} {\bfseries 51} (2023) 2360002}.

\bibitem{Altmannshofer:2019zhy}
W.~Altmannshofer, S.~Gori, J.~Mart\'\i{}n-Albo, A.~Sousa and M.~Wallbank,
  \emph{{Neutrino Tridents at DUNE}},
  \href{https://doi.org/10.1103/PhysRevD.100.115029}{\emph{Phys. Rev. D}
  {\bfseries 100} (2019) 115029}
  [\href{https://arxiv.org/abs/1902.06765}{{\ttfamily 1902.06765}}].

\bibitem{Conrad:1997ne}
J.M.~Conrad, M.H.~Shaevitz and T.~Bolton, \emph{{Precision measurements with
  high-energy neutrino beams}},
  \href{https://doi.org/10.1103/RevModPhys.70.1341}{\emph{Rev. Mod. Phys.}
  {\bfseries 70} (1998) 1341}
  [\href{https://arxiv.org/abs/hep-ex/9707015}{{\ttfamily hep-ex/9707015}}].

\bibitem{Sakumoto:1990py}
W.K.~Sakumoto et~al., \emph{{Calibration of the CCFR Target Calorimeter}},
  \href{https://doi.org/10.1016/0168-9002(90)91832-V}{\emph{Nucl. Instrum.
  Meth. A} {\bfseries 294} (1990) 179}.

\bibitem{King:1991gs}
B.J.~King et~al., \emph{{Measuring Muon Momenta with the CCFR Neutrino
  Detector}}, \href{https://doi.org/10.1016/0168-9002(91)90408-I}{\emph{Nucl.
  Instrum. Meth. A} {\bfseries 302} (1991) 254}.

\bibitem{Spentzouris:1998pf}
{\scshape CCFR/NuTeV} collaboration, \emph{{Precision electroweak results from
  neutrino nucleon scattering at CCFR/NuTeV}},
  \href{https://doi.org/10.1016/S0920-5632(98)00022-X}{\emph{Nucl. Phys. B
  Proc. Suppl.} {\bfseries 66} (1998) 112}.

\bibitem{Magill:2016hgc}
G.~Magill and R.~Plestid, \emph{{Neutrino Trident Production at the Intensity
  Frontier}}, \href{https://doi.org/10.1103/PhysRevD.95.073004}{\emph{Phys.
  Rev. D} {\bfseries 95} (2017) 073004}
  [\href{https://arxiv.org/abs/1612.05642}{{\ttfamily 1612.05642}}].

\bibitem{vonWeizsacker:1934nji}
C.F.~von Weizsacker, \emph{{Radiation emitted in collisions of very fast
  electrons}}, \href{https://doi.org/10.1007/BF01333110}{\emph{Z. Phys.}
  {\bfseries 88} (1934) 612}.

\bibitem{Williams:1934ad}
E.J.~Williams, \emph{{Nature of the high-energy particles of penetrating
  radiation and status of ionization and radiation formulae}},
  \href{https://doi.org/10.1103/PhysRev.45.729}{\emph{Phys. Rev.} {\bfseries
  45} (1934) 729}.

\bibitem{Belusevic:1987cw}
R.~Belusevic and J.~Smith, \emph{{W - Z Interference in Neutrino - Nucleus
  Scattering}}, \href{https://doi.org/10.1103/PhysRevD.37.2419}{\emph{Phys.
  Rev. D} {\bfseries 37} (1988) 2419}.

\bibitem{Vysotsky:2002ix}
M.I.~Vysotsky, I.V.~Gaidaenko and V.A.~Novikov, \emph{{On lepton pair
  production in neutrino nucleus collisions}},
  \href{https://doi.org/10.1134/1.1508695}{\emph{Phys. Atom. Nucl.} {\bfseries
  65} (2002) 1634}.

\bibitem{Altmannshofer:2014pba}
W.~Altmannshofer, S.~Gori, M.~Pospelov and I.~Yavin, \emph{{Neutrino Trident
  Production: A Powerful Probe of New Physics with Neutrino Beams}},
  \href{https://doi.org/10.1103/PhysRevLett.113.091801}{\emph{Phys. Rev. Lett.}
  {\bfseries 113} (2014) 091801}
  [\href{https://arxiv.org/abs/1406.2332}{{\ttfamily 1406.2332}}].

\bibitem{Denton:2024glz}
P.B.~Denton and J.~Gehrlein, \emph{{A Modern Look at the Oscillation Physics
  Case for a Neutrino Factory}},
  \href{https://arxiv.org/abs/2407.02572}{{\ttfamily 2407.02572}}.

\bibitem{ICAL:2015stm}
{\scshape ICAL} collaboration, \emph{{Physics Potential of the ICAL detector at
  the India-based Neutrino Observatory (INO)}},
  \href{https://doi.org/10.1007/s12043-017-1373-4}{\emph{Pramana} {\bfseries
  88} (2017) 79} [\href{https://arxiv.org/abs/1505.07380}{{\ttfamily
  1505.07380}}].

\bibitem{ParticleDataGroup:2024cfk}
{\scshape Particle Data Group} collaboration, \emph{{Review of particle
  physics}}, \href{https://doi.org/10.1103/PhysRevD.110.030001}{\emph{Phys.
  Rev. D} {\bfseries 110} (2024) 030001}.

\bibitem{Alloul:2013bka}
A.~Alloul, N.D.~Christensen, C.~Degrande, C.~Duhr and B.~Fuks, \emph{{FeynRules
  2.0 - A complete toolbox for tree-level phenomenology}},
  \href{https://doi.org/10.1016/j.cpc.2014.04.012}{\emph{Comput. Phys. Commun.}
  {\bfseries 185} (2014) 2250}
  [\href{https://arxiv.org/abs/1310.1921}{{\ttfamily 1310.1921}}].

\bibitem{Alwall:2011uj}
J.~Alwall, M.~Herquet, F.~Maltoni, O.~Mattelaer and T.~Stelzer, \emph{{MadGraph
  5 : Going Beyond}},
  \href{https://doi.org/10.1007/JHEP06(2011)128}{\emph{JHEP} {\bfseries 06}
  (2011) 128} [\href{https://arxiv.org/abs/1106.0522}{{\ttfamily 1106.0522}}].

\bibitem{Hahn:2000kx}
T.~Hahn, \emph{{Generating Feynman diagrams and amplitudes with FeynArts 3}},
  \href{https://doi.org/10.1016/S0010-4655(01)00290-9}{\emph{Comput. Phys.
  Commun.} {\bfseries 140} (2001) 418}
  [\href{https://arxiv.org/abs/hep-ph/0012260}{{\ttfamily hep-ph/0012260}}].

\bibitem{MERTIG1991345}
R.~Mertig, M.~Böhm and A.~Denner, \emph{Feyncalc - computer-algebraic
  calculation of feynman amplitudes},
  \href{https://doi.org/https://doi.org/10.1016/0010-4655(91)90130-D}{\emph{Computer
  Physics Communications} {\bfseries 64} (1991) 345}.

\bibitem{Shtabovenko:2016sxi}
V.~Shtabovenko, R.~Mertig and F.~Orellana, \emph{{New Developments in FeynCalc
  9.0}}, \href{https://doi.org/10.1016/j.cpc.2016.06.008}{\emph{Comput. Phys.
  Commun.} {\bfseries 207} (2016) 432}
  [\href{https://arxiv.org/abs/1601.01167}{{\ttfamily 1601.01167}}].

\bibitem{Shtabovenko:2020gxv}
V.~Shtabovenko, R.~Mertig and F.~Orellana, \emph{{FeynCalc 9.3: New features
  and improvements}},
  \href{https://doi.org/10.1016/j.cpc.2020.107478}{\emph{Comput. Phys. Commun.}
  {\bfseries 256} (2020) 107478}
  [\href{https://arxiv.org/abs/2001.04407}{{\ttfamily 2001.04407}}].

\end{thebibliography}

\providecommand{\href}[2]{#2}\begingroup\raggedright\endgroup
\end{document}